\documentclass[10pt,preprint]{aastex}

\newcommand{\myemail}{keivan.stassun@vanderbilt.edu}
\newcommand{\bbmax}{BB$_{\rm max}$}
\newcommand{\bbmin}{BB$_{\rm min}$}
\newcommand{\ewca}{EW[\ion{Ca}{2}]}
\newcommand{\prot}{P_{\rm opt}}

\shorttitle{PMS X-ray and Optical Variability}
\shortauthors{Stassun et al.}

\slugcomment{To appear in the Astrophysical Journal}

\begin{document}

\title{A Simultaneous Optical and X-ray Variability Study of
the Orion Nebula Cluster. I. Incidence of Time-Correlated 
X-ray/Optical Variations\footnote{This study is part of the {\it Chandra} 
Orion Ultradeep Project (COUP).}}

\author{Keivan G.\ Stassun\altaffilmark{2},
M.\ van den Berg\altaffilmark{3},
Eric Feigelson\altaffilmark{4},
Ettore Flaccomio\altaffilmark{5}}

\altaffiltext{2}{Department of Physics \& Astronomy,
Vanderbilt University, Nashville, TN 37235; \myemail}
\altaffiltext{3}{Harvard-Smithsonian Center for Astrophysics, 
60 Garden St., Cambridge, MA 02138}
\altaffiltext{4}{Department of Astronomy \& Astrophysics, 
Penn State University, University Park, PA 16802}
\altaffiltext{5}{INAF-Osservatorio Astronomico di Palermo Giuseppe
S.\ Vaiana, Piazza del Parlamento 1, 91034 Palermo, Italy}

\begin{abstract}
We present a database of $BVRI$ time-series photometry of the
Orion Nebula Cluster obtained with two ground-based telescopes
at different longitudes to provide simultaneous coverage with the
13-d {\it Chandra} observation of the cluster. The resulting database
of simultaneous optical and X-ray light curves for some 800
pre--main-sequence (PMS) stars represents, by a factor of 
hundreds, the largest synoptic, multi--wavelength-regime, time-series
study of young stars to date. This database will permit detailed
analyses of the relationship between optical and X-ray variability 
among a statistically significant ensemble of PMS stars, 
with the goal of elucidating the
origins of PMS X-ray production. In this first paper,
we present the optical observations, describe the combined X-ray/optical
database, and perform an analysis of time-correlated variability in 
the optical and X-ray light curves. We identify
40 stars (representing 5\% of our study sample) with possible
time-correlated optical and X-ray variability. 
Examples of both positive and
negative time-correlations are found, possibly representing X-ray
flares and persistent coronal features associated with both cool and
hot surface spots (i.e.\ magnetically active regions and accretion
shocks). We also find two possible examples of ``white-light" flares
coincident with X-ray flares; these may correspond to the impulsive
heating phase in solar-analog flares.  However, though interesting,
these represent unusual cases.  More generally, we find
very little evidence to suggest a direct causal link between 
the sources of optical and X-ray variability in PMS stars. The 
conclusion that accretion is a primary driver of X-ray production 
in PMS stars is not supported by our findings.
\end{abstract}


\section{Introduction\label{intro}}
The X-ray luminosities of solar-type stars, when they are very young,
are prodigious. Indeed, it is now well established that low-mass
pre--main-sequence (PMS) stars can emit X-rays at levels up to $\sim
10^4$ times that of the present-day Sun \citep[e.g.][]{feig-mont}. But
after two decades of research since T~Tauri stars (TTSs) were first
identified as strong X-ray sources \citep[e.g.][]{feig-decamp},
we still do not know the origins of X-ray production in 
young stars. 

While flares observed in the X-ray light curves suggest that the X-rays
are produced mainly in solar-analog magnetic
reconnection flares, the absence of an obvious rotation-activity
relation in these stars prevents a clear association of the X-rays
to dynamo-generated fields \citep{flac-basic,feig03,stass04}. The flares 
can be so powerful that non-solar geometries of the magnetic fields
must be considered, including field lines linking the star with the
protoplanetary disk \citep{favata05}.
In addition, the role of accretion is unclear, and may affect the
production of X-rays either directly or indirectly. For example,
energetic shocks associated with accretion flows at or near the stellar 
surface may enhance X-ray production, as has been suggested in the detailed
spectroscopic analyses of TW~Hya \citep{kastner02,stelzer04} and
BP~Tau \citep{schmitt05}. At the same time, the mass-loading of magnetic
field lines may actually inhibit X-ray production in accreting
systems, as suggested by \citet{preibisch05}.
These issues are reviewed more fully in \citet{feig-ppv}.

We have undertaken a synoptic study of 
the Orion Nebula
Cluster (ONC) combining a nearly continuous, 13-d {\it Chandra}
observation with simultaneous, multi-wavelength, time-series
photometry in the optical. 
A specific motivation is to
establish the incidence of time-correlated X-ray and optical variability
in order to begin disentangling the roles of magnetic and accretion activity
in the production of X-rays.
For example, accretion shocks on the surfaces of
classical TTSs (CTTSs; TTSs that still possess accretion disks)
produce strong variability at optical wavelengths, caused by time-variable
accretion, rotational modulation of accretion ``hot spots," or both
\citep{herbst94,bouvier97,stasswood99}. 
If X-rays are produced near the sites of accretion shocks,
then one might expect that changes
in the strength of these shocks will induce changes in the strength
of the X-ray emission, and that X-ray and optical variability might
therefore be correlated in time. Alternatively, X-rays may originate
in coronal structures associated with magnetically active regions on
these stars, in which case we might expect that X-ray emission will
be more pronounced when the magnetically active regions (which appear
as dark photospheric spots in the optical) face toward us, and least
pronounced when these regions face away from us; that is, optical
and X-ray variability will be anti-correlated in time. 
More complex behavior may result from the superposition of these
effects, emphasizing the need for time-resolved measurements.
More generally,
by establishing the frequency with which X-ray and optical variations
are co-temporal, we seek to constrain the extent to which the sources
of the X-ray and optical variability may be co-spatial or otherwise
causally linked.

Detailed analyses of simultaneous, time-resolved X-ray and optical variability
have previously been reported for three PMS stars: the CTTSs
V773~Tau \citep{feig94} and BP~Tau \citep{gullbring97}, and the
weak-lined TTS (WTTS) V410~Tau \citep{stelzer03}. In
none of these were the optical and X-ray variability found to
be time-correlated. Unfortunately, in all three cases the statistical
significance of the results were limited by the short duration of
the X-ray observations: V773~Tau was observed by ROSAT for 6.7~hr,
BP~Tau was observed by ROSAT for 8~hr, and V410~Tau
was observed by {\it Chandra} for 7.2~hr.

The combined X-ray/optical dataset reported here includes some 
800 members of the ONC and thus represents, by a factor
of hundreds, the largest attempt to date to study the relationship
between X-ray and optical variability in PMS stars. 
We will report results from this database in stages, focusing first
on statistical properties of the study sample as a whole.
In this first paper, we establish the incidence of
{\it time-correlated} optical and X-ray variability. 
A companion paper 
\citep[][hereafter Paper~II]{stass06} focuses on
{\it time-averaged} measures of optical and X-ray variability, and
on the relationship between these variability indicators and basic
stellar properties.
This database will 
then form the basis for follow-up studies focusing on fine-grained analyses
of individual objects. 

In \S \ref{data}, we present the X-ray and optical
time-series database that forms the basis for these studies and 
describe the methods that we employ in this paper
to search for time-correlated X-ray/optical variability. In \S
\ref{corr-time}, we identify 40 candidate stars with possible
time-correlated optical and X-ray variability, and classify them
according to common light-curve characteristics to guide follow-up
study of these interesting objects.
Importantly, these candidates
represent at most $\sim 5$\% of our study sample; $\sim 95$\%
of the stars in our sample exhibit no evidence for
time-correlated optical and X-ray variability. 
As we discuss in \S \ref{discussion}, these results imply that
the sites of optical and X-ray variability in PMS stars are not---in 
the vast majority of cases---instantaneously one and the same. A direct
causal link between the sources of optical and X-ray variability in
PMS stars is not supported by our study.
We summarize our findings in \S \ref{conclusions}.

\section{A Database of Simultaneous, Optical and X-ray 
Light Curves of the ONC\label{data}}

\subsection{X-ray observations: The COUP\label{data-xray}}
As the foundation for the synoptic X-ray/optical database used in this 
and follow-up studies, we use the deep X-ray observation of
the ONC obtained by \dataset[ADS/Sa.CXO#obs/COUP]{the {\it Chandra}
Orion Ultradeep Project (COUP)} collaboration. Details of the COUP
dataset are described elsewhere \citep{getman05}.  Briefly,
the ONC was observed by the {\it Chandra} ACIS detector 
\citep[energy passband of 0.5--10 keV;][]{getman05} from 8 Jan
2003 to 21 Jan 2003 for a total of 849~ks (9.8 days), representing
the deepest X-ray observation of a star-forming region ever
obtained. There were five small gaps of $\sim 29$ ks each (when the
spacecraft passed through the Van Allen belts), but the data stream is
otherwise continuous over a total timespan of 13.2 days.  Within the
ACIS 17' $\times$ 17' field of view centered on the Trapezium, 1616
point sources were identified, of which $\simeq 1400$ are associated
ONC stars \citep{getman05b}. For each of these, \citet{getman05}
performed a spectral extraction and analysis from which they derived
the source's X-ray luminosity $(L_X)$ and other parameters. 
Most importantly for our present purposes, they also
performed an analysis of photon arrival times, 
providing measures of each source's X-ray variability and
yielding for each source a nearly continuous X-ray light curve 
spanning 13.2-d.

The COUP database also includes a compilation of a number of
other physical characteristics for each source from photometric and
spectroscopic catalogs already in the literature. For our purposes
in this and follow-up studies, the most relevant of these is 
the equivalent width (EW) of the \ion{Ca}{2}
infrared triplet from \citet{hill98}, which serves as a measure
of the mass accretion rate in each object.

\subsection{Simultaneous optical observations\label{data-opt}}
To complement this unique and rich database of X-ray light curves and
ancillary data, we organized a simultaneous, multi-band, ground-based
monitoring campaign, with the aim of obtaining well-sampled optical
light curves that overlap well with the X-ray light curves. Using small
telescopes at two observing sites to minimize diurnal observing gaps,
we repeatedly imaged the same field as that observed by {\it Chandra}
ACIS, cycling through the $BVRI$ filters, and alternating between
short and long exposure times in each filter to maximize the dynamic
range of the observations. Constraints in scheduling the ground-based
observations together with uncertainties in the scheduling of the {\it
Chandra} observations prevented us from obtaining optical data at the
start of the {\it Chandra} exposure. The optical observations begin 6
days after the start of the 13.2-d X-ray observation, and continue
for 2 days beyond, for a total overlap of 7.2~d.

We observed with the WIYN 0.9-m telescope at the Kitt Peak National
Observatory (KPNO) in Arizona, USA, over the period UT 2003 Jan
15--25. Due to poor weather, no data were obtained on the night of UT
2003 Jan 18. 
At the same time, we observed
with the 1.5-m Cassini telescope in Loiano, Italy, over the period
UT 2003 Jan 15--22, with no data obtained on the
nights of UT 2003 Jan 16, 17, 20, 21. 
Exposure times were varied slightly to compensate for variable
observing conditions, but were otherwise similar at both 
sites. Typical exposure times were 5~s in $BVRI$ for short
exposures, 420~s for long exposures in $V$ and $R$, and 720~s for
long exposures in $B$ and $I$.  We typically obtained a total of $\sim 8$
measurements per filter per night, in both short and long exposures,
with a cadence of $\sim 1$~hr$^{-1}$. 
A log of the optical observations is presented in Table \ref{obs-table}.

The images were reduced, and instrumental magnitudes
for all point sources extracted, using standard IRAF\footnote{IRAF
is distributed by the National Optical Astronomy Observatory,
which is operated by the Association of Universities for Research
in Astronomy, Inc., under cooperative agreement with the National
Science Foundation.} procedures. Differential light curves were
determined from PSF photometry using an algorithm for
inhomogeneous ensemble photometry \citep{honeycutt}
as implemented in \citet{stass99,stass02} for observations of
high-nebulosity regions such as the ONC. 
Systematic offsets between the KPNO and Loiano observations were adjusted
by comparing the instrumental magnitudes of non-variable stars. 
All differential light curves are available in electronic 
format from the authors. Calibrated mean optical magnitudes for each star
are reported in the COUP database.

In order
to cross-identify the optical sources with the X-ray sources in the
COUP database, we first determined the optical identifications of the
stars from the catalogs of \citet{hill97} and \citet{herbst02} using a
1\farcs5 matching radius. We then used these optical identifications to
associate each object with its X-ray counterpart in the COUP database
using the table of cross-identifications provided by \citet{getman05}.
In this way, we were able to unambiguously associate 814 optical
sources with X-ray counterparts in the COUP database.

These 814 objects span a very large range of apparent magnitudes
and colors, with $7.3 < I < 18.4$ and $-0.1 < (V-I) < 5.7$, and a
correspondingly large range of spectral types and extinctions,
B3 $<$ SpTy $<$ M7 and $0.0 < A_V < 10.8$ \citep{hill97}. This sample
is thus representative of the underlying ONC population; only
the very massive, heavily obscured, and sub-stellar populations are excluded.

An unavoidable consequence of this breadth of stellar properties is 
that objects do not appear
with uniform frequency in our optical light curves. For example,
the brightest objects appear only in the short-exposure light curves,
being saturated in the long-exposure frames. Similarly, the coolest and
the most heavily reddened objects appear only in the long-wavelength
passbands.  In addition, some objects near the bright and faint
extremes of our survey have more sparsely sampled light curves, as
variations in observing conditions (e.g.\ seeing, sky transparency,
background) cause these objects to be lost from some exposures. Thus,
any individual object in our final database possesses between 1
and 8 different optical light curves ($BVRI$ filters, long
and short exposures) with time sampling patterns that are very similar
but, for the reasons already noted, not identical. Of course, objects
whose characteristics are intermediate to the extremes discussed
above possess the most complete set of optical light curves and the
most complete time sampling.

\subsection{Measures of variability\label{methods}}
The stars in our database exhibit
a large variety of behaviors in their optical and X-ray variability.
Figs.\ \ref{coup1499}--\ref{coup132} show
the optical and X-ray light curves for a selection of objects
as a representative visual summary of the range and types of 
variability observed. 
Here we describe the methods we use to measure and to correlate
variability in the X-ray and optical light curves in our database.

\subsubsection{Time-averaged variability indicators}
\begin{figure}[ht]
\epsscale{0.9}
\plotone{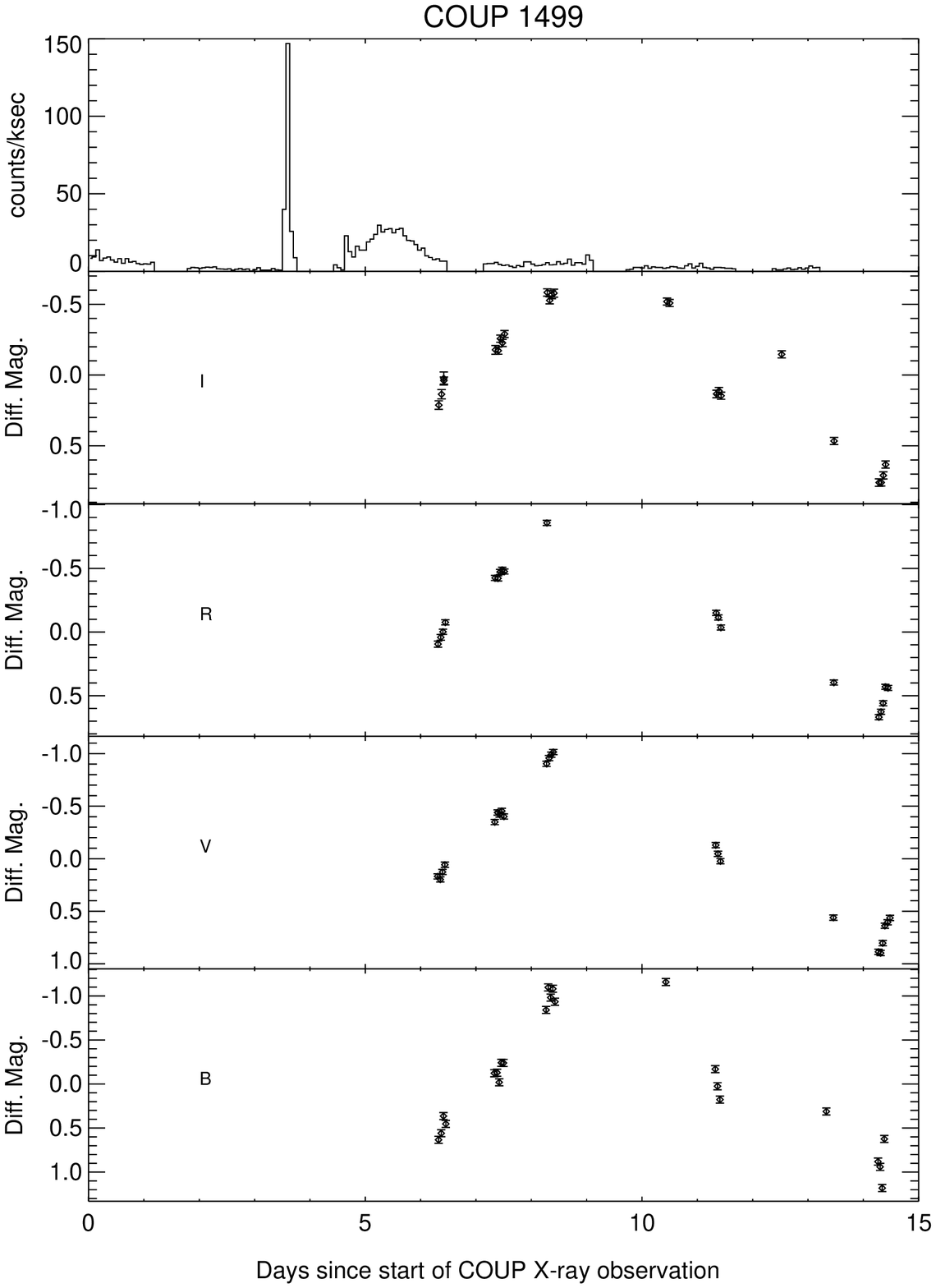}
\caption{\label{coup1499}
X-ray and optical light curves for COUP 1499. 
The top panel shows the COUP X-ray light curve, while the other panels 
show the optical light curves. 
This star has a $J$ statistic value of 45.4, corresponding
to a peak-to-peak $V$-band amplitude of $\approx 2.1$ mag.
The X-ray variability of this star is $\log$(\bbmax/\bbmin) $= 2.45$.
See \S \ref{methods} for a description of these variability measures.}
\end{figure}

\begin{figure}[ht]
\plotone{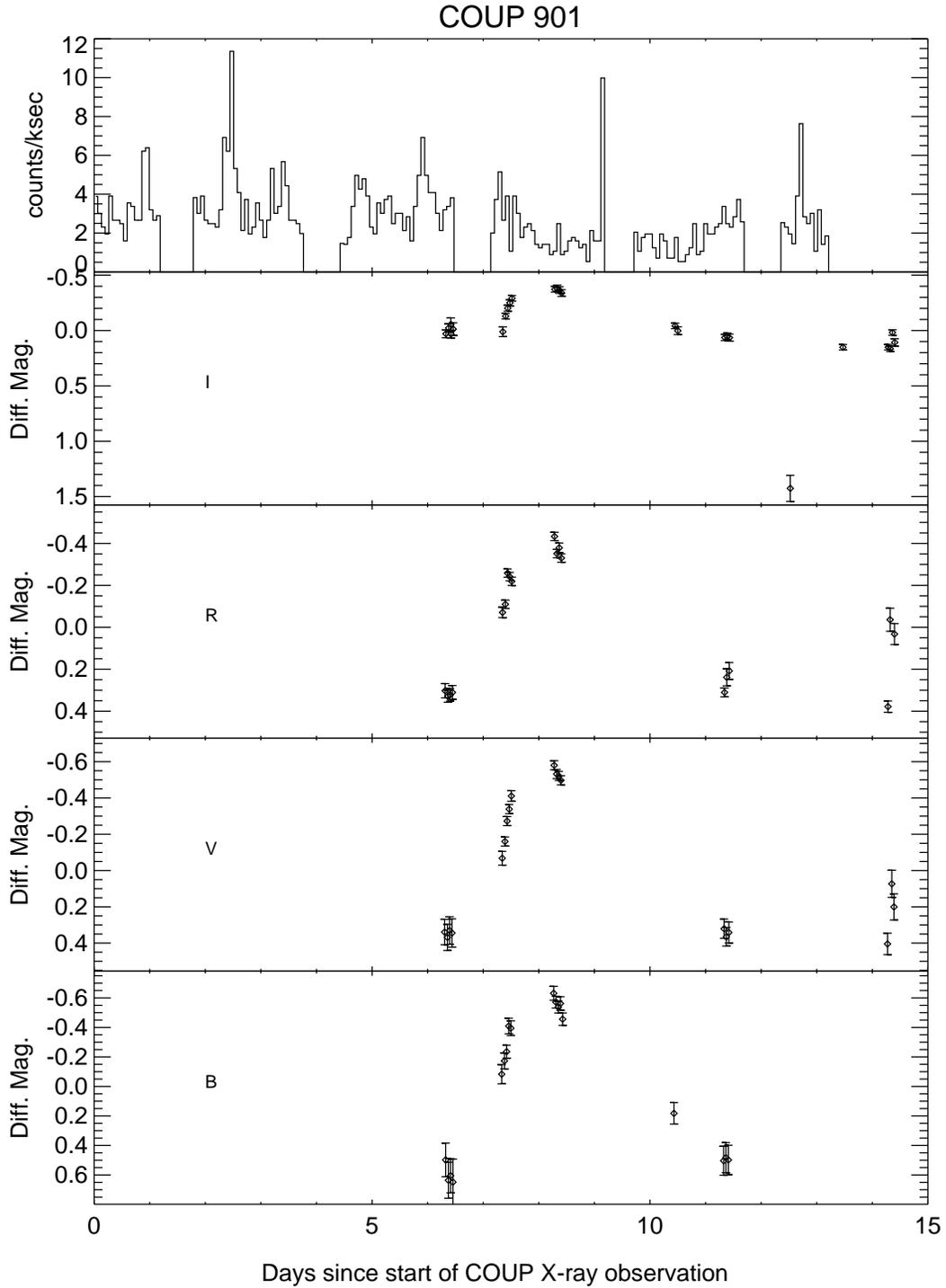}
\caption{\label{coup901}
Same as Fig.\ \ref{coup1499}, but for 
COUP 901, which has a $J$ statistic value of 28.3,
corresponding to a peak-to-peak $V$-band amplitude of $\approx 1.4$ mag.
The X-ray variability of this star is $\log$(\bbmax/\bbmin) $=0.38$. 
In the optical light curves, small circles represent
KPNO data, large circles represent Loiano data; see \S \ref{data-opt}
for details about the optical observations.}
\end{figure}

\begin{figure}[ht]
\plotone{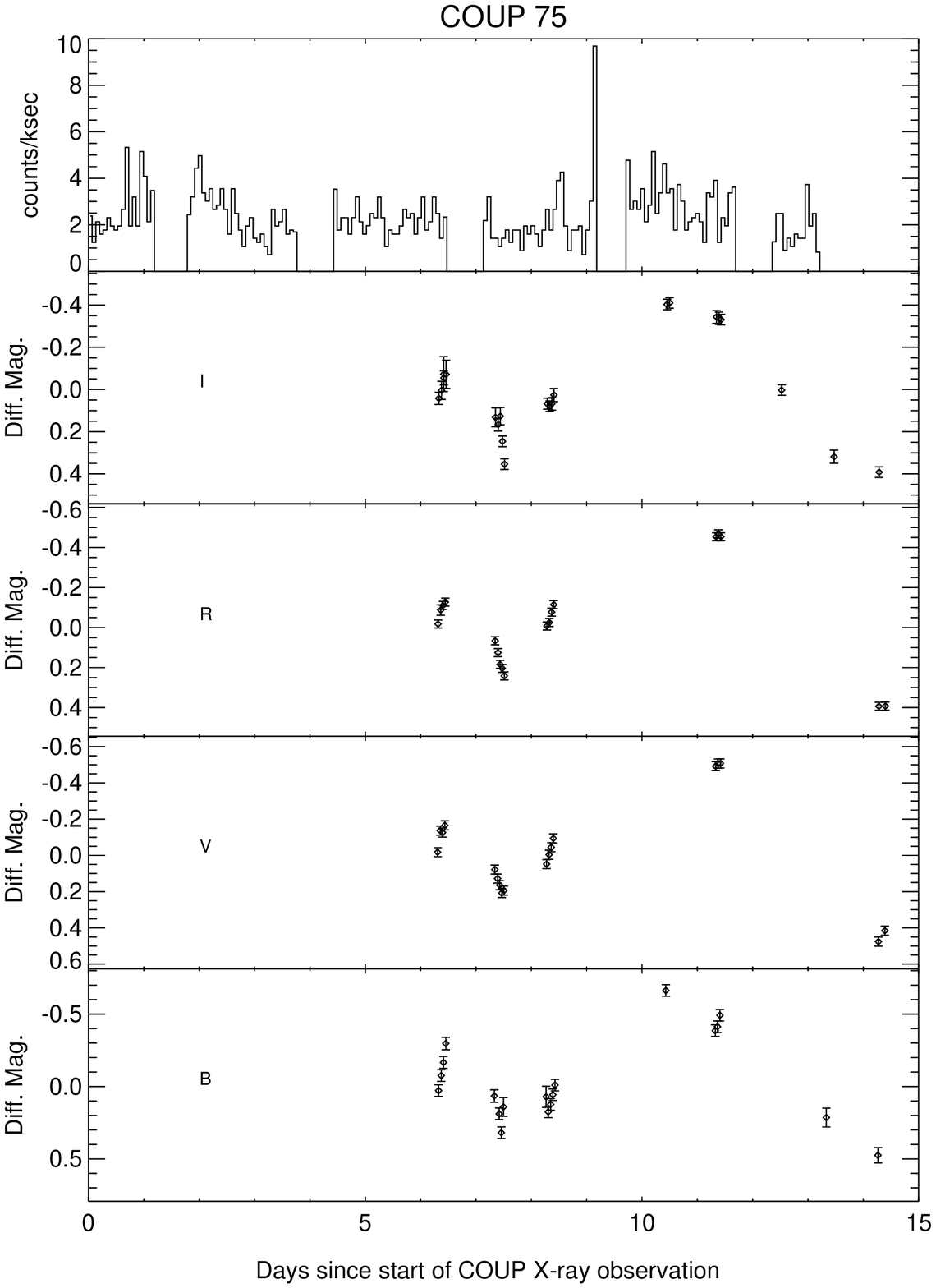}
\caption{\label{coup75}
Same as Fig.\ \ref{coup901}, but for
COUP 75, which has a $J$ statistic value of 21.9,
corresponding to a peak-to-peak $V$-band amplitude of $\approx 1.0$ mag.
The X-ray variability of this star is $\log$(\bbmax/\bbmin) $=0.26$. }
\end{figure}

\begin{figure}[ht]
\plotone{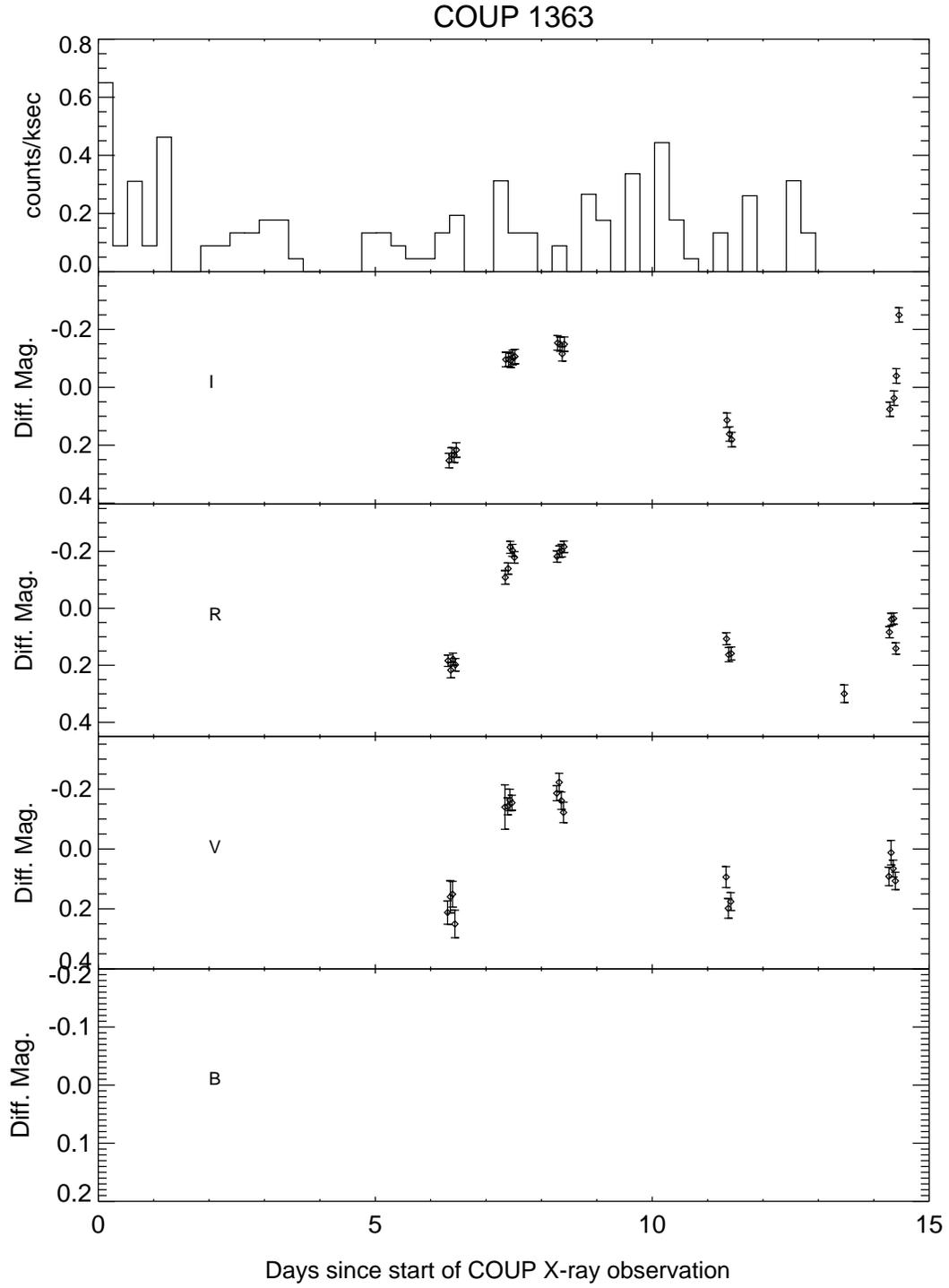}
\caption{\label{coup1363}
Same as Fig.\ \ref{coup901}, but for
COUP 1363, which has a $J$ statistic value of 13.1,
corresponding to a
peak-to-peak $V$-band amplitude of $\approx 0.4$ mag.
The X-ray variability of this star is $\log$(\bbmax/\bbmin) $=0.81$. 
No $B$-band light curve is available for this star.}
\end{figure}

\begin{figure}[ht]
\plotone{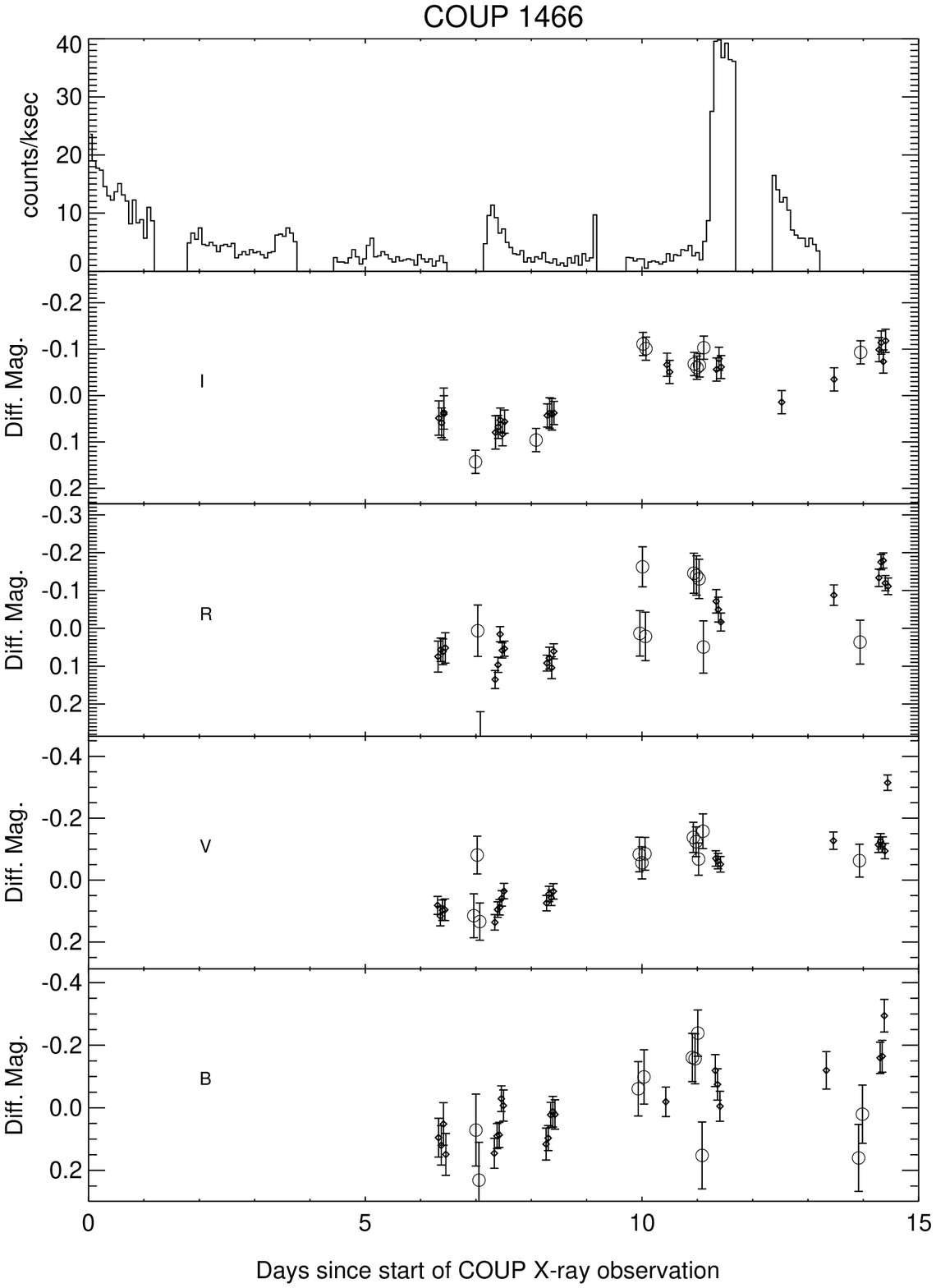}
\caption{\label{coup1466}
Same as Fig.\ \ref{coup901}, but for
COUP 1466, which has a $J$ statistic value of 7.7,
corresponding to a peak-to-peak $V$-band amplitude of $\approx 0.2$ mag.
The X-ray variability of this star is $\log$(\bbmax/\bbmin) $=1.36$. }
\end{figure}

\begin{figure}[ht]
\plotone{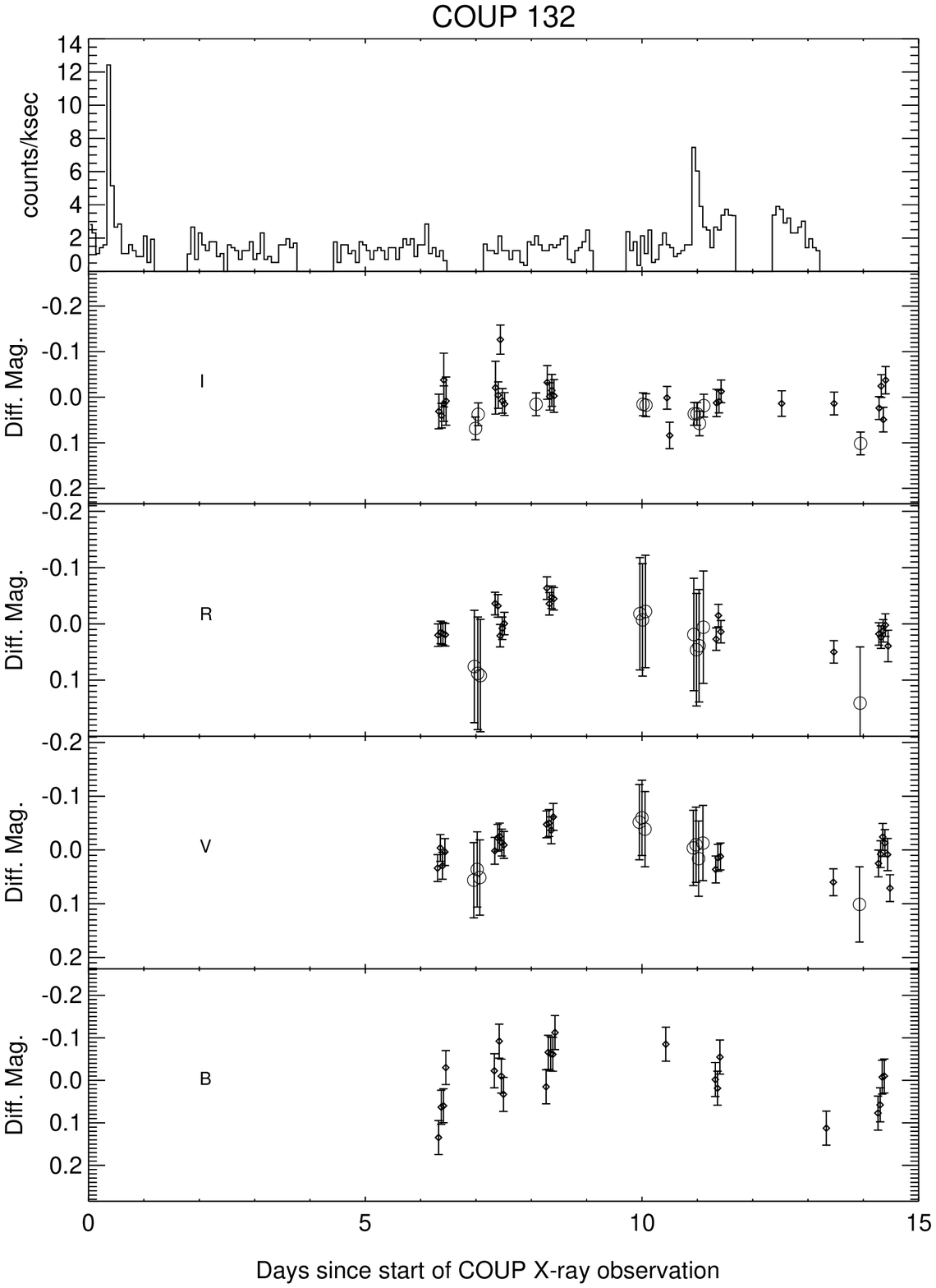}
\caption{\label{coup132}
Same as Fig.\ \ref{coup901}, but for
COUP 132, which has a $J$ statistic value of 1.7,
corresponding to a
peak-to-peak $V$-band amplitude of $\approx 0.1$ mag.
The X-ray variability of this star is $\log$(\bbmax/\bbmin) $=1.53$. }
\end{figure}

To quantify the time-averaged X-ray variability of the stars in
our database, we use the
Bayesian Block (BB) analysis provided by the COUP 
\citep{getman05}. Briefly, the BB analysis \citep{scargle98}
segments each light curve into
the maximum number of time blocks such that the differences in the
mean flux levels of the blocks are statistically significant. The
BB analysis thus yields a robust measure of the maximum and minimum
flux levels in the source's segmented X-ray light curve (\bbmax\ and \bbmin,
respectively); we use the statistic $\log($\bbmax/\bbmin$)$
to quantify the magnitude of variability in the X-ray light curve.
As discussed in Paper~II, 556 (68\%) of the 814 stars in our 
database are X-ray variable at greater than 99.9\%
confidence according to the BB statistic, with a range 
of amplitudes $0.12 < \log($\bbmax/\bbmin$) < 2.89$. The six objects
displayed in Figs.\ \ref{coup1499}--\ref{coup132} are representative
of this range, with $0.25 < \log($\bbmax/\bbmin$) < 2.45$.

To characterize the time-averaged optical variability of these stars,
we use the $J$ statistic described by \citet{stetson96}. Briefly, the
$J$ statistic is similar to the standard $\chi^2$ statistic,
however
the $J$ statistic simultaneously
considers the measurements from all of the light curves of a given
object. 
As such, the $J$ statistic is more robust against outliers,
especially when multiple light curves are available in a given
passband (i.e.\ long and short exposures).
As discussed in Paper~II, 
358/814 (44\%) of the stars in our database are 
optically variable at greater than 99.9\% confidence according to
the $J$ statistic, with a range $1.5 < J < 45.5$, 
corresponding to peak-to-peak variability in the $V$ 
band ranging from a maximum of $\gtrsim 2$ mag to a minimum of 
$\lesssim 0.05$ mag (upper limit). 
The six objects displayed in Figs.\ \ref{coup1499}--\ref{coup132}
are representative of this range, with $1.7 < J < 45.4$.
With this range of variability
amplitudes, and with variability timescales ranging from hours to days, 
the optical variability 
of our sample is typical of what has been observed before in optical
variability studies of TTSs \citep[e.g.][]{herbst94}.

We defer to Paper~II a full analysis of the relationships between
these time-averaged variability indicators, but note here that the light
curves shown in Figs.\ \ref{coup1499}--\ref{coup132} are
representative of the behavior observed in our database as a whole.
In general, we find examples of optical and X-ray variability in
every possible permutation: 
We find stars that are highly variable in both optical and X-ray
light (e.g.\ Fig.\ \ref{coup1499}), stars that are strongly
variable in the optical but non-variable in X-rays (e.g.\ Fig.\ 
\ref{coup75}) and vice-versa (e.g.\ Fig.\ \ref{coup132}), and 
stars that are non-variable altogether. 

\subsubsection{Time-correlated variability\label{corr}}
For each of the stars in our database, we calculate 
Kendall's $\tau$ statistic to quantify the degree to which
variability in the X-ray and optical light curves are
correlated in time.
Kendall's $\tau$ is a standard non-parametric statistical
test for the degree of association between two variables, and has a
known probability distribution. Kendall's $\tau$ test rank-orders
the time-matched optical magnitudes and X-ray fluxes, 
computes the degree to which the optical and X-ray ``ranks"
are correlated, and calculates the probability that the measured
correlation is consistent with the null hypothesis of no correlation.
Measurement errors are ignored in computing $\tau$.

To allow for the possibility that time-correlated variability
may be a function of wavelength,
we calculate Kendall's $\tau$ and
its associated null-hypothesis probability by correlating 
each star's X-ray light curve separately against each of its 
associated optical light curves. Where
multiple light curves are available in a given passband (i.e.\ long
and short exposures), we select the light curve with the smallest
ratio of mean photometric error to total number of measurements, 
$\left<\sigma\right>/N$, i.e.\ the light curve with the best combination
of signal-to-noise and time sampling.
For a given optical light curve, we match each optical data point
with the corresponding time bin in the star's X-ray light curve.
In total, our search for time-correlated variability in our sample
of 814 stars involves 1944 different $BVRI$ light curves.

In the analysis that follows,
we take Kendall's $\tau$ probabilities of less than 0.01 as
suggestive of time-correlated optical and X-ray variability. The sign
of the $\tau$ statistic indicates whether the optical and X-ray light 
curves are positively or negatively time-correlated.

None of the examples shown in Figs.\ \ref{coup1499}--\ref{coup132} 
exhibit statistically significant time-correlated optical and
X-ray variability. For instance, at the same time that COUP 1499 
(Fig.\ \ref{coup1499}) shows variability of $\gtrsim 2$ mag in the 
optical, its X-ray light curve is almost completely non-variable.
To be sure, this star is one of the strongest X-ray variables in
our database, with $\log$(\bbmax/\bbmin) $ =2.45$.
%but this X-ray variability does not occur
%{\it simultaneously} with the optical variability.
%Indeed, this star's X-ray output is not only roughly constant while
%its optical flux is extremely variable, it is nearly non-existent;
The sources of X-ray and optical variability in this star are
evidently not simultaneously one and the same.
Figs.\ \ref{coup901}--\ref{coup1363} show similar examples,
while Figs.\ \ref{coup1466}--\ref{coup132} exhibit the reverse behavior,
with strong X-ray flaring events that are not simultaneously reflected
in the optical.
Such a lack of time-correlated optical and X-ray variability is a
general feature of the $\sim 800$ stars in our database, as we now
discuss.

\section{Results: Time-correlations of optical and X-ray light curves
\label{corr-time}}
In this section we present the results of our analysis
of time-correlated optical and X-ray variability in our database of 814
COUP sources with simultaneous optical and X-ray light curves. 
We identify 40 candidate stars whose optical and X-ray
light curves appear to be correlated or anti-correlated in time.
These represent the first such cases yet reported for PMS stars,
and we classify these objects according to
possible physical origins for the observed variability.
More generally, however, time-correlated variability evidently 
occurs very rarely in our study sample. 

\subsection{Incidence of time-correlated variability\label{corr-incidence}}
\begin{figure}[ht]
\figurenum{7a}
\epsscale{0.9}
\plotone{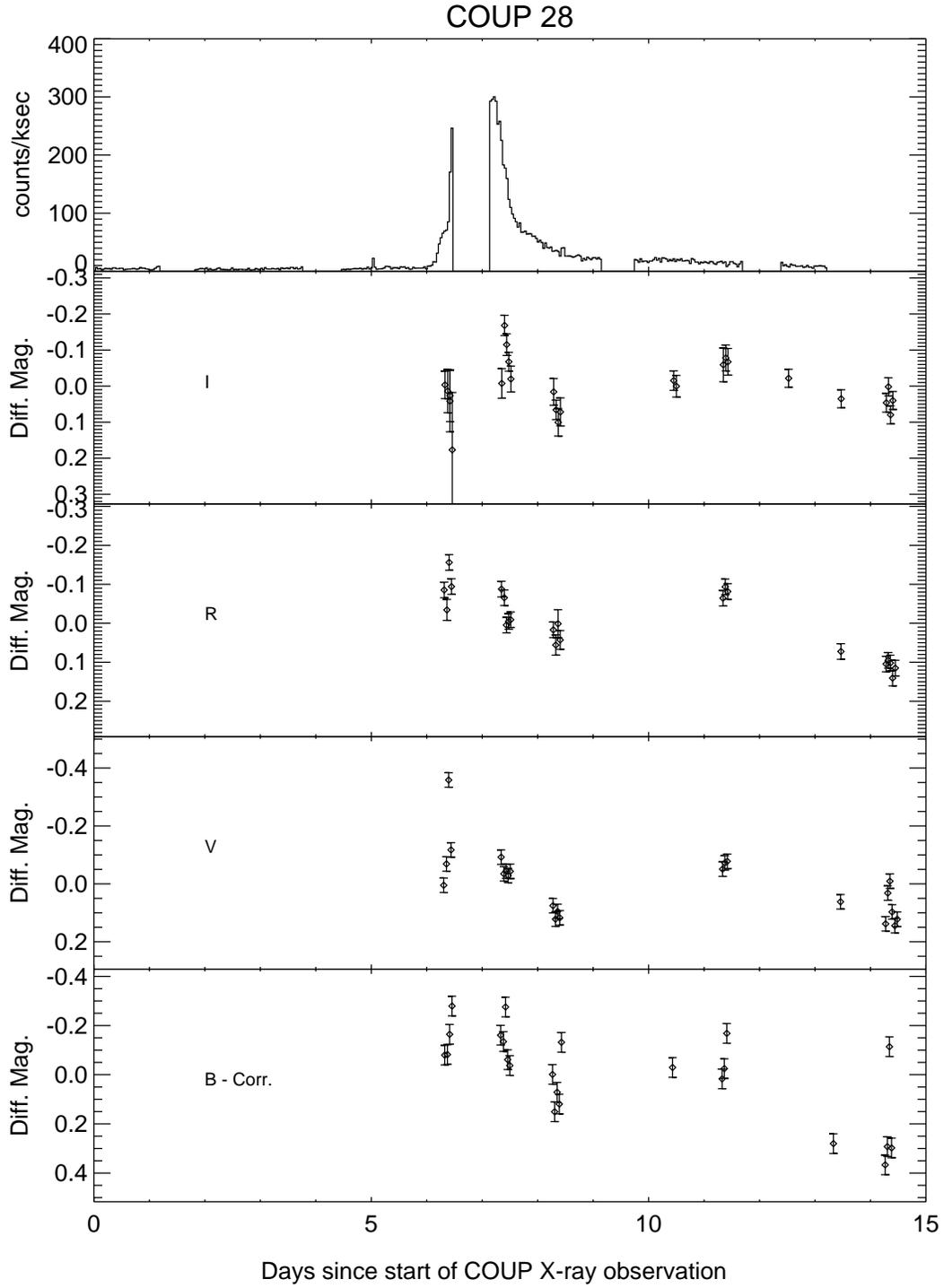}
\caption{\label{coup28}
Same as Fig.\ \ref{coup132}, but for COUP 28. 
Optical light curves
possibly time-correlated with the X-ray light curve (as indicated by
a Kendall's $\tau$ test; see \S \ref{corr}) are identified
with `Corr'. 
}
\end{figure}

\begin{figure}[ht]
\figurenum{7b}
\epsscale{0.9}
\plotone{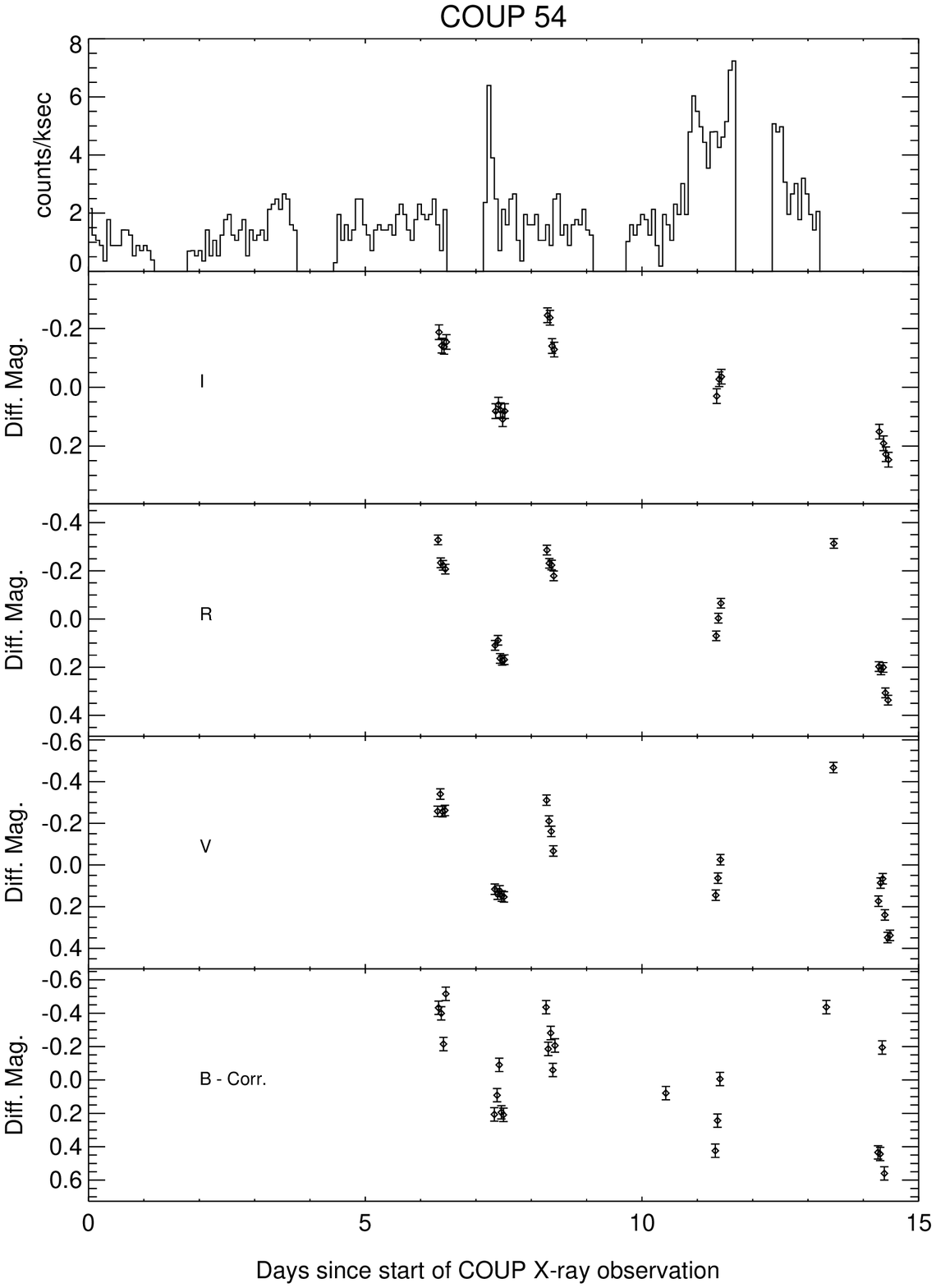}
\caption{\label{coup54}
Same as Fig.\ \ref{coup28}, but for COUP 54.
This figure appears in the electronic edition of the journal only.
}
\end{figure}

\begin{figure}[ht]
\figurenum{7c}
\epsscale{0.9}
\plotone{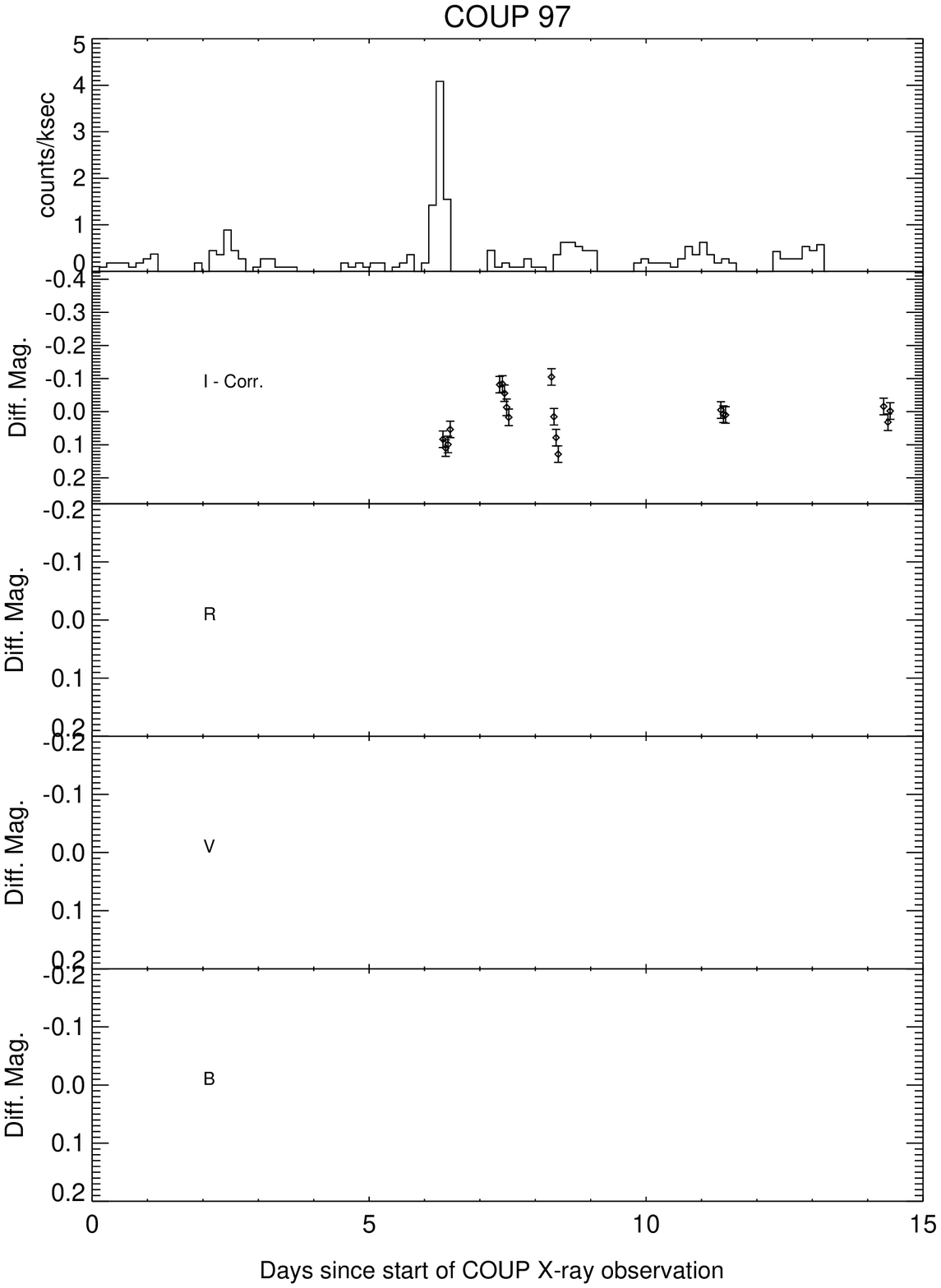}
\caption{\label{coup97}
Same as Fig.\ \ref{coup28}, but for COUP 97.
This figure appears in the electronic edition of the journal only.
}
\end{figure}

\begin{figure}[ht]
\figurenum{7d}
\epsscale{0.9}
\plotone{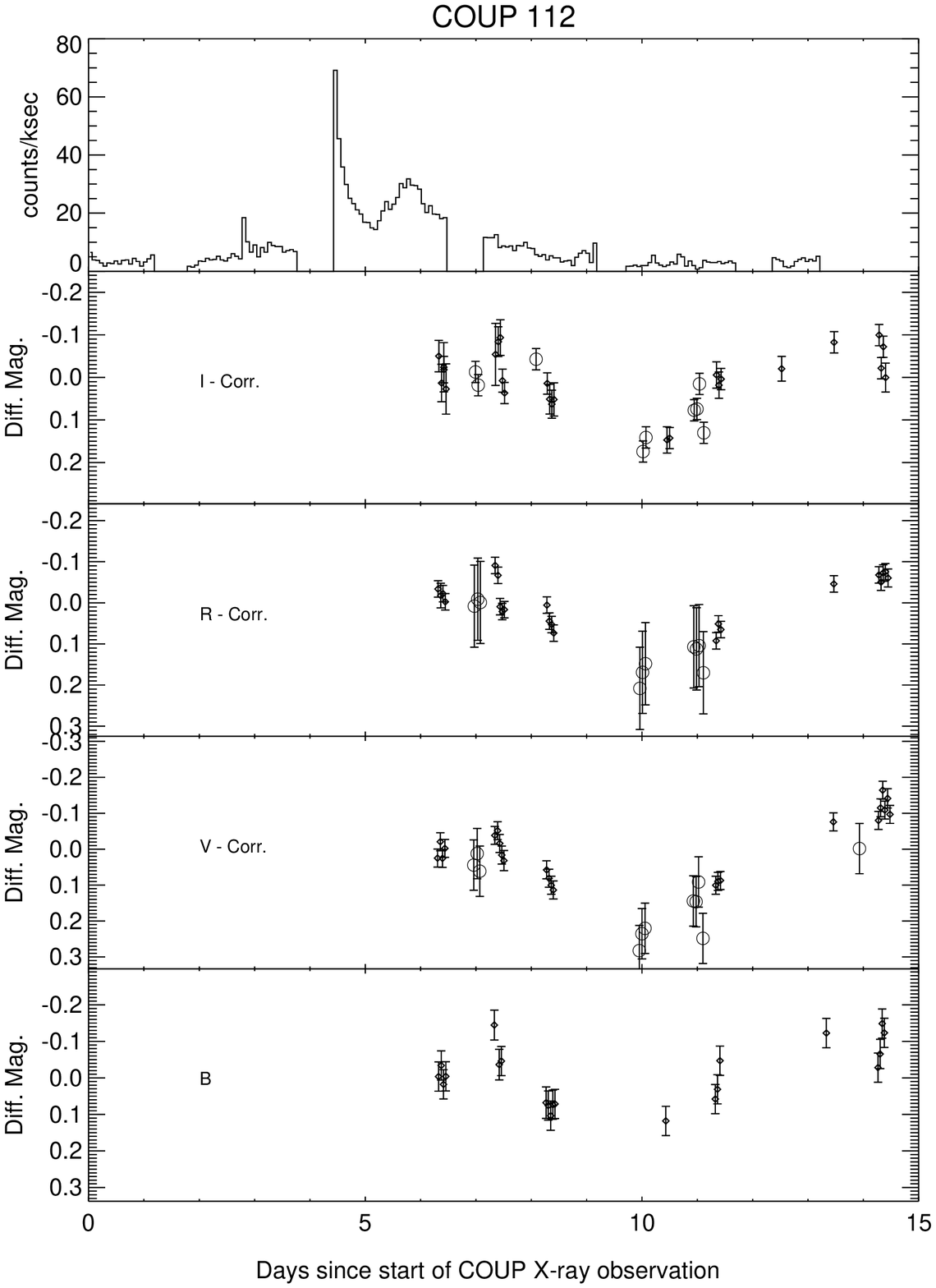}
\caption{\label{coup112}
Same as Fig.\ \ref{coup28}, but for COUP 112.
}
\end{figure}

\begin{figure}[ht]
\figurenum{7e}
\epsscale{0.9}
\plotone{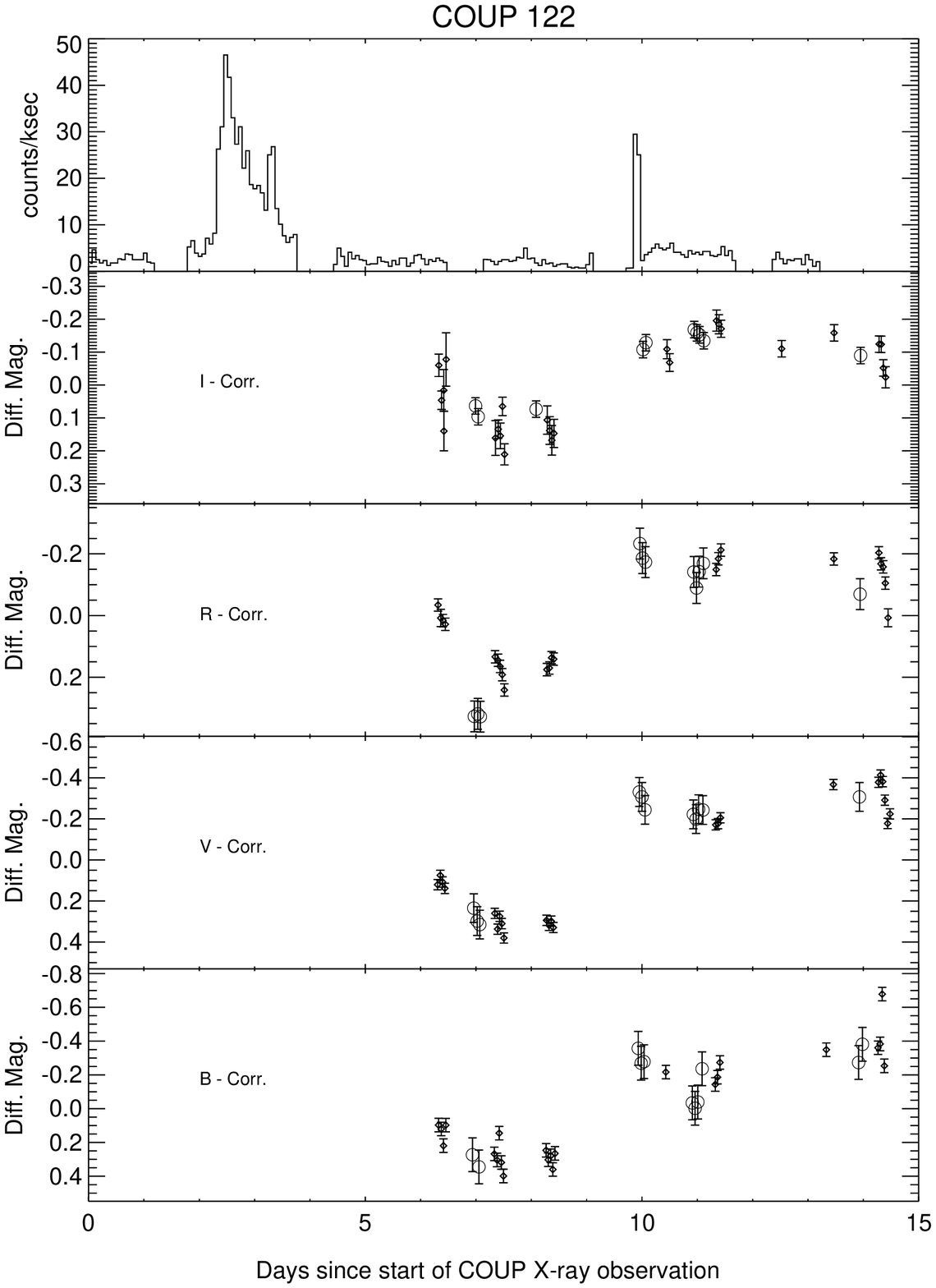}
\caption{\label{coup122}
Same as Fig.\ \ref{coup28}, but for COUP 122.
}
\end{figure}

\begin{figure}[ht]
\figurenum{7f}
\epsscale{0.9}
\plotone{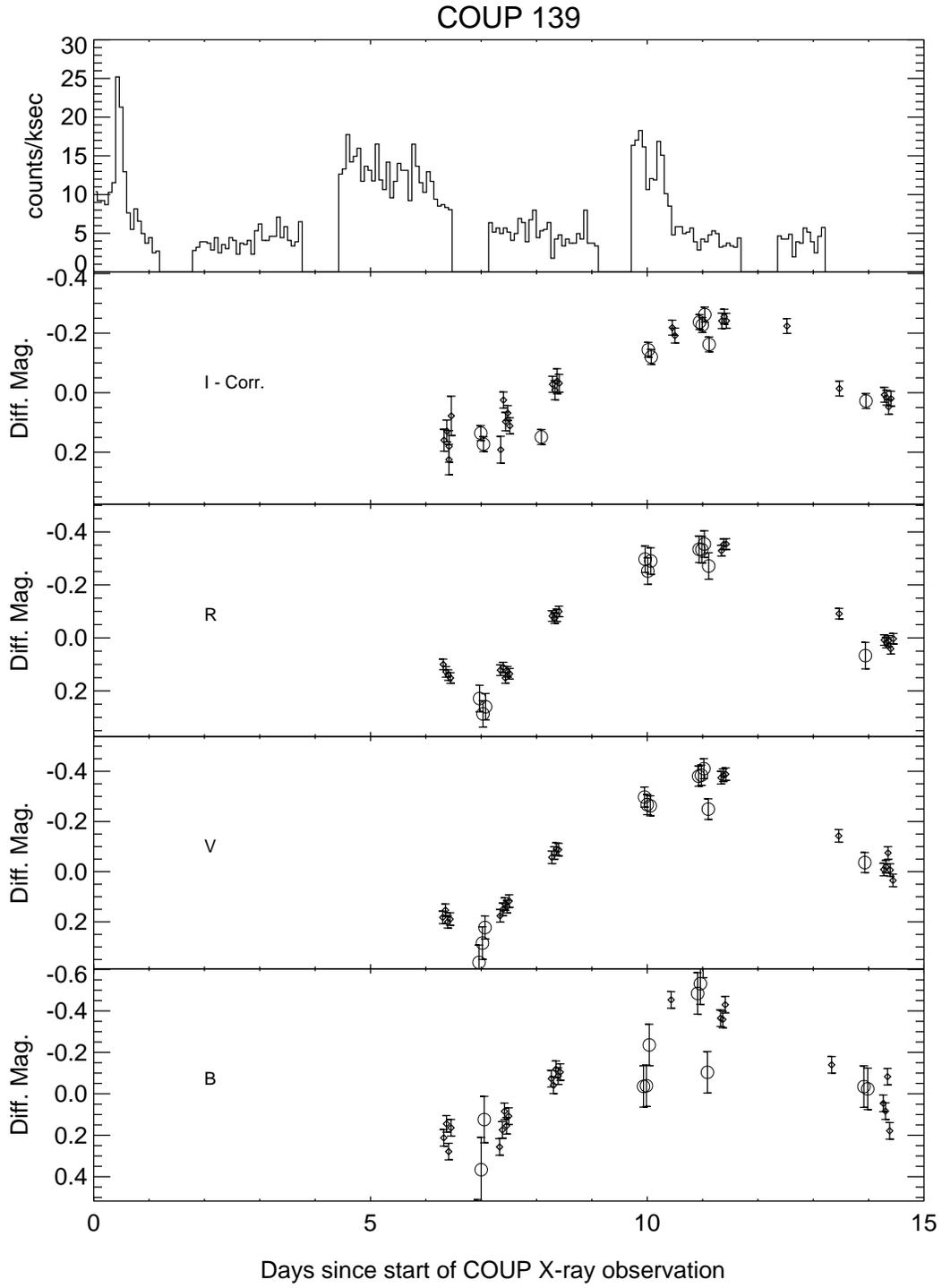}
\caption{\label{coup139}
Same as Fig.\ \ref{coup28}, but for COUP 139.
}
\end{figure}

\begin{figure}[ht]
\figurenum{7g}
\epsscale{0.9}
\plotone{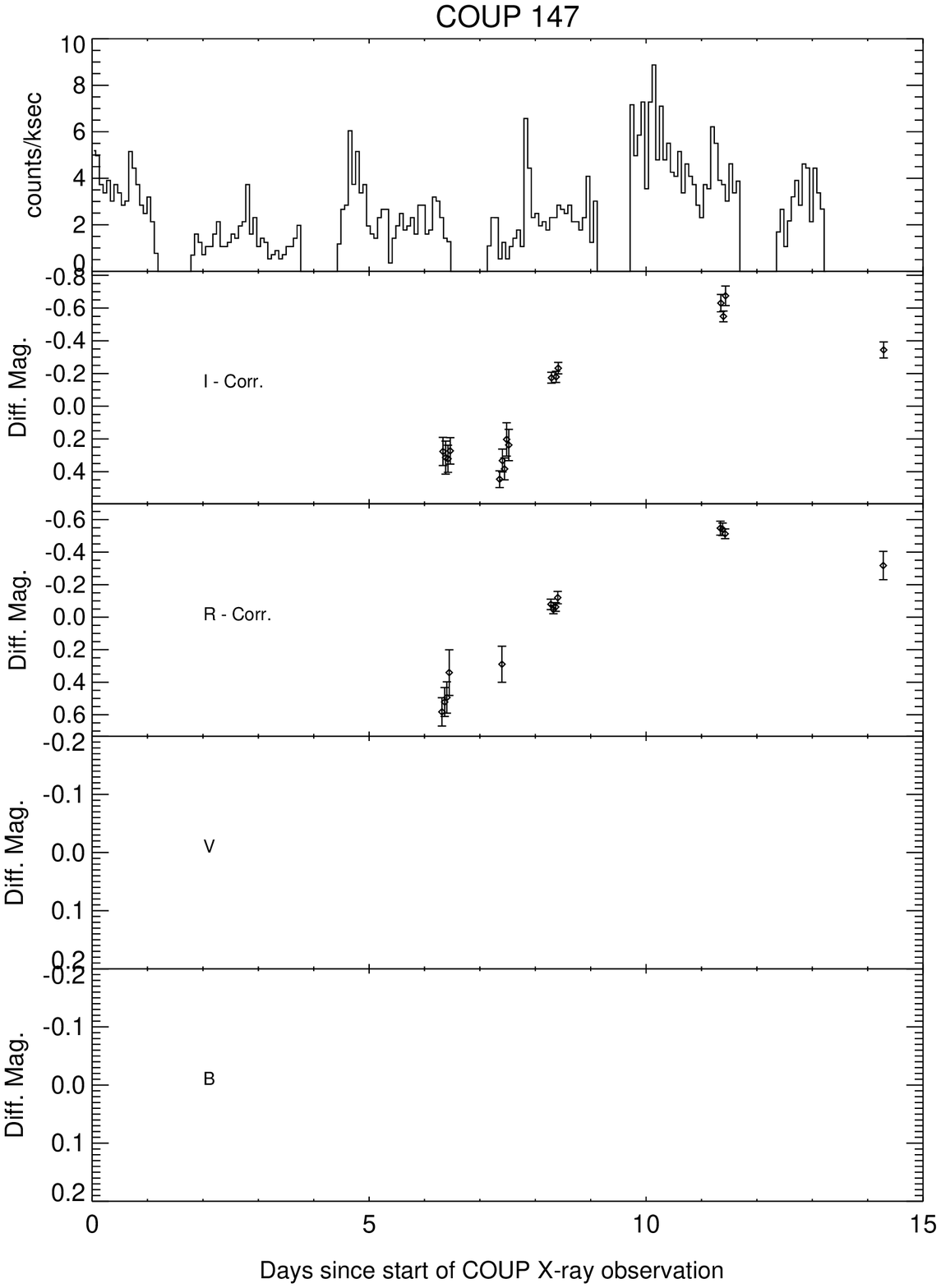}
\caption{\label{coup147}
Same as Fig.\ \ref{coup28}, but for COUP 147.
This figure appears in the electronic edition of the journal only.
}
\end{figure}

\begin{figure}[ht]
\figurenum{7h}
\epsscale{0.9}
\plotone{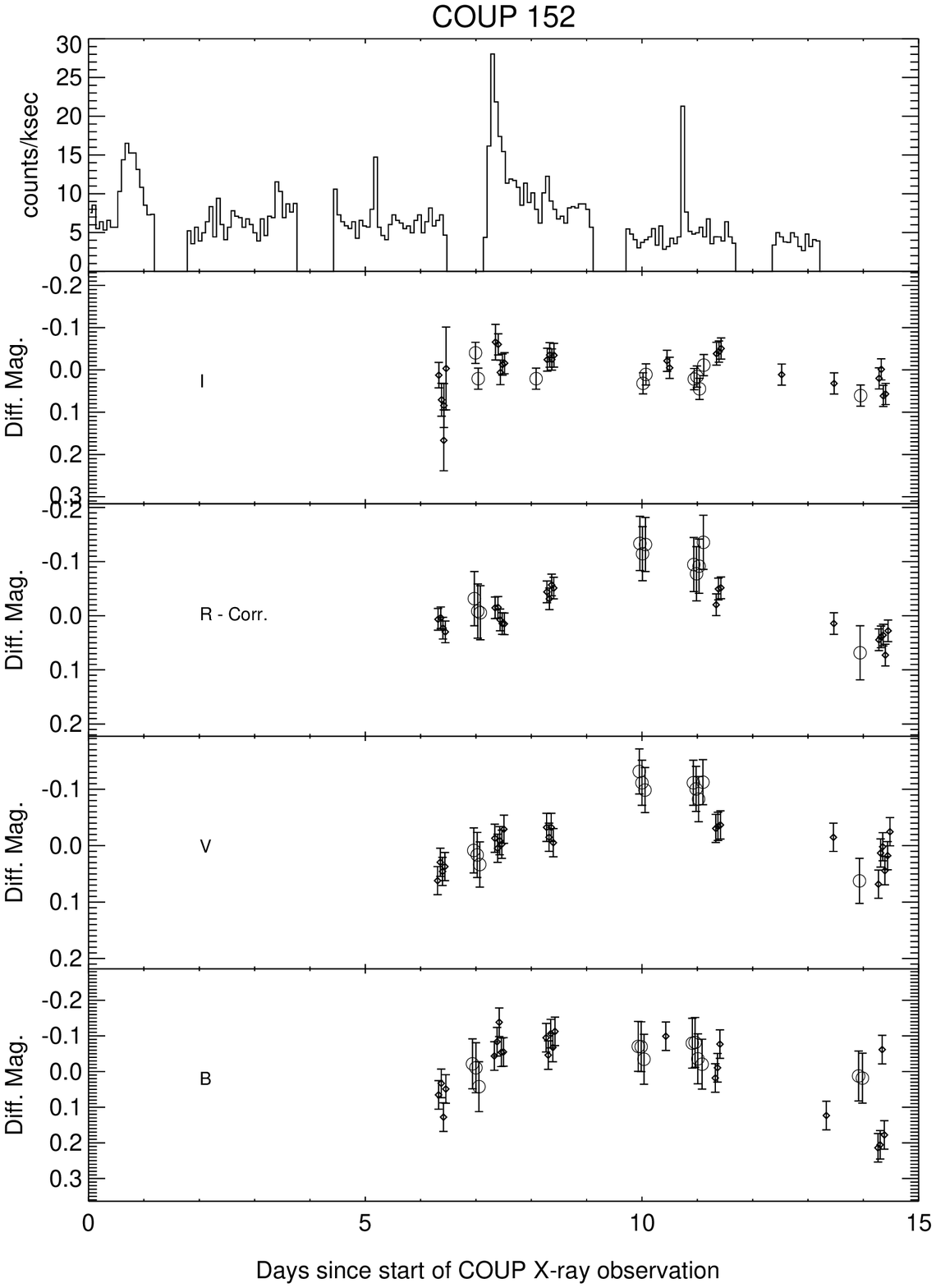}
\caption{\label{coup152}
Same as Fig.\ \ref{coup28}, but for COUP 152.
This figure appears in the electronic edition of the journal only.
}
\end{figure}

\begin{figure}[ht]
\figurenum{7i}
\epsscale{0.9}
\plotone{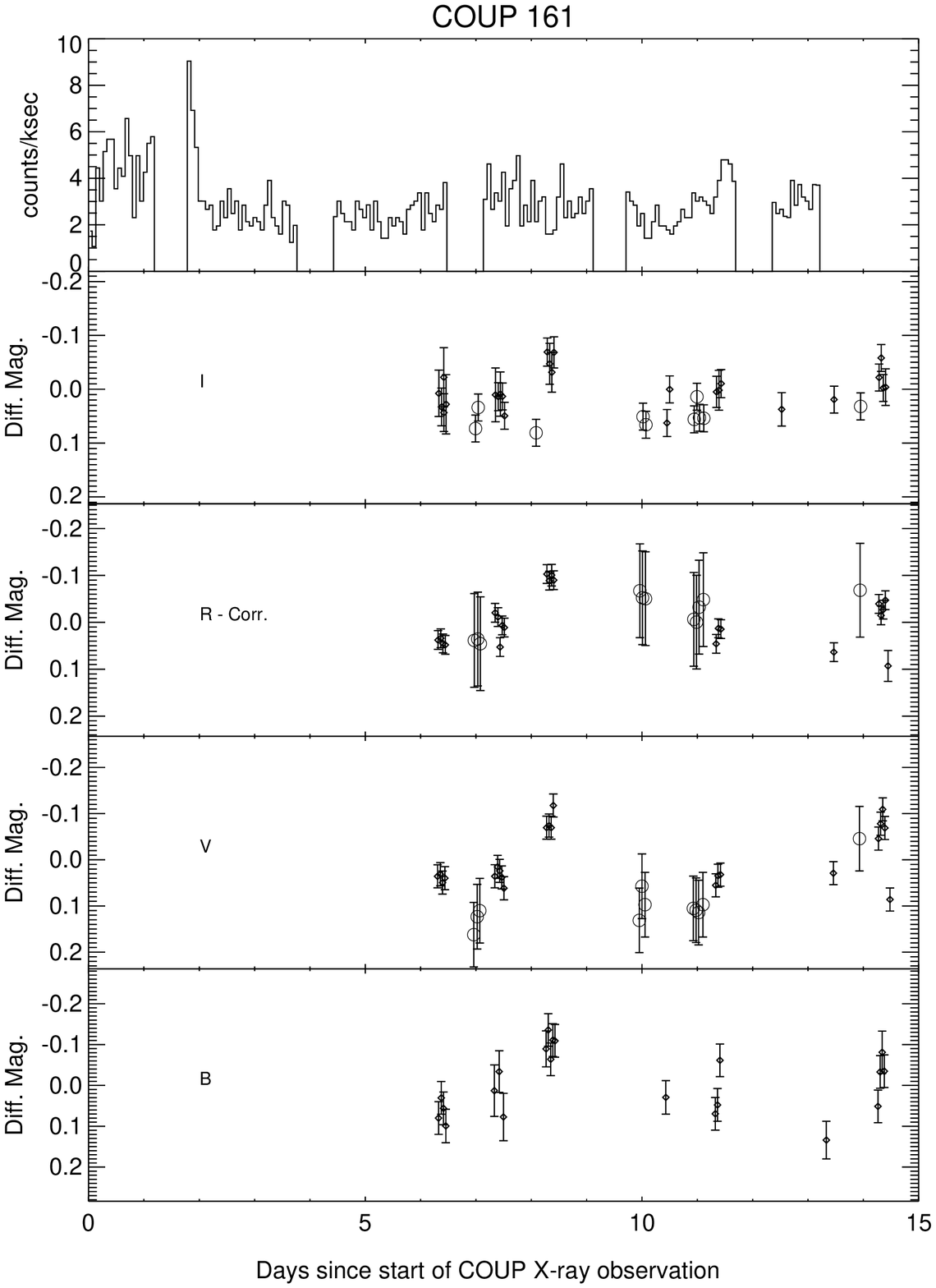}
\caption{\label{coup161}
Same as Fig.\ \ref{coup28}, but for COUP 161.
}
\end{figure}

\begin{figure}[ht]
\figurenum{7j}
\epsscale{0.9}
\plotone{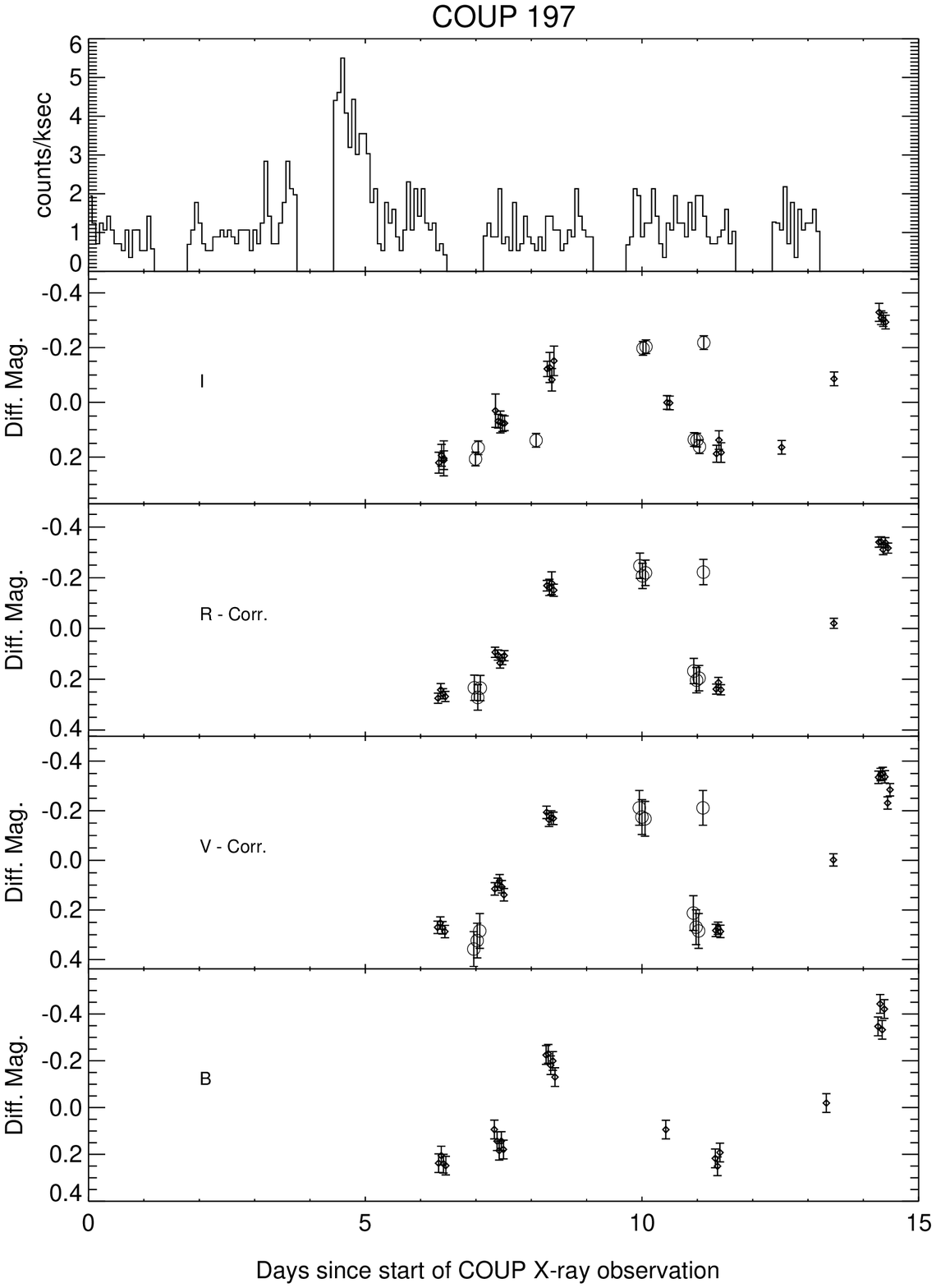}
\caption{\label{coup197}
Same as Fig.\ \ref{coup28}, but for COUP 197.
This figure appears in the electronic edition of the journal only.
}
\end{figure}

\begin{figure}[ht]
\figurenum{7k}
\epsscale{0.9}
\plotone{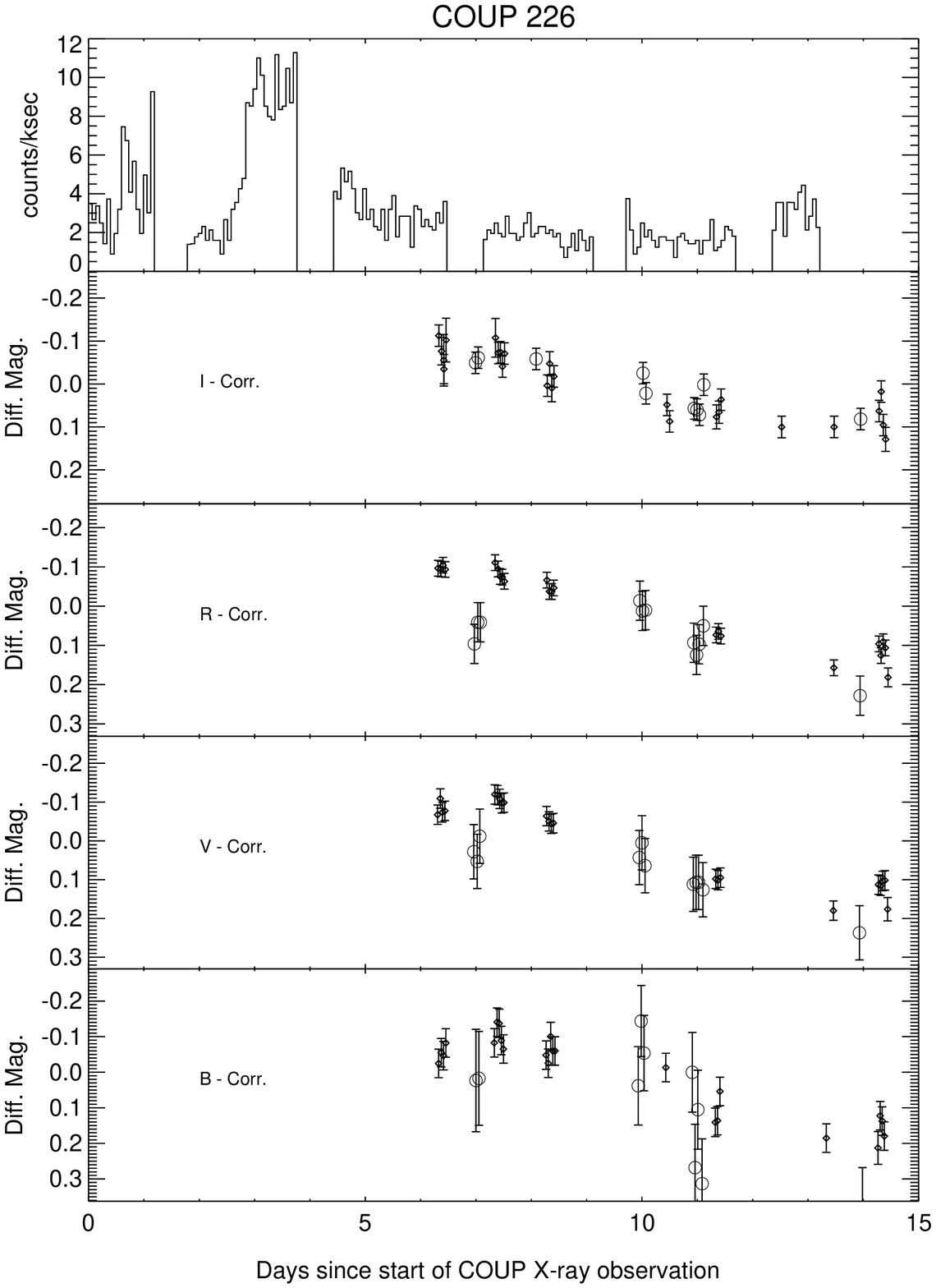}
\caption{\label{coup226}
Same as Fig.\ \ref{coup28}, but for COUP 226.
}
\end{figure}

\begin{figure}[ht]
\figurenum{7l}
\epsscale{0.9}
\plotone{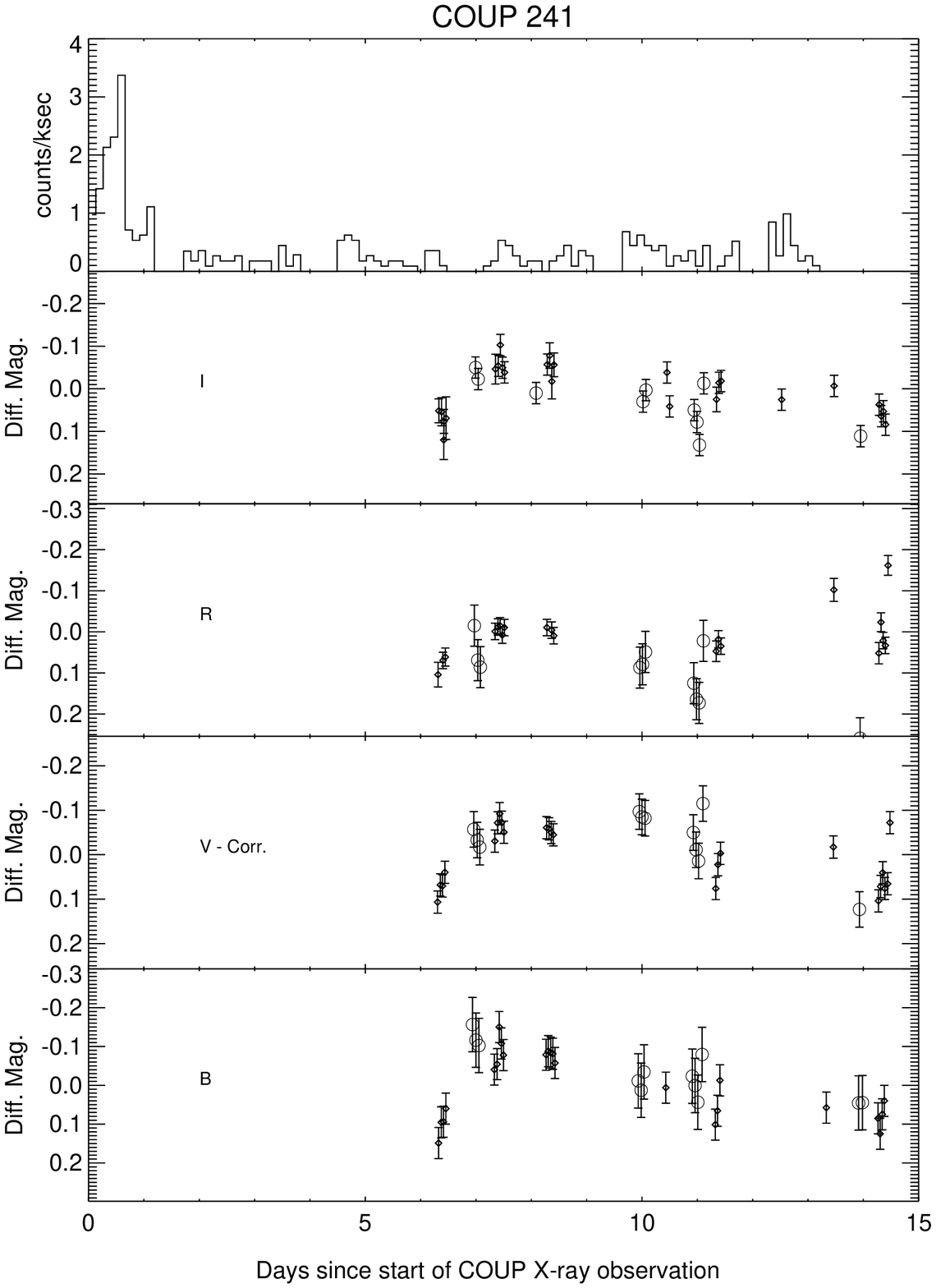}
\caption{\label{coup241}
Same as Fig.\ \ref{coup28}, but for COUP 241.
This figure appears in the electronic edition of the journal only.
}
\end{figure}

\begin{figure}[ht]
\figurenum{7m}
\epsscale{0.9}
\plotone{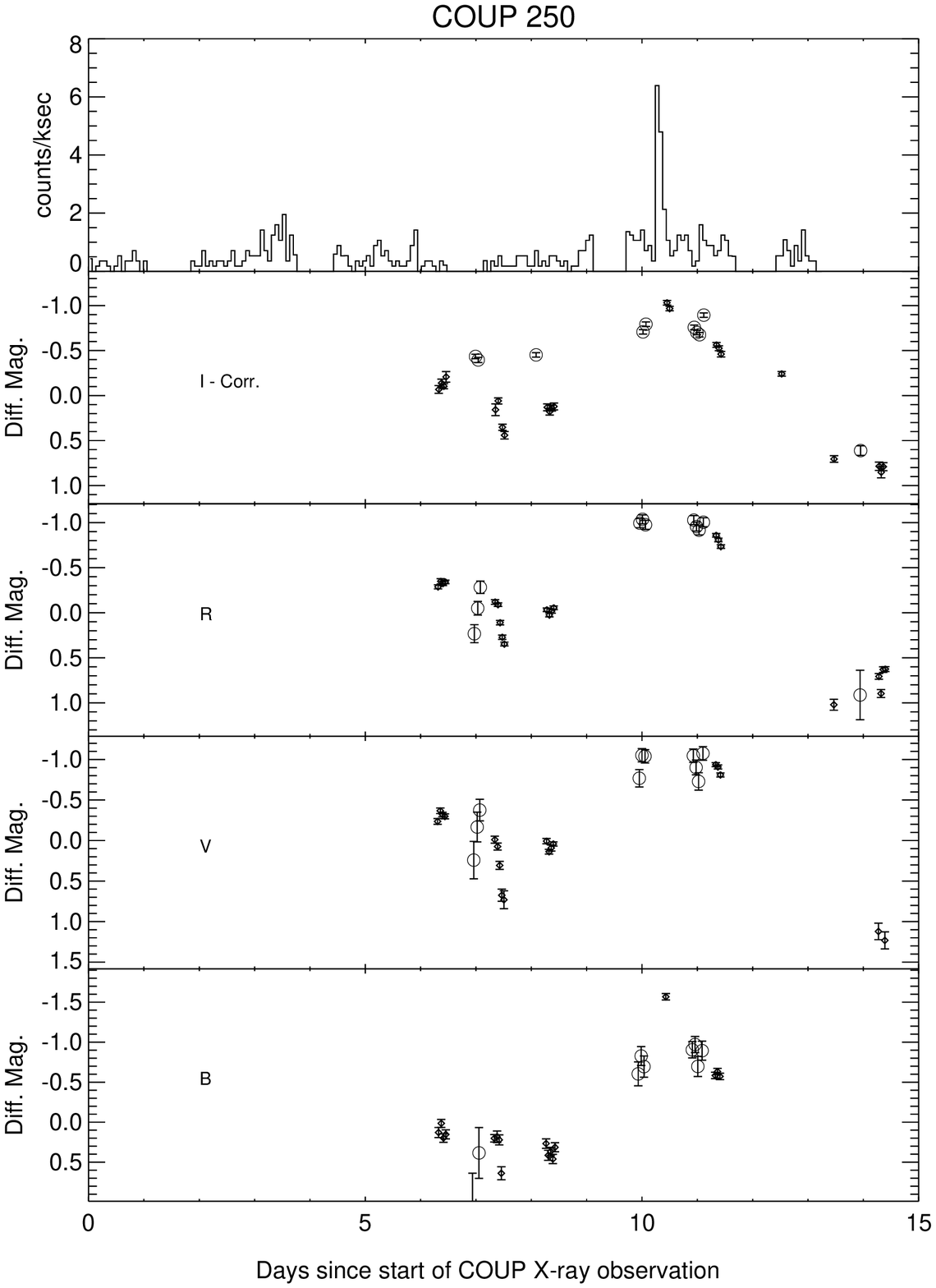}
\caption{\label{coup250}
Same as Fig.\ \ref{coup28}, but for COUP 250.
}
\end{figure}

\begin{figure}[ht]
\figurenum{7n}
\epsscale{0.9}
\plotone{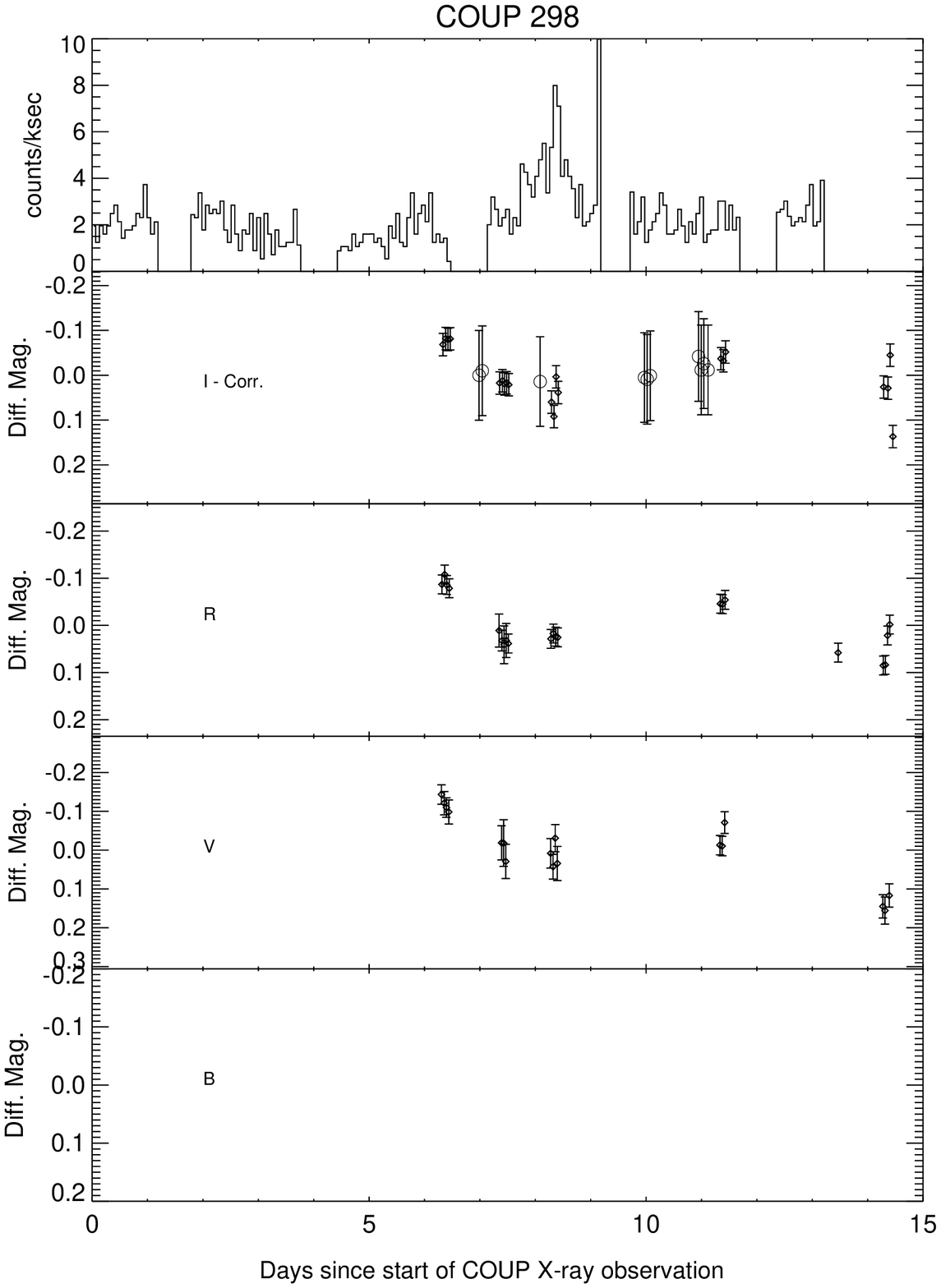}
\caption{\label{coup298}
Same as Fig.\ \ref{coup28}, but for COUP 298.
This figure appears in the electronic edition of the journal only.
}
\end{figure}

\begin{figure}[ht]
\figurenum{7o}
\epsscale{0.9}
\plotone{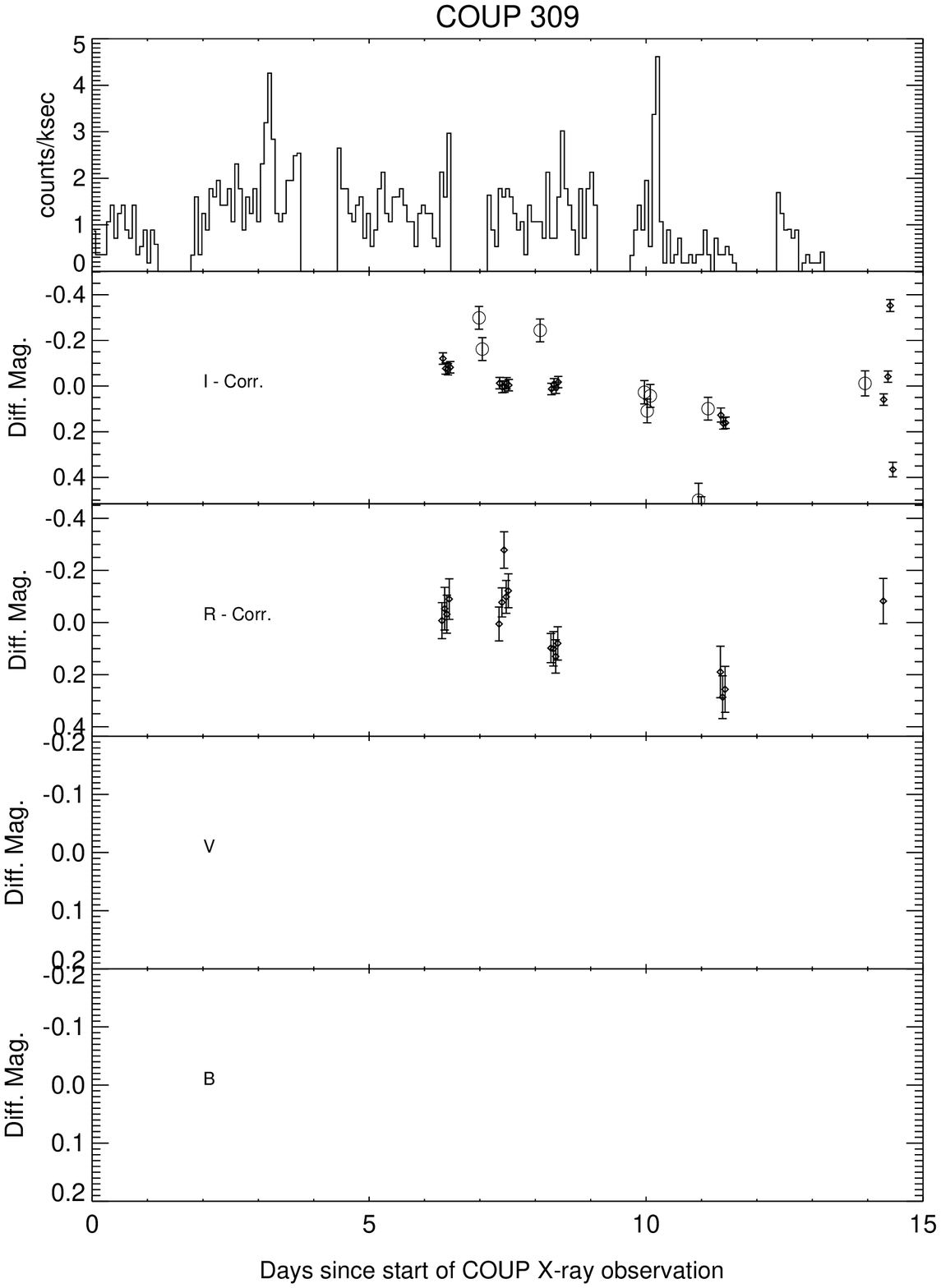}
\caption{\label{coup309}
Same as Fig.\ \ref{coup28}, but for COUP 309.
This figure appears in the electronic edition of the journal only.
}
\end{figure}

\begin{figure}[ht]
\figurenum{7p}
\epsscale{0.9}
\plotone{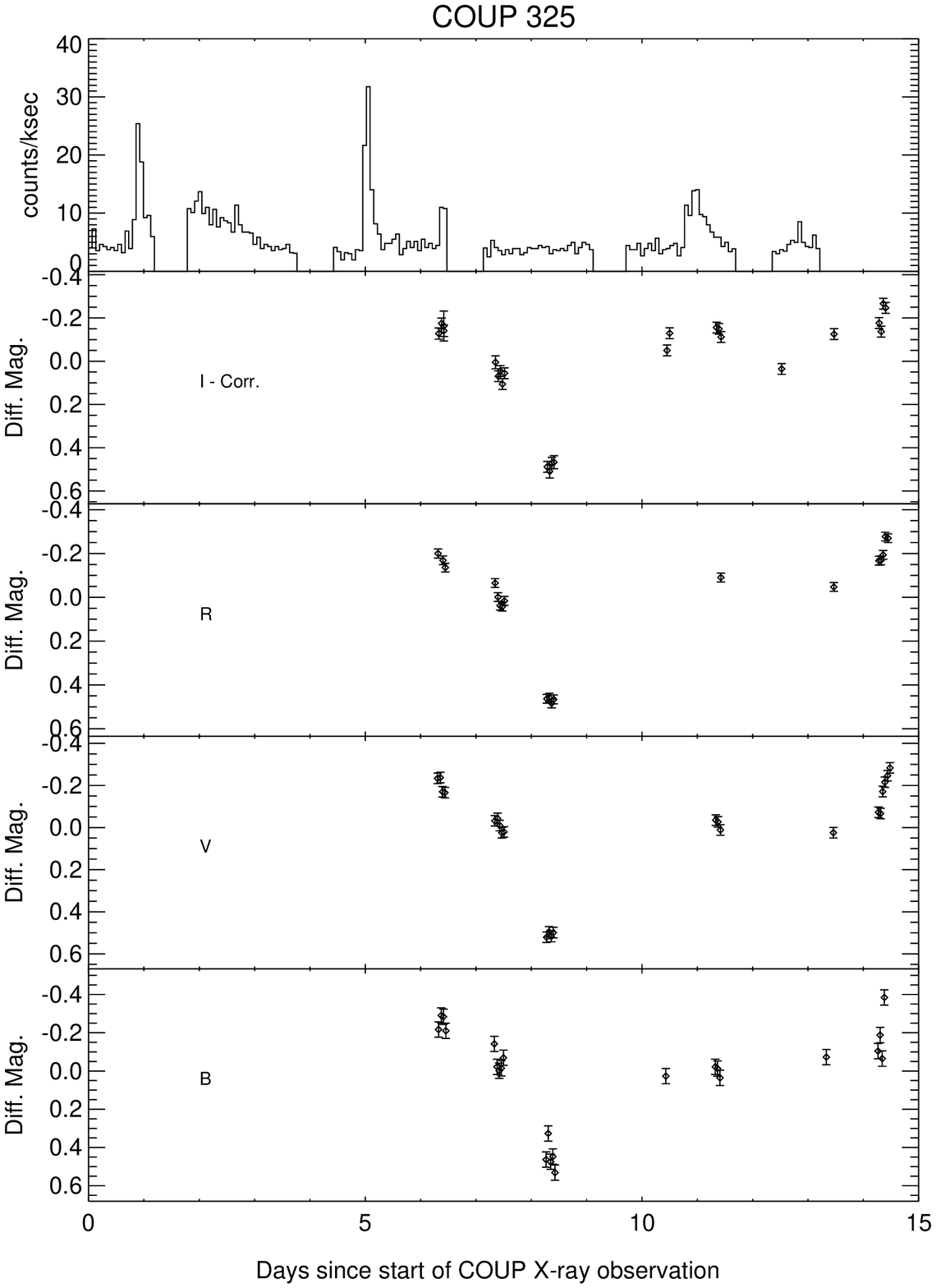}
\caption{\label{coup325}
Same as Fig.\ \ref{coup28}, but for COUP 325.
This figure appears in the electronic edition of the journal only.
}
\end{figure}

\begin{figure}[ht]
\figurenum{7q}
\epsscale{0.9}
\plotone{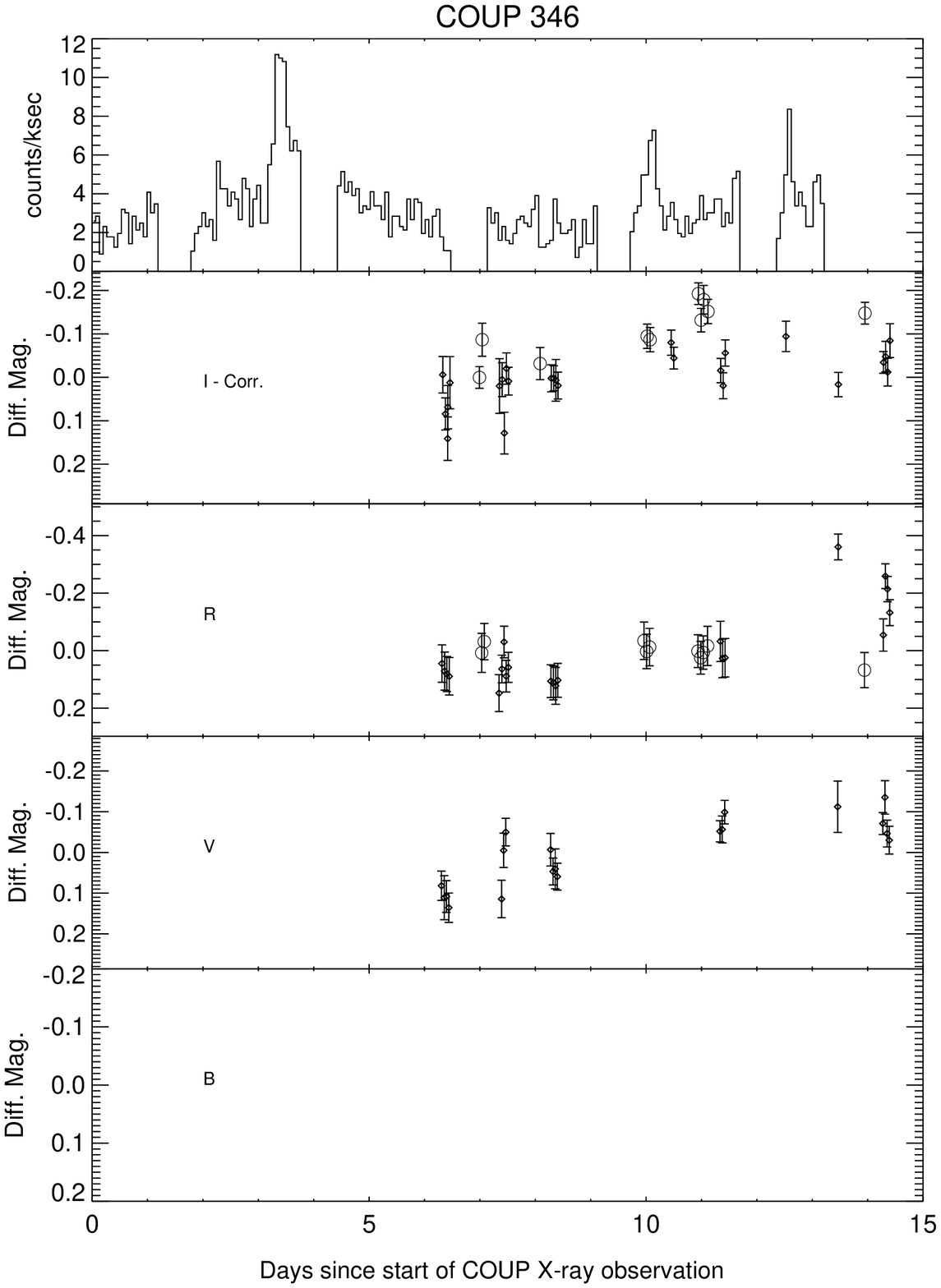}
\caption{\label{coup346}
Same as Fig.\ \ref{coup28}, but for COUP 346.
This figure appears in the electronic edition of the journal only.
}
\end{figure}

\begin{figure}[ht]
\figurenum{7r}
\epsscale{0.9}
\plotone{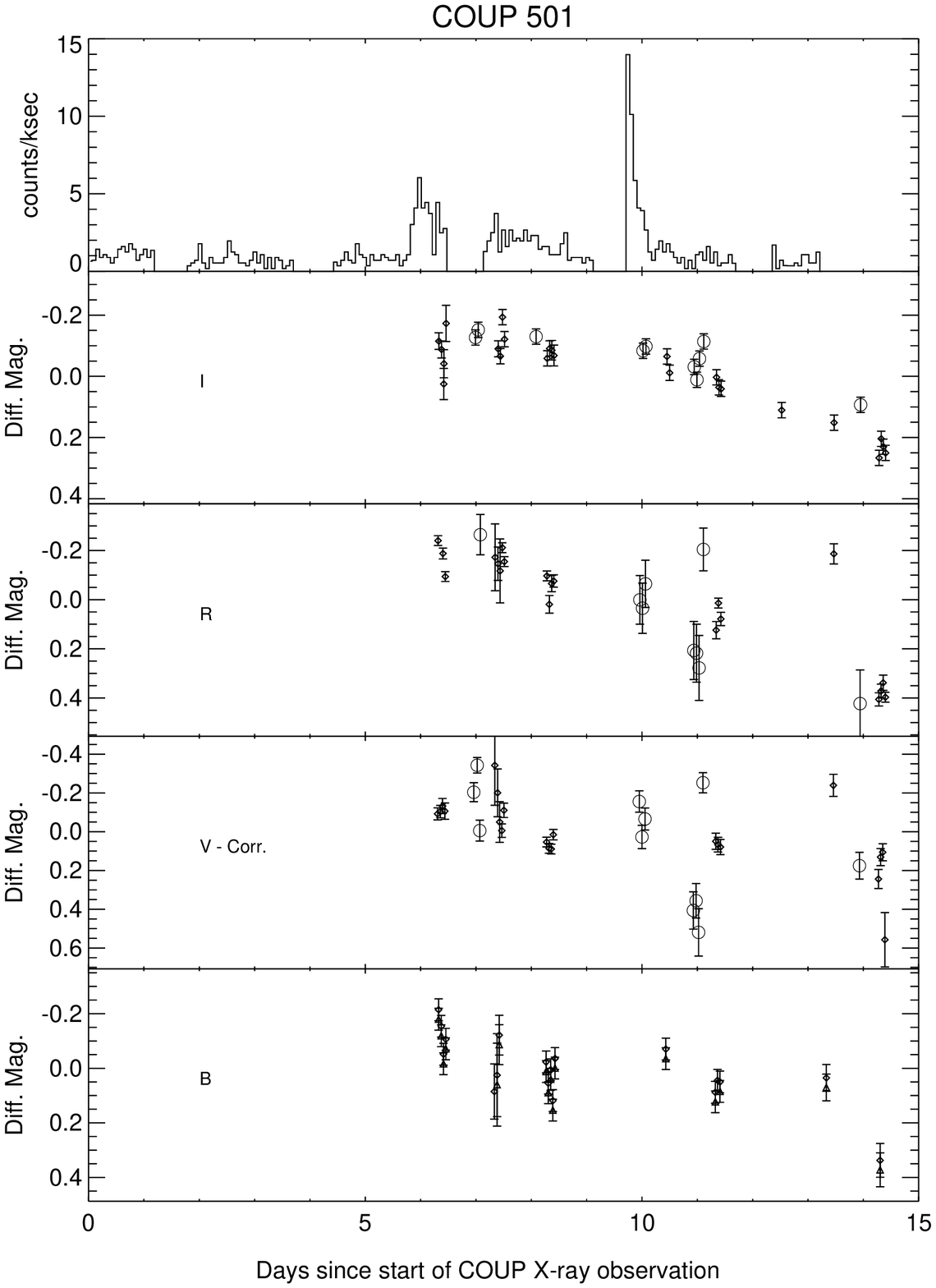}
\caption{\label{coup501}
Same as Fig.\ \ref{coup28}, but for COUP 501.
}
\end{figure}

\begin{figure}[ht]
\figurenum{7s}
\epsscale{0.9}
\plotone{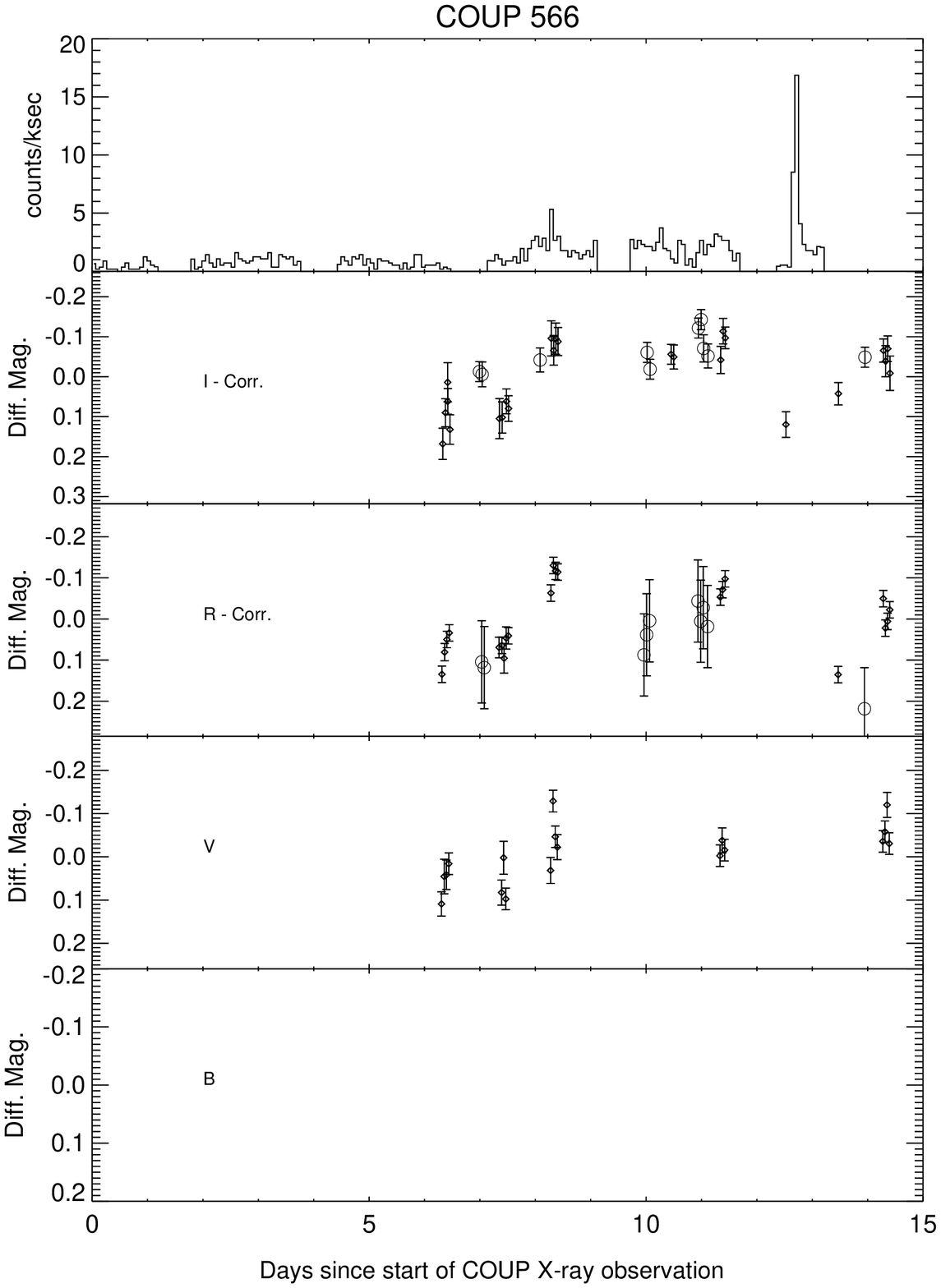}
\caption{\label{coup566}
Same as Fig.\ \ref{coup28}, but for COUP 566.
This figure appears in the electronic edition of the journal only.
}
\end{figure}

\begin{figure}[ht]
\figurenum{7t}
\epsscale{0.9}
\plotone{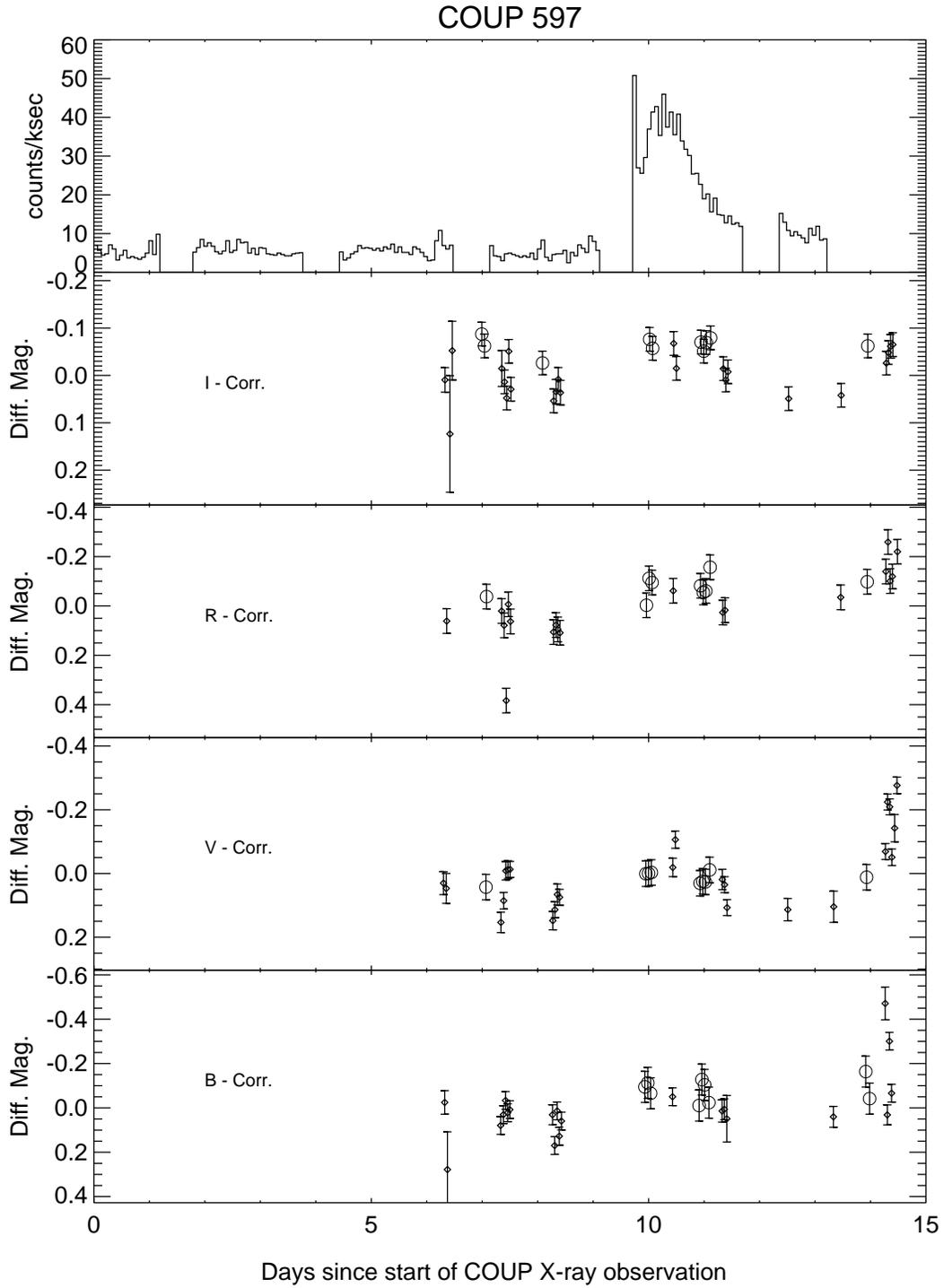}
\caption{\label{coup597}
Same as Fig.\ \ref{coup28}, but for COUP 597.
}
\end{figure}

\begin{figure}[ht]
\figurenum{7u}
\epsscale{0.9}
\plotone{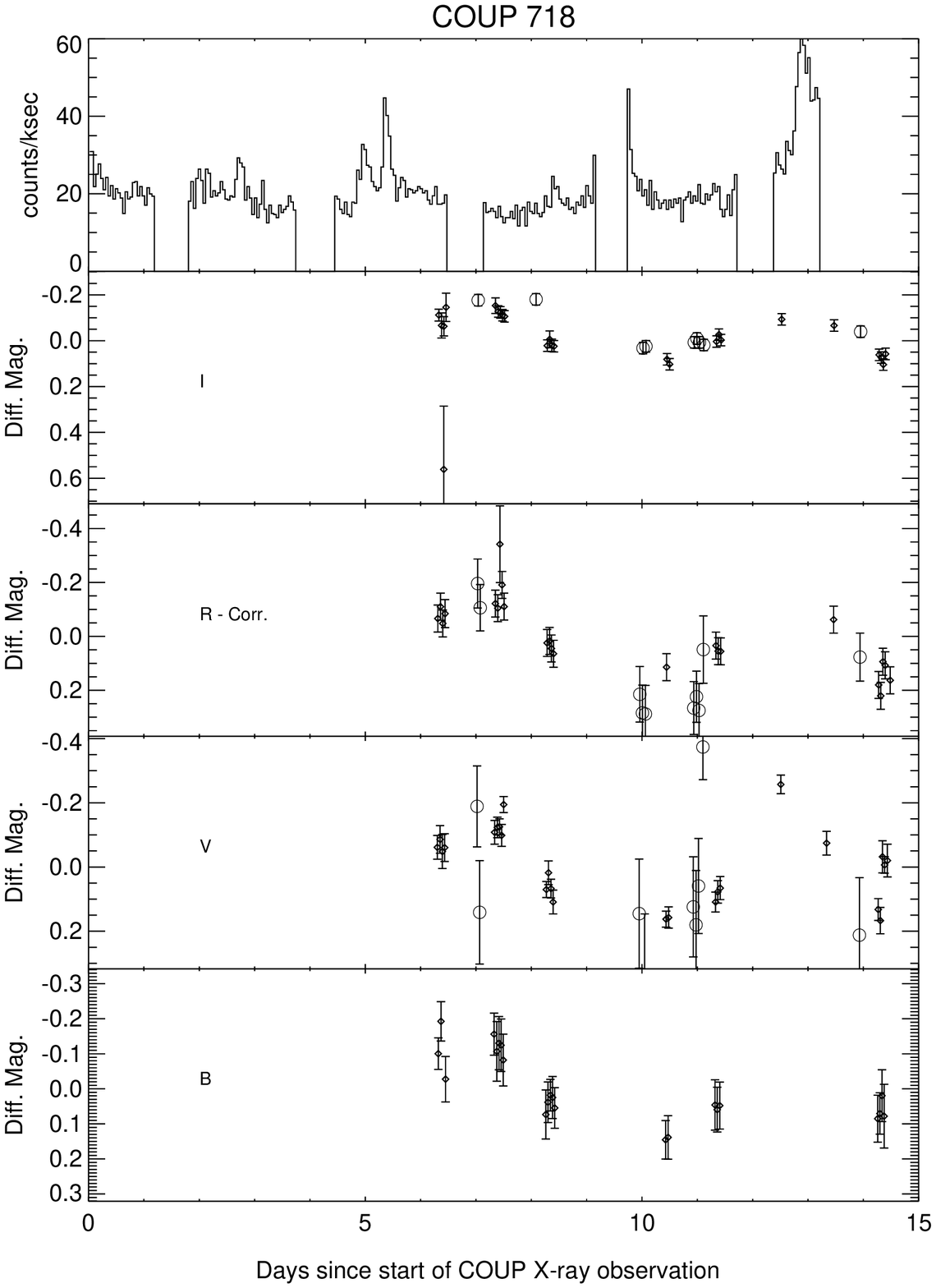}
\caption{\label{coup718}
Same as Fig.\ \ref{coup28}, but for COUP 718.
This figure appears in the electronic edition of the journal only.
}
\end{figure}

\begin{figure}[ht]
\figurenum{7v}
\epsscale{0.9}
\plotone{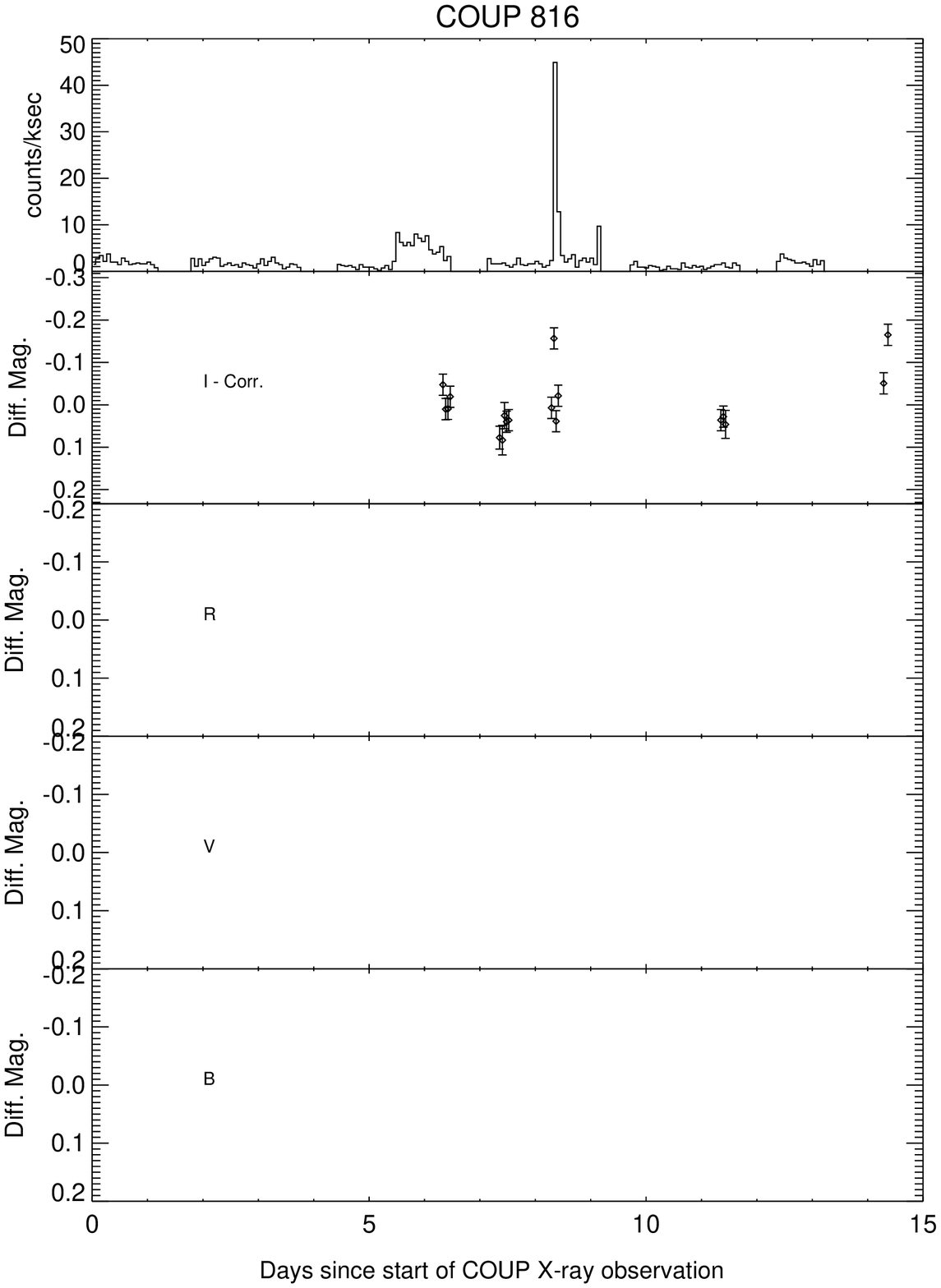}
\caption{\label{coup816}
Same as Fig.\ \ref{coup28}, but for COUP 816.
}
\end{figure}

\begin{figure}[ht]
\figurenum{7w}
\epsscale{0.9}
\plotone{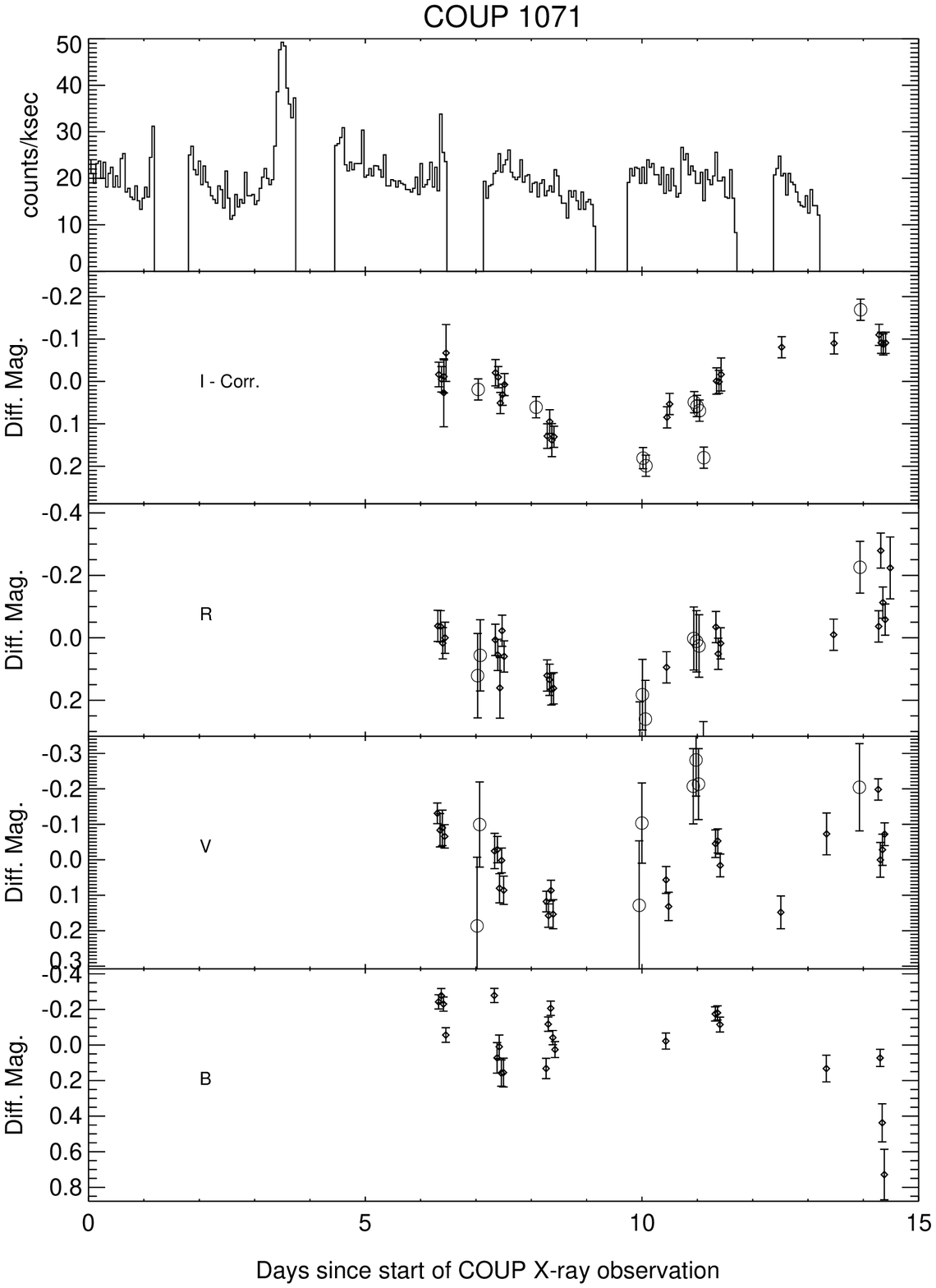}
\caption{\label{coup1071}
Same as Fig.\ \ref{coup28}, but for COUP 1071.
This figure appears in the electronic edition of the journal only.
}
\end{figure}

\begin{figure}[ht]
\figurenum{7x}
\epsscale{0.9}
\plotone{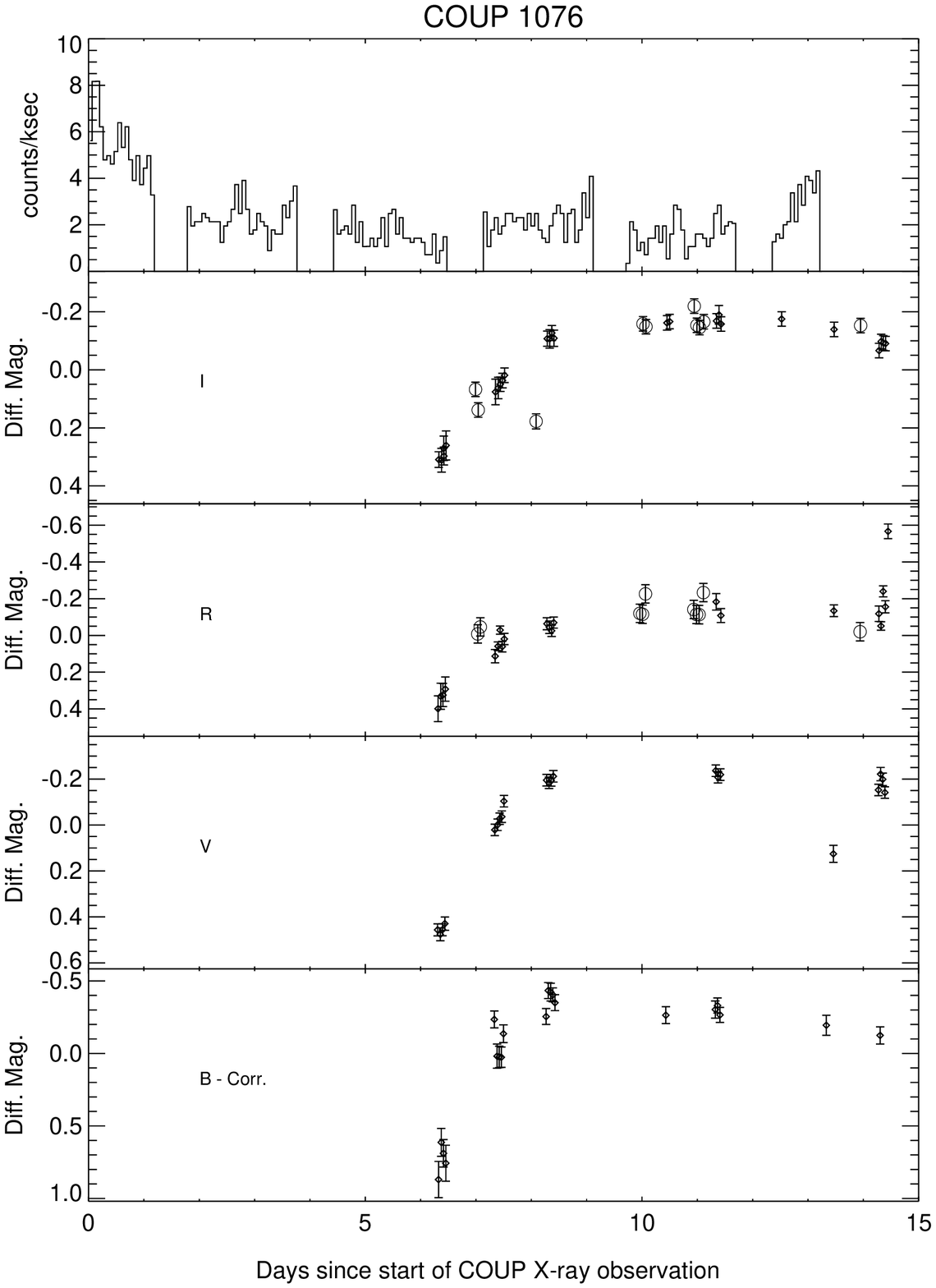}
\caption{\label{coup1076}
Same as Fig.\ \ref{coup28}, but for COUP 1076.
This figure appears in the electronic edition of the journal only.
}
\end{figure}

\begin{figure}[ht]
\figurenum{7y}
\epsscale{0.9}
\plotone{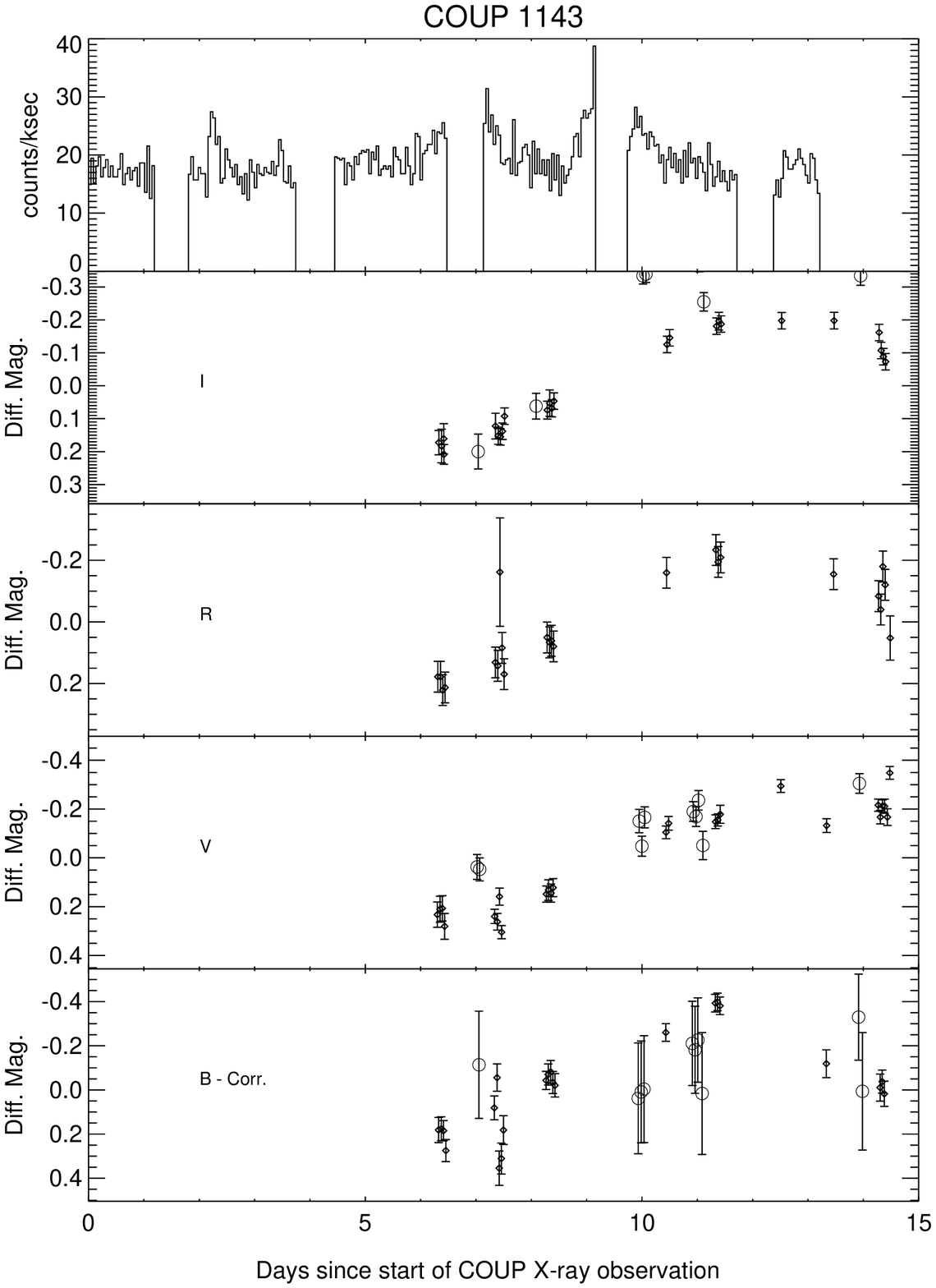}
\caption{\label{coup1143}
Same as Fig.\ \ref{coup28}, but for COUP 1143.
This figure appears in the electronic edition of the journal only.
}
\end{figure}

\begin{figure}[ht]
\figurenum{7z}
\epsscale{0.9}
\plotone{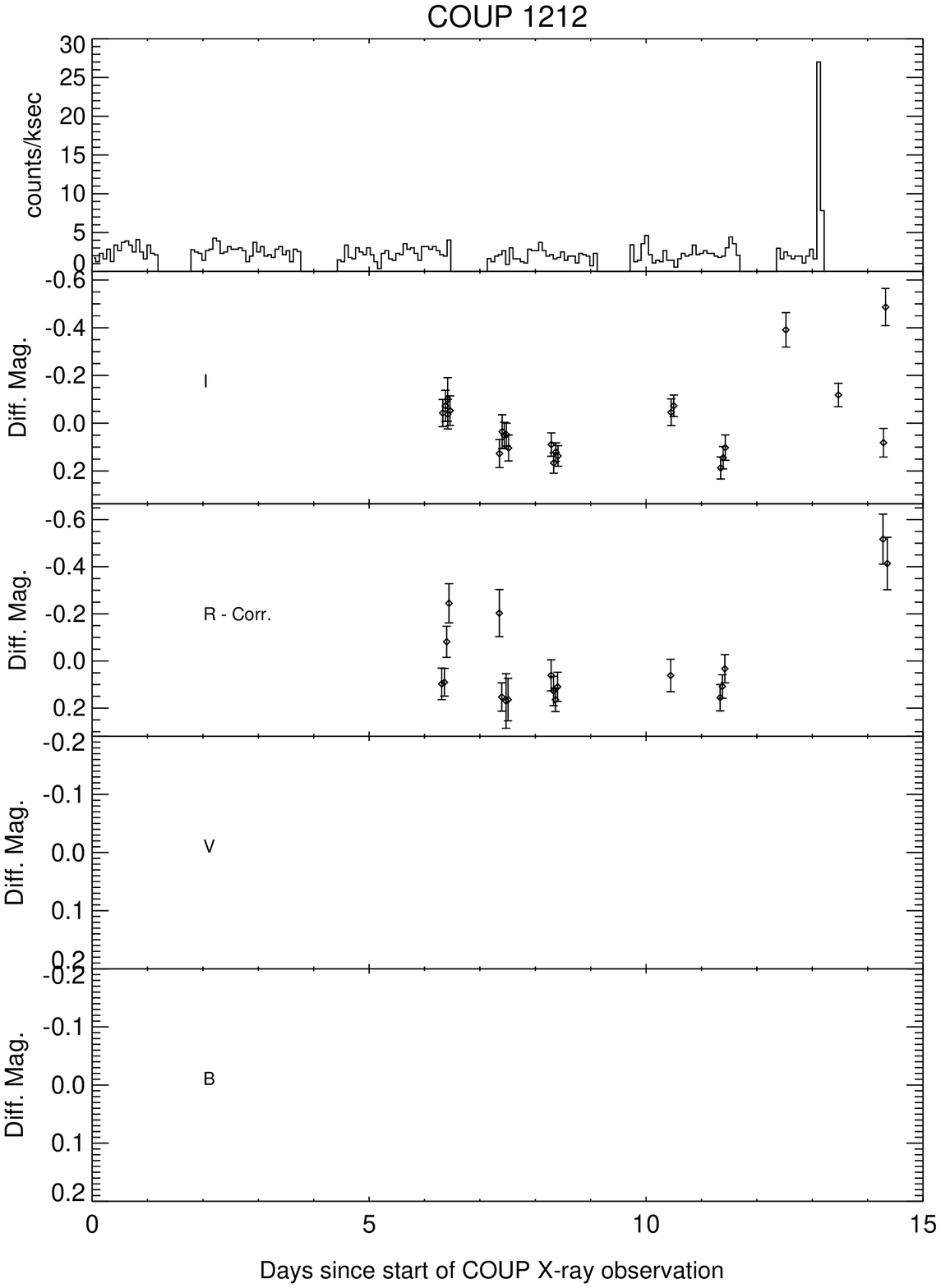}
\caption{\label{coup1212}
Same as Fig.\ \ref{coup28}, but for COUP 1212.
This figure appears in the electronic edition of the journal only.
}
\end{figure}

\begin{figure}[ht]
\figurenum{7aa}
\epsscale{0.9}
\plotone{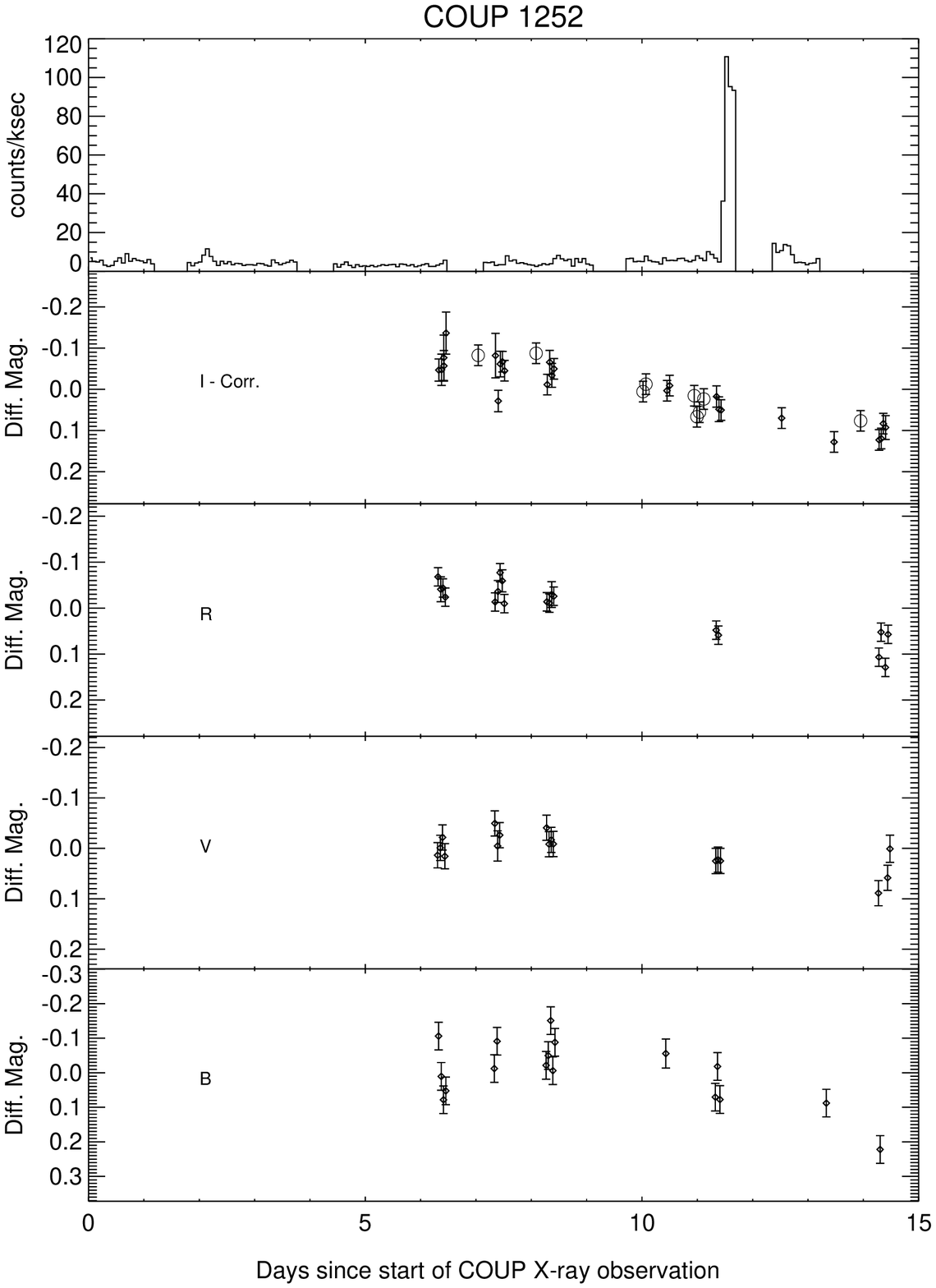}
\caption{\label{coup1252}
Same as Fig.\ \ref{coup28}, but for COUP 1252.
This figure appears in the electronic edition of the journal only.
}
\end{figure}

\begin{figure}[ht]
\figurenum{7ab}
\epsscale{0.9}
\plotone{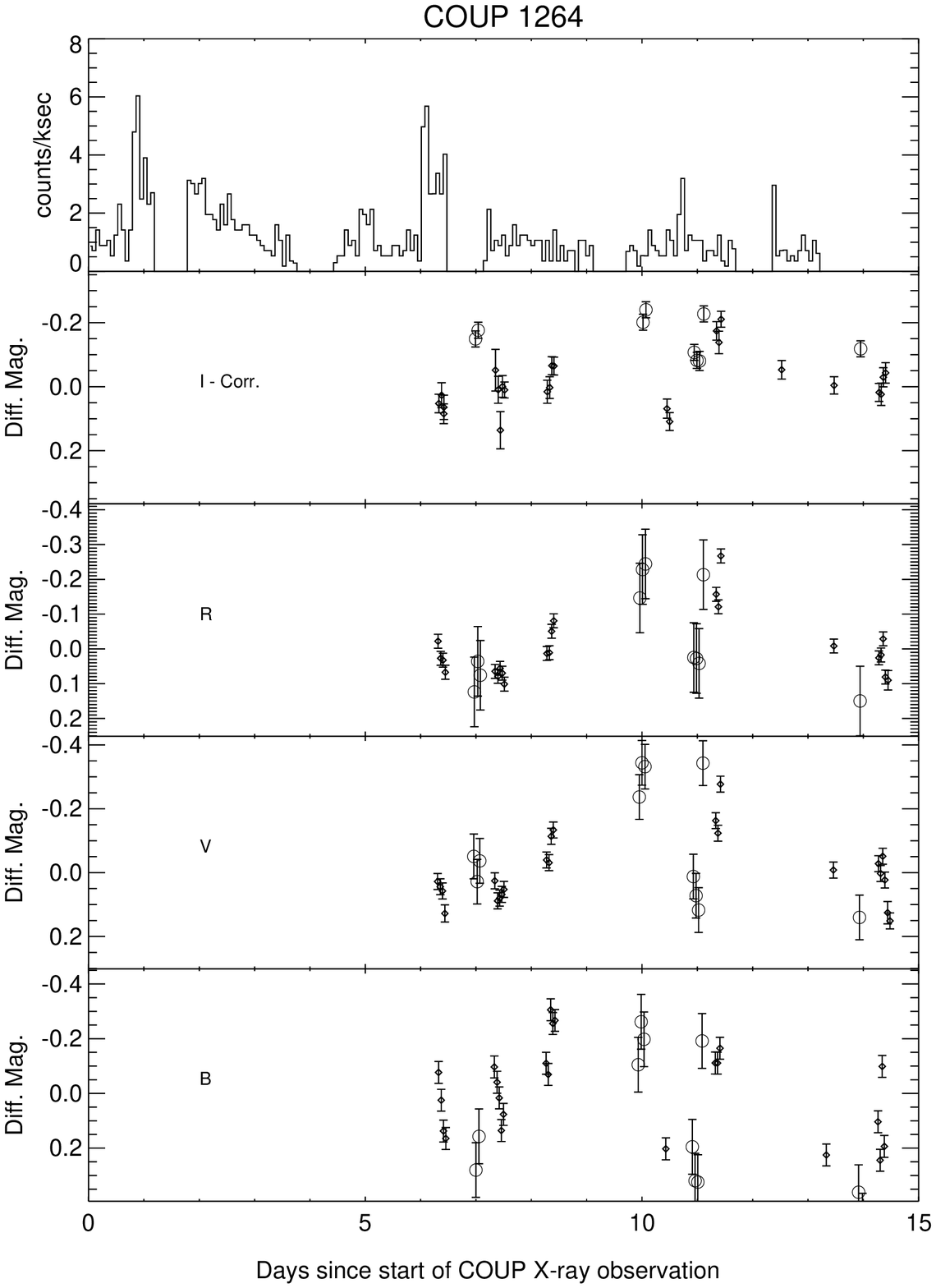}
\caption{\label{coup1264}
Same as Fig.\ \ref{coup28}, but for COUP 1264.
This figure appears in the electronic edition of the journal only.
}
\end{figure}

\begin{figure}[ht]
\figurenum{7ac}
\epsscale{0.9}
\plotone{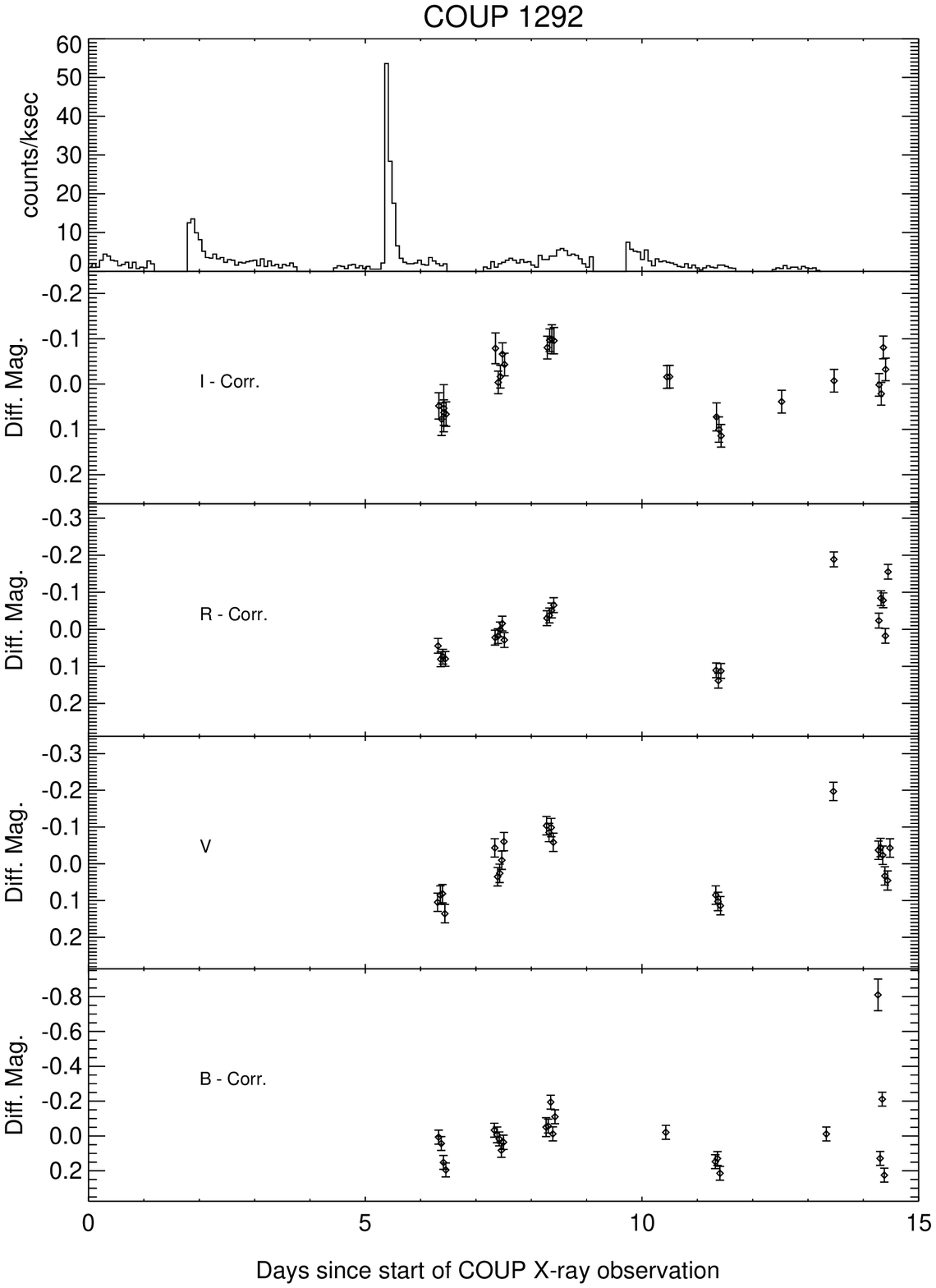}
\caption{\label{coup1292}
Same as Fig.\ \ref{coup28}, but for COUP 1292.
This figure appears in the electronic edition of the journal only.
}
\end{figure}

\begin{figure}[ht]
\figurenum{7ad}
\epsscale{0.9}
\plotone{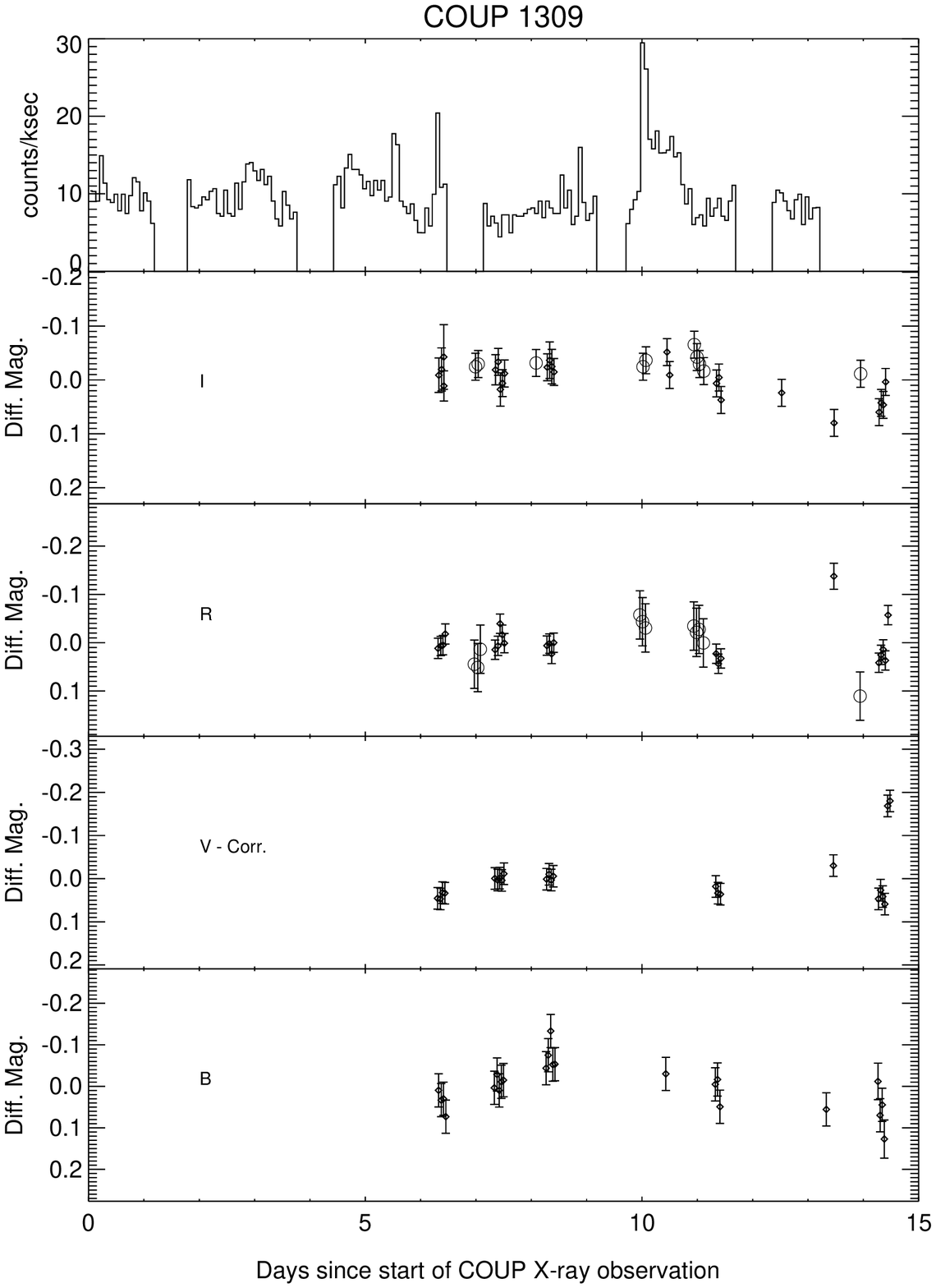}
\caption{\label{coup1309}
Same as Fig.\ \ref{coup28}, but for COUP 1309.
This figure appears in the electronic edition of the journal only.
}
\end{figure}

\begin{figure}[ht]
\figurenum{7ae}
\epsscale{0.9}
\plotone{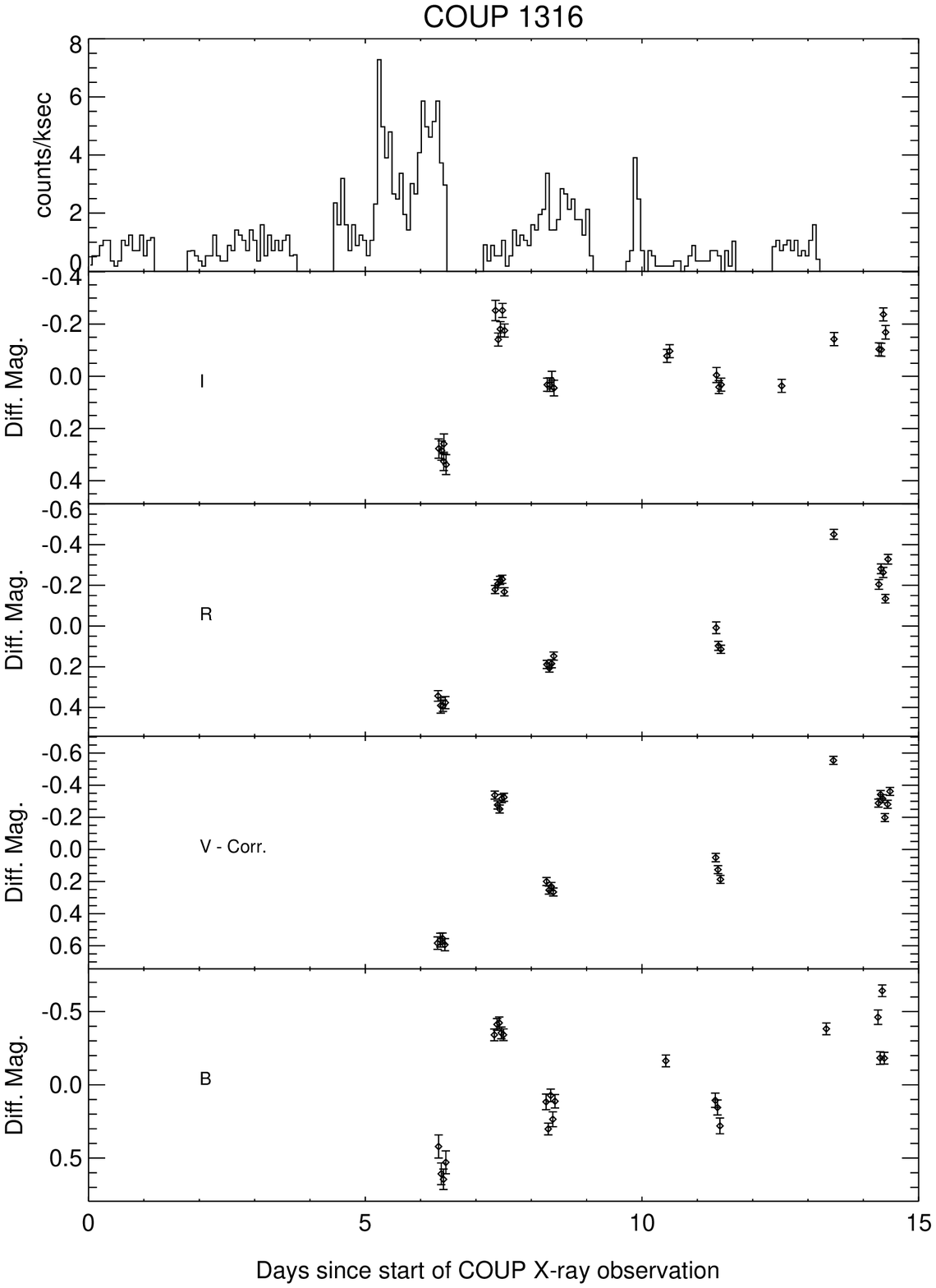}
\caption{\label{coup1316}
Same as Fig.\ \ref{coup28}, but for COUP 1316.
This figure appears in the electronic edition of the journal only.
}
\end{figure}

\begin{figure}[ht]
\figurenum{7af}
\epsscale{0.9}
\plotone{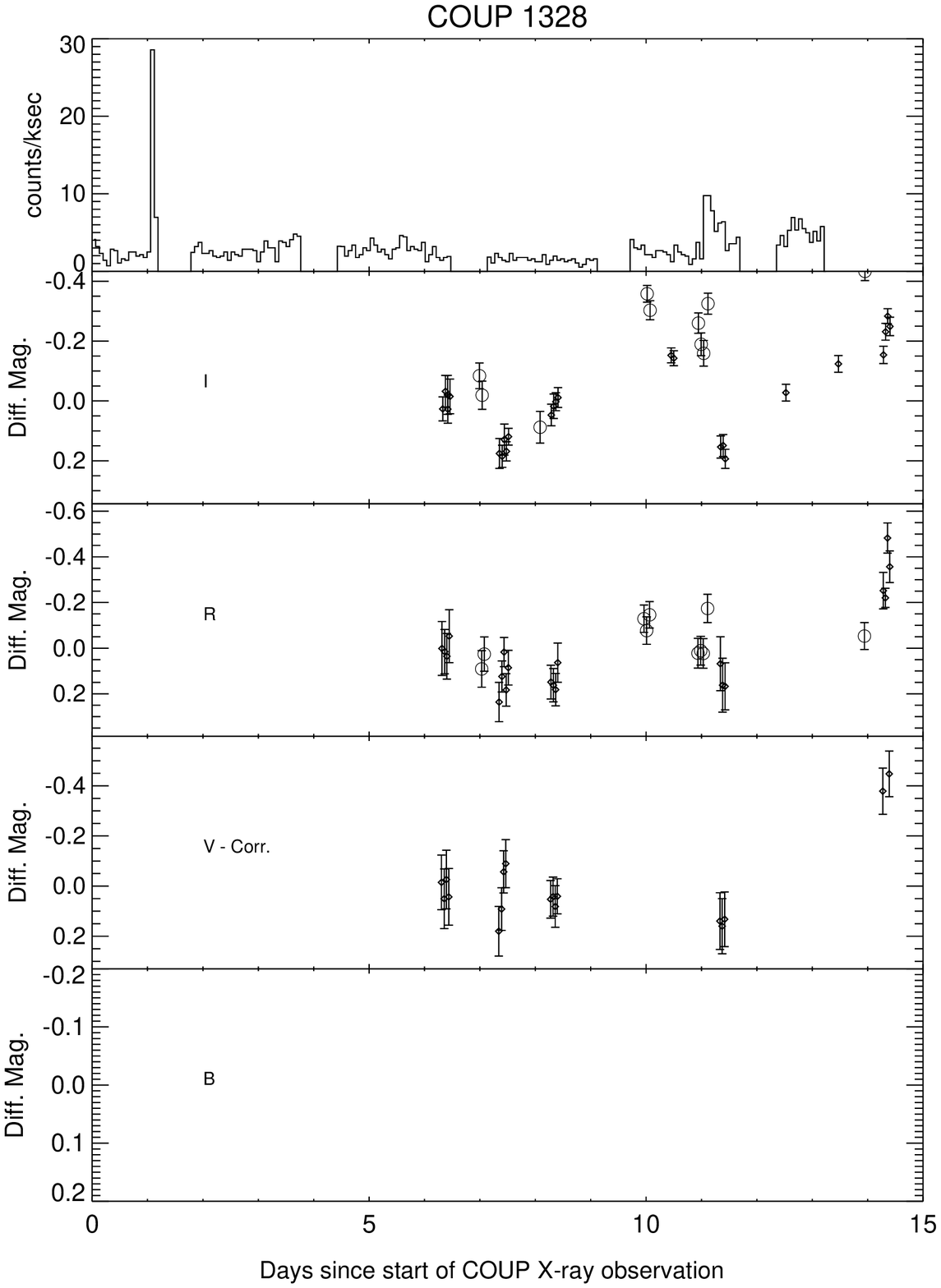}
\caption{\label{coup1328}
Same as Fig.\ \ref{coup28}, but for COUP 1328.
This figure appears in the electronic edition of the journal only.
}
\end{figure}

\begin{figure}[ht]
\figurenum{7ag}
\epsscale{0.9}
\plotone{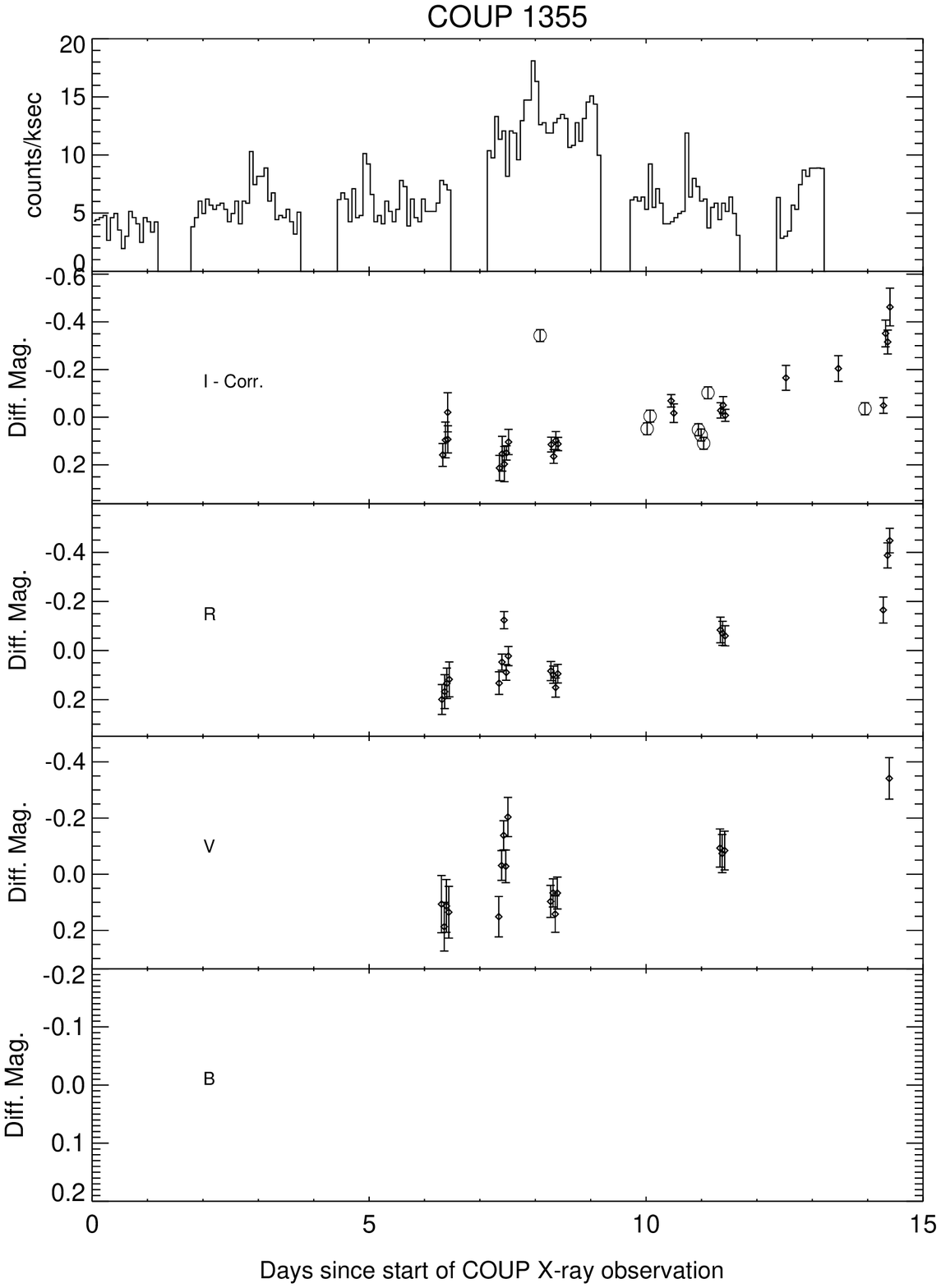}
\caption{\label{coup1355}
Same as Fig.\ \ref{coup28}, but for COUP 1355.
}
\end{figure}

\begin{figure}[ht]
\figurenum{7ah}
\epsscale{0.9}
\plotone{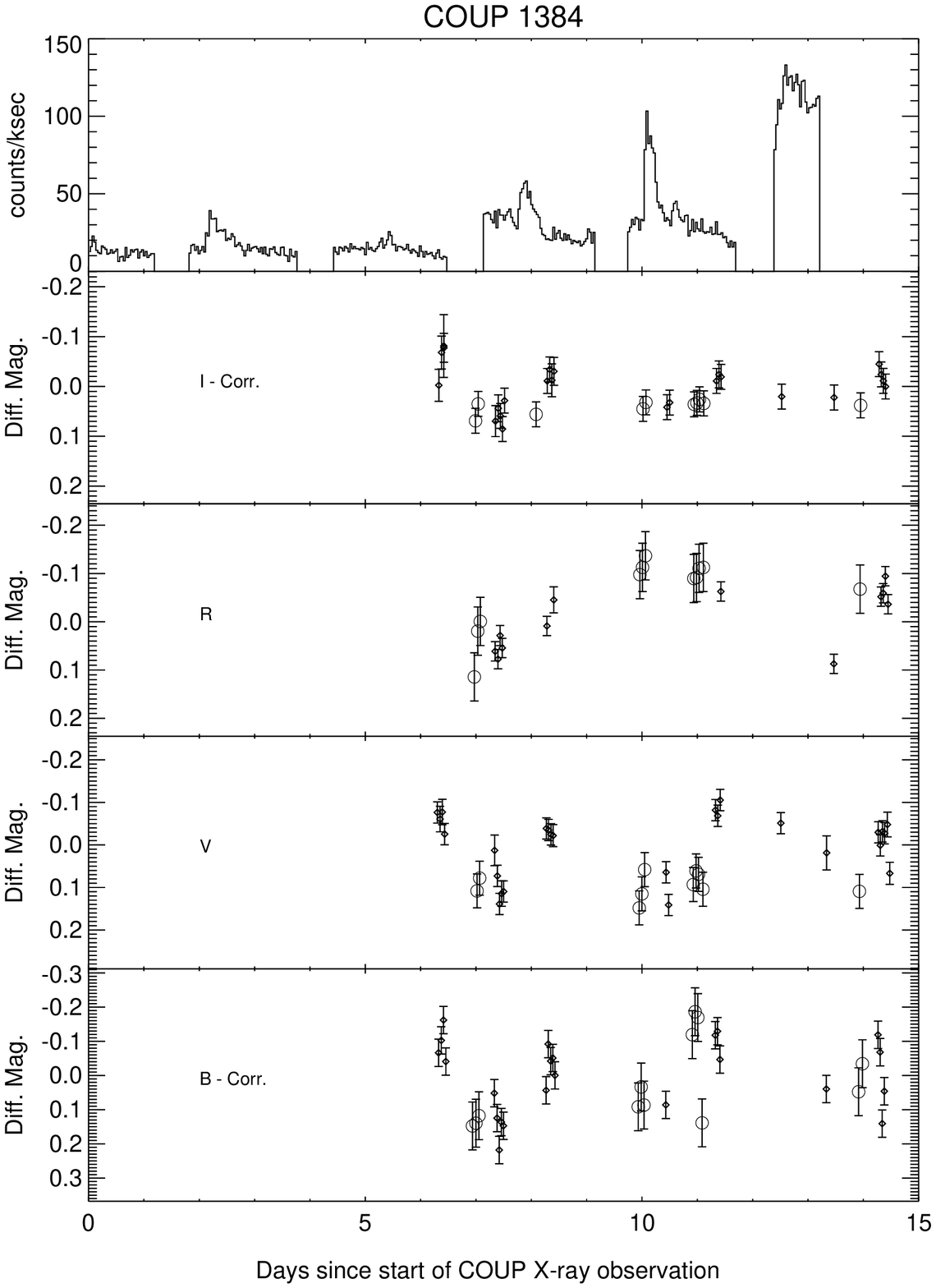}
\caption{\label{coup1384}
Same as Fig.\ \ref{coup28}, but for COUP 1384.
}
\end{figure}

\begin{figure}[ht]
\figurenum{7ai}
\epsscale{0.9}
\plotone{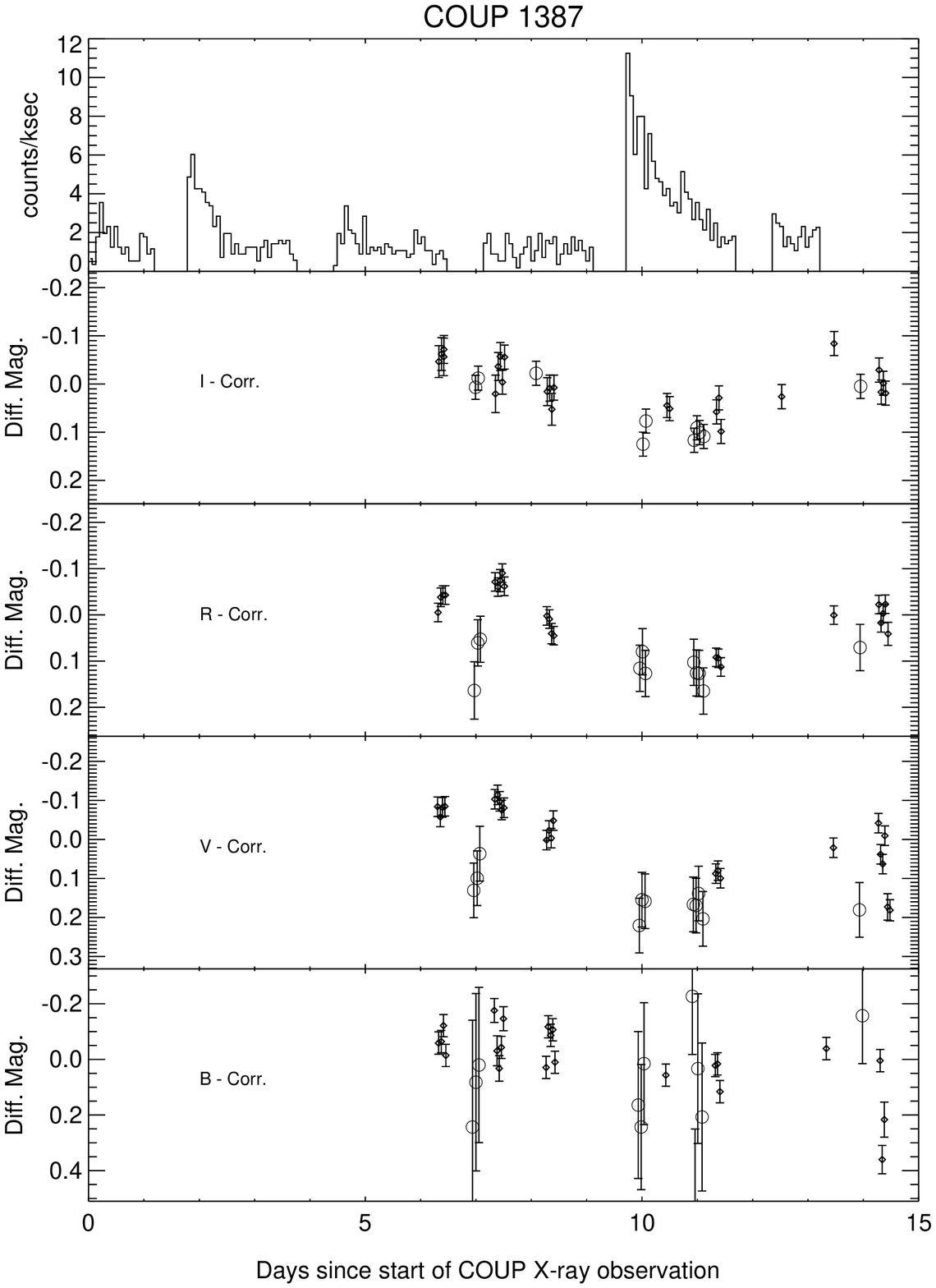}
\caption{\label{coup1387}
Same as Fig.\ \ref{coup28}, but for COUP 1387.
This figure appears in the electronic edition of the journal only.
}
\end{figure}

\begin{figure}[ht]
\figurenum{7aj}
\epsscale{0.9}
\plotone{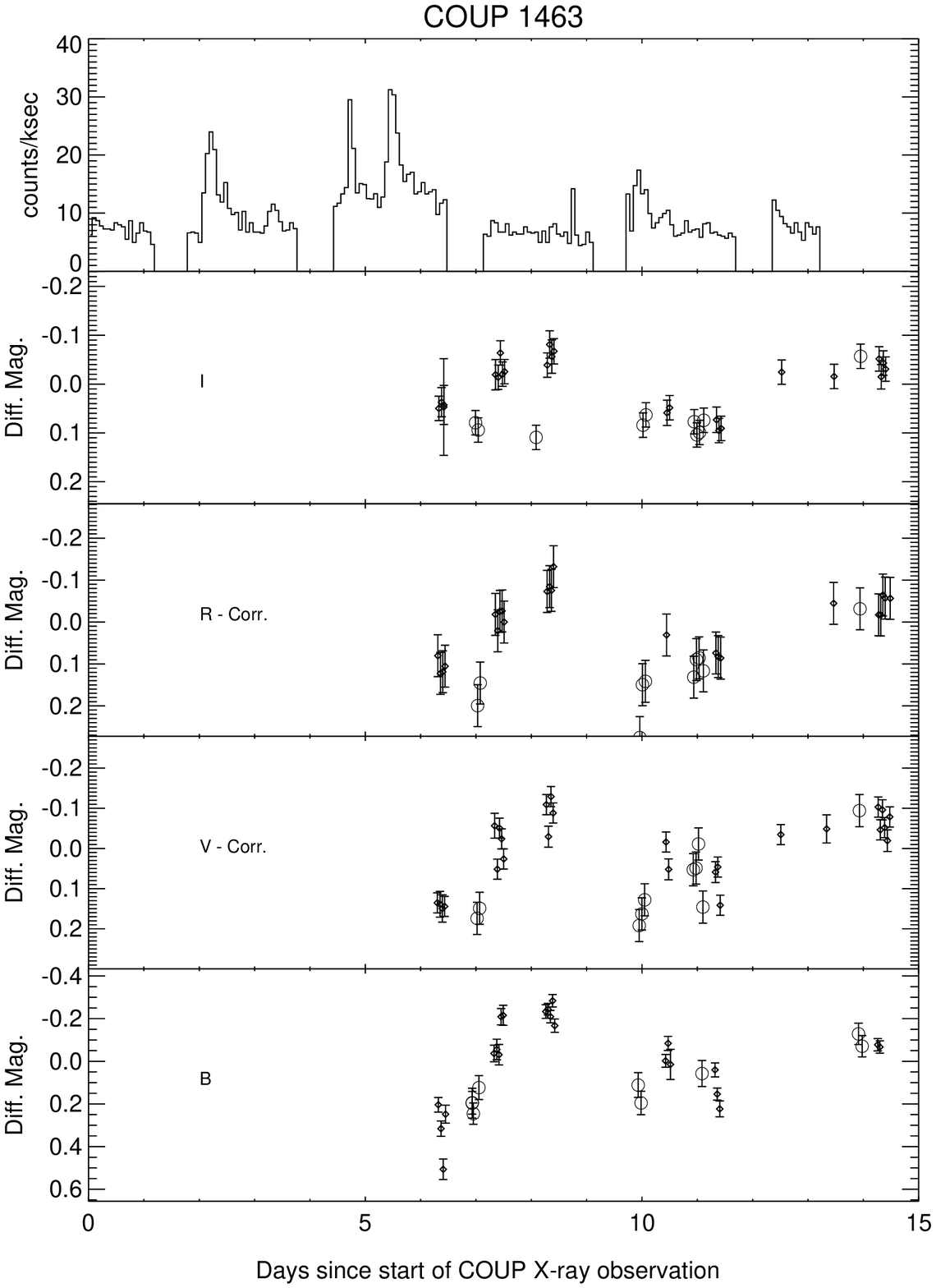}
\caption{\label{coup1463}
Same as Fig.\ \ref{coup28}, but for COUP 1463.
}
\end{figure}

\begin{figure}[ht]
\figurenum{7ak}
\epsscale{0.9}
\plotone{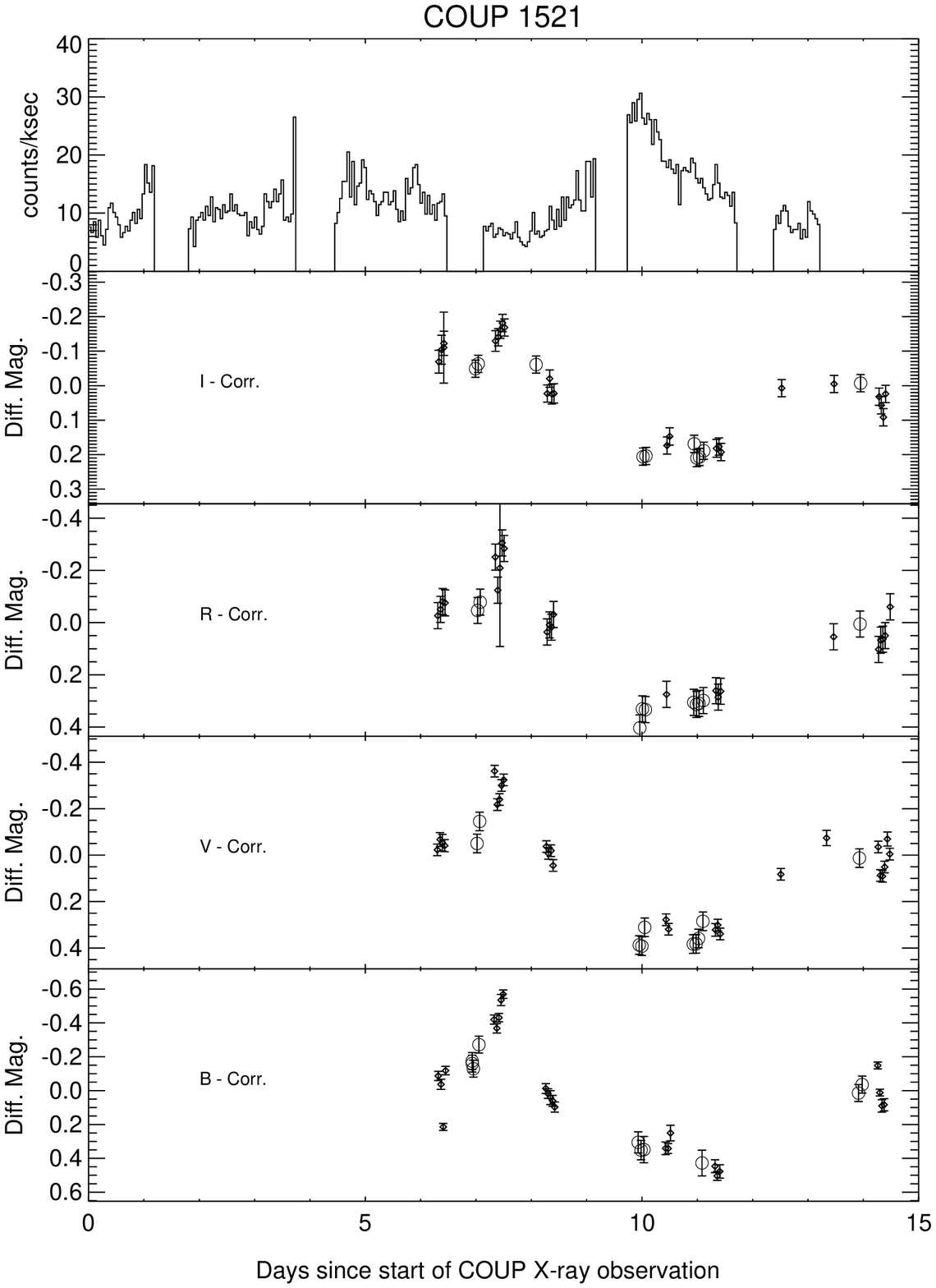}
\caption{\label{coup1521}
Same as Fig.\ \ref{coup28}, but for COUP 1521.
This figure appears in the electronic edition of the journal only.
}
\end{figure}

\begin{figure}[ht]
\figurenum{7al}
\epsscale{0.9}
\plotone{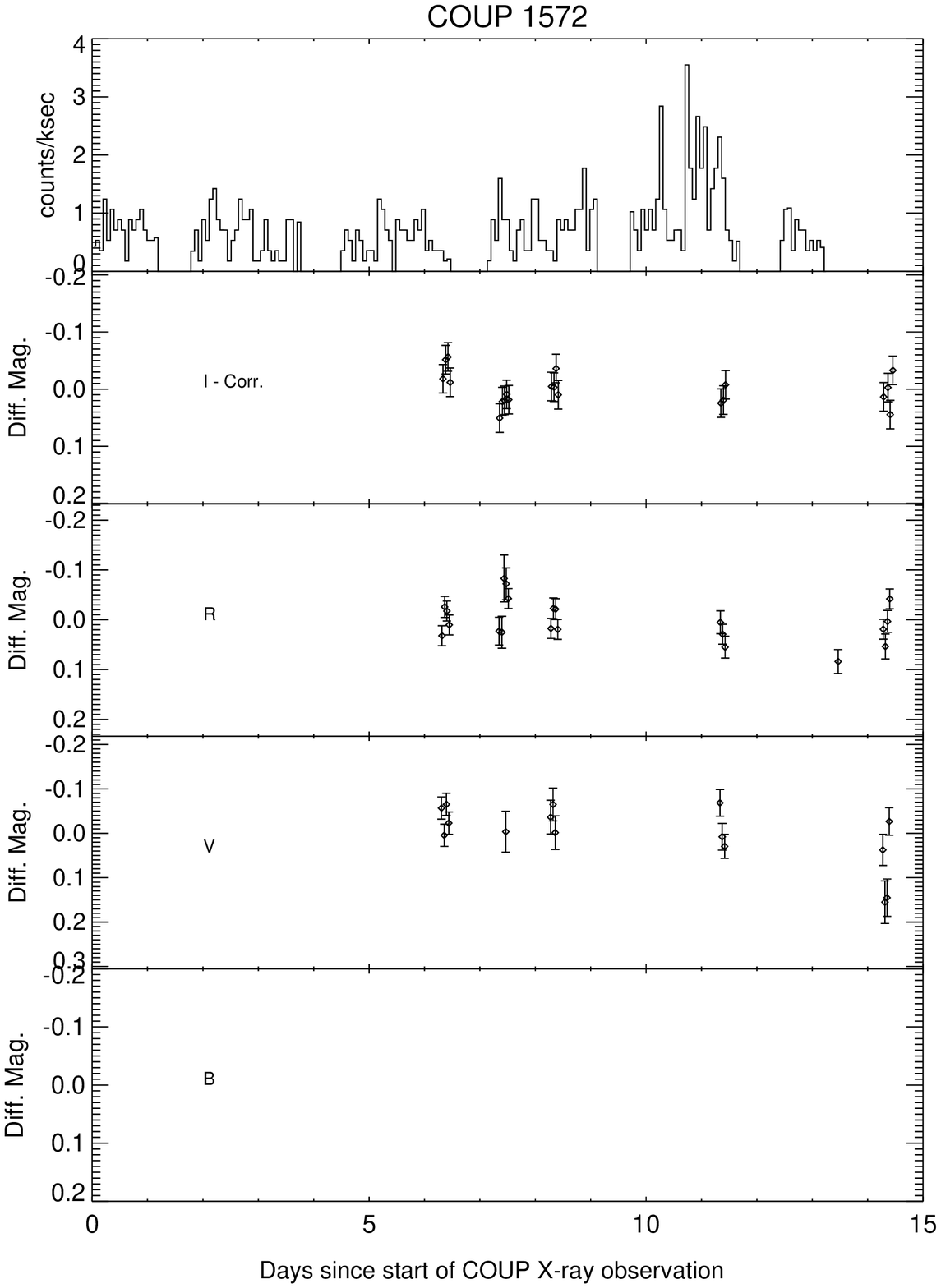}
\caption{\label{coup1572}
Same as Fig.\ \ref{coup28}, but for COUP 1572.
This figure appears in the electronic edition of the journal only.
}
\end{figure}

\begin{figure}[ht]
\figurenum{7am}
\epsscale{0.9}
\plotone{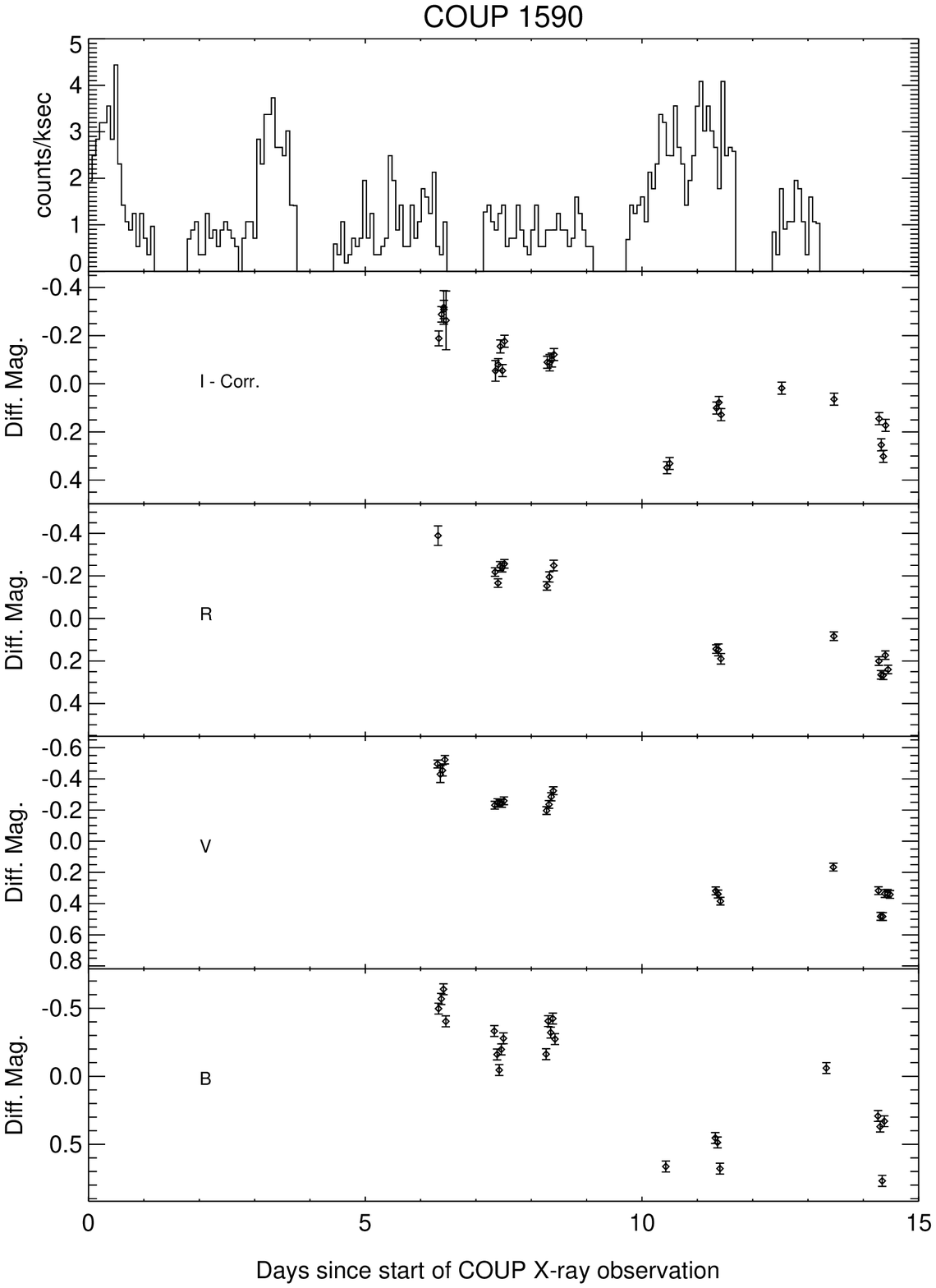}
\caption{\label{coup1590}
Same as Fig.\ \ref{coup28}, but for COUP 1590.
}
\end{figure}

\begin{figure}[ht]
\figurenum{7an}
\epsscale{0.9}
\plotone{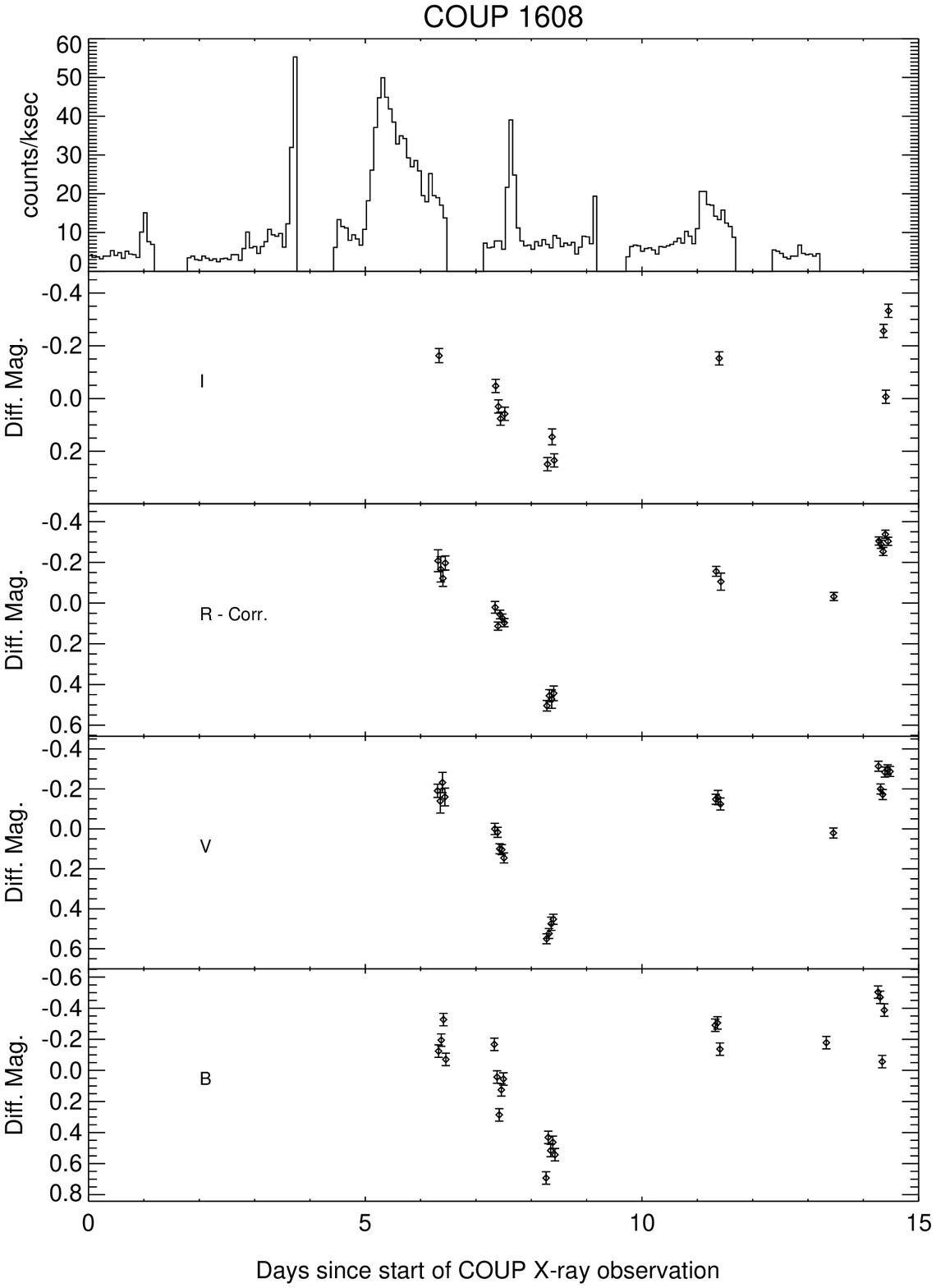}
\caption{\label{coup1608}
Same as Fig.\ \ref{coup28}, but for COUP 1608.
This figure appears in the electronic edition of the journal only.
}
\end{figure}

Following the procedure described in \S\ref{corr}, we have searched
for time-correlated variability in the optical and X-ray light curves
of the 814 stars in our database. 
Forty stars show evidence for the X-ray light curve
being time-correlated with at least one optical light
curve on the basis of a non-parametric
Kendall's $\tau$ test (\S \ref{corr}). These 40 stars are listed
in Table \ref{correlated-table}, and their X-ray and optical light
curves shown in Figs.\ \ref{coup28}--\ref{coup1608}.

Relative to our entire study sample of 814 stars,
the fractional incidence of time-correlated variability represented
by these 40 stars is $40/814 = 5\%$.
%depends on how one defines the comparison sample.
%Of course, time-correlated variability is less likely to be observed
%among those stars that are {\it neither} optically nor X-ray variable
%(i.e.\ stars that are non-variable altogether).
In our study sample, 178 stars are neither optically nor X-ray variable
according to the
BB and $J$ statistical measures discussed in \S\ref{methods} 
(see also Paper~II for a thorough analysis and discussion of these 
time-averaged variability indicators). If we exclude these 
178 stars from the comparison sample (leaving 636 stars), 
the 40 stars exhibiting time-correlated variability then represent
a fractional incidence of $40/636=6$\%.
In any case, time-correlated optical and X-ray variability is evidently 
an uncommon phenomenon in our sample. 

From statistical arguments,
it is probable that a number of these candidates will be spurious.
Estimating the number of expected false positives 
accurately is complicated by the
fact that a given star may possess multiple optical light curves
that are not statistically independent. We may, however, estimate an
upper limit to the expected number of false positives, given that
we adopted a Kendall's $\tau$ null-hypothesis probability 
of 0.01 as our criterion for identifying these candidates, and that
our search for time-correlated variability in our sample of 814 
stars involves 1944 different $BVRI$ optical light curves (\S\ref{corr}).
This implies 19--20 false positives. 
As a reality check on the Kendall's $\tau$ statistic,
we have additionally performed a Monte Carlo simulation in which we
applied the Kendall's $\tau$ test to one million random 
pairings of our optical and X-ray light curves, and we
find that Kendall's $\tau$ probabilities
of less than 0.01 occur with a frequency of 1.4\%, very close to the
expected value of 1.0\%.
The candidates listed in Table \ref{correlated-table} include
a total of 63 $BVRI$ light curves that are found to be time-correlated 
with their associated X-ray light curves according to Kendall's
$\tau$, and thus represent 
a $\sim 3 \sigma$ detection of time-correlated variability over the
expected number of false positives.

It is not possible on the basis of these statistical
arguments alone to determine which of these candidates are the false
positives and which are the true time-correlated variables. 
In many cases, the appearance of the X-ray and optical variations 
are quite dissimilar, so the statistical correlation may often 
not arise from a physical correlation.
Some candidates might nonetheless be regarded as more credible than 
others based on the number of
their optical light curves that pass the Kendall's $\tau$ test. For
example, COUP~122 (Fig.\ \ref{coup122})
has four optical light curves ($BVRI$) and all four
of them are identified as time-correlated with the X-ray light 
curve (see Table \ref{correlated-table}). 
In contrast, COUP~1309 (Fig.\ \ref{coup1309}) also has four optical
light curves, none of which appear to be convincingly correlated with
the X-ray light curve, and indeed only one of them is identified as
time-correlated by the Kendall's $\tau$ statistic. In other cases,
only one optical light curve is identified as being time-correlated
with the X-ray light curve even though the other optical light
curves share a similar morphology. For example, in COUP~152 (Fig.\
\ref{coup152}) the $V$- and $B$-band light curves are not identified
as time-correlated, despite a very similar morphology to the $R$-band
light curve. Common sense suggests that this is not a physically
meaningful result; instead, it is likely that small, random noise
differences among the optical light curves allow just one of them to pass
the 0.01 Kendall's $\tau$ probability threshold. Such a precarious
correlation is less credible than, e.g., the case of COUP~122
(Fig.\ \ref{coup122}) noted above.
The reader may wish to apply other subjective criteria to the light
curves shown in Figs.\ \ref{coup28}--\ref{coup1608}.

We thus do not attempt to identify specific candidates 
in Table \ref{correlated-table} as definitive. 
Rather, we emphasize that these 40 candidates have been identified 
on an objective, statistical basis and that
an estimated upper limit on the number of false positives 
suggests that $\sim \frac{1}{2}$ of these
candidates may represent {\it bona fide} cases of time-correlated
optical and X-ray variability. This implies an overall incidence 
of time-correlated variability in our sample of $\sim 3$\%.
Time-correlated optical and X-ray variability is evidently
an uncommon phenomenon in our sample.

\subsection{Relation to accretion\label{corr-accretion}}
If X-ray production in PMS stars is connected to accretion
related processes, we might expect time-correlated optical/X-ray
variability to be more common among actively accreting stars, as
these stars may be more likely to have their optical variability
dominated by accretion.
Following previous studies \citep[e.g.][]{flac-basic,stass04}, we take the
``accretors" to be those stars with \ion{Ca}{2} strongly in emission,
i.e., \ewca\ $\le -1$ \AA, and ``non-accretors" to be
those with \ion{Ca}{2} clearly in absorption, i.e., \ewca\ $\ge 1$
\AA. \ewca\ measurements have been reported for 493 stars in our
study sample. Of these, 151 stars are accretors and 145 stars are
non-accretors, as defined here (the remainder show indeterminate
values close to 0 \AA)\footnote{As a caveat, we note that \ewca\ is
an imperfect indicator of accretion. As shown by \citet{sicilia05},
some PMS stars with clear accretion signatures in H$\alpha$
do not manifest this accretion activity in \ion{Ca}{2}. It is
therefore possible or even likely that some low-level accretors
will go undetected in \ion{Ca}{2}. The criterion that we adopt
for identifying the presence of accretion is thus in practice a
conservative one; stars with \ewca\ $\le -1$ \AA\ and \ewca\ $\ge 1$
\AA\ are securely identified as accretors and non-accretors, while
some stars with indeterminate values close to 0 \AA\ may be weakly
accreting objects that will go unidentified as such.}.

Eighteen of the 40 stars in Table~\ref{correlated-table}
have \ewca\ measurements allowing them to be clearly
classified as accretors or non-accretors (Table~\ref{correlated-table}).
These 18 stars are roughly
evenly divided in terms of their accretion properties (10 accretors,
8 non-accretors). Thus there is no indication that accretors are
more likely than non-accretors to exhibit time-correlated optical
and X-ray variability.

\subsection{Nature of time-correlated variability\label{credible}}
In this section, we discuss the 40 stars that we have identified 
as candidates for time-correlated X-ray/optical variability 
(Table \ref{correlated-table}). 
Classifying them according to common light
curve characteristics, we suggest physical interpretations for the
origins of the observed variability. Note that, due to the often
complex nature of the observed variability, some stars are not
classified while others may be included
in more than one category (see Table~\ref{correlated-table} for
an overview). This discussion is presented to highlight what may
represent the first examples in the literature of time-correlated
optical and X-ray variability among PMS stars, and to guide
follow-up investigation of these specific objects.  We remind the reader 
that in fact the
overwhelming majority of the stars in our study sample do not exhibit
evidence for time-correlated optical and X-ray variability.

\subsubsection{Group A: Fast X-ray with slow optical variations:
Flares and star spots?}
Several stars show periodic or quasi-periodic modulations in their
optical light curve(s) on time scales that appear to be consistent
with optical variability periods ($\prot$) reported for them in the
literature. The persistence of the periodicity suggests that it is
related to the stars' rotation periods and that we are seeing either
dark spots in regions of magnetic activity, or accretion hot spots,
rotating in and out of view. At the same time, these stars' X-ray
light curves exhibit typical coronal flare-like morphologies that are
coincident in time with extrema in the optical light curves.

This type of behavior is found both in cases for which the
X-ray/optical correlations are positive and negative, and among
both accretors and non-accretors. Sources for which an X-ray flare
coincides with optical maximum (positive correlations) include
\object{COUP 28}, \object[COUP 112]{112}, \object[COUP 250]{250},
\object[COUP 501]{501}, and \object[COUP 1608]{1608}. Sources
for which an X-ray flare coincides with optical minimum
(anti-correlations) include \object{COUP 298}, 
\object[COUP 1384]{1384}, and \object[COUP 1387]{1387}.

Additional stars that can be included in this category are those
with no previously reported $\prot$ but which still show gradual
(and possibly periodic) modulations in at least one optical light
curve. Sources for which an X-ray flare coincides with optical maximum
(positive correlations) include \object{COUP 152}, \object[COUP 325]{325}, 
and \object[COUP 597]{597}. Sources for which an 
X-ray flare coincides with optical minimum (anti-correlations) include
\object{COUP 54}, 
\object[COUP 152]{152}, and \object[COUP 1316]{1316}. 

Finally, there are some stars whose time-correlations are dominated
by other behavior (see Groups B/C below) but which in
addition exhibit X-ray flares near optical maximum or minimum. Sources
for which an X-ray flare coincides with optical maximum include
\object{COUP 122} and \object[COUP 718]{718}. 
Sources for which an
X-ray flare coincides with optical minimum include \object{COUP 566}
and \object[COUP 1292]{1292}.

It is possible that these stars represent examples of
physical associations between regions of X-ray flaring and stellar
spots---either hot accretion spots (positive correlations) or dark
magnetic spots (anti-correlations). 
However, it is also possible that these stars represent
a more general class of objects that show periodic rotational
modulation and X-ray flares with no preferred correlation with
rotational phase. For example, if we assume a typical $\prot$ of
6~d, and that X-ray flares occur randomly in time with a typical
duration of 0.3~d, then the probability of finding either a 
correlation or anti-correlation is $2\times 0.3/6 = 10$\%. 
Moreover, among the ten
cases of positive correlations included above, only three show clear
evidence of active accretion.
In any case, detailed follow-up of these stars in
comparison to the full sample of COUP sources showing X-ray flares
should help resolve these interpretations.

\subsubsection{Group B: Slow X-ray with slow optical variations: 
Coronal structure and star spots?}
We have found several stars that show gradual and periodic or
quasi-periodic variations in both
their optical and X-ray light curves; in particular, the X-ray light
curves do not exhibit the flare-like features that characterize
the objects in Group~A. Assuming that modulation in the X-ray light
curve indicates the presence of a stable structure (within
the time span of the {\it Chandra} observation) that moves in and out
of view as the star rotates, its origin could be a bright accretion
spot or a magnetically active region on the stellar surface. In
the first case, a positive correlation with the optical light
curves might be expected. In the second, an X-ray--bright coronal
feature accompanied by a photospheric dark spot would give rise to
anti-correlated behavior.

There are nine cases for which the X-ray and optical light curves
produce a net anti-correlation: \object{COUP 139}, \object[COUP
161]{161}, \object[COUP 718]{718}, \object[COUP 1143]{1143}, 
\object[COUP 1355]{1355},
\object[COUP 1463]{1463}, \object[COUP 1521]{1521}, and \object[COUP
1590]{1590}. With the exception of the last object, which showed weak
signatures of active accretion in the observations of \citet{hill98},
the available \ewca\ measurements are consistent with these stars
all being non-accretors, and therefore implying that the photometric
and X-ray variations are not related to accretion.

All but one star (COUP 1521) have previously published
optical periods, and the variability observed in our optical light
curves is fully consistent with those periods. In addition, the
X-ray light curves of three of these stars (COUP 161, 1355, and
1521) have been independently found
by \citet{flac-spots} to be periodic, with X-ray periods $(P_X)$
that are equal to the optical periods (i.e.\ $P_X\approx\prot$; see
Table~\ref{correlated-table}), corroborating the interpretation that
the X-ray variability arises from rotational modulation. As discussed
by \citet{flac-spots}, this rotational modulation implies that the
X-ray emission is produced in close proximity to the stellar surface,
such that the X-ray emitting region is occulted as it rotates to
the back side of the star.  These objects thus represent compelling
cases of stars having stable coronal structures that are in close
spatial proximity to---and likely to be physically linked with---dark,
magnetically active regions on their surfaces.

Interestingly, for \object{COUP 139} and \object[COUP
1590]{1590}, \citet{flac-spots} find $P_X=\frac{1}{2}\prot$ (see
Table~\ref{correlated-table}). That is, the light curves
exhibit two X-ray modulations for each optical modulation. This is
particularly evident in \object{COUP 139} (Fig.~\ref{coup139}). Of
course, this behavior may simply be the result of randomly repeating
flares and/or randomly elevated X-ray emission levels, as discussed
for the Group~A stars. Another interpretation is that
the stellar coronae in these cases have two X-ray bright
regions, separated by $\pi$ in longitude, but that we only see a dark,
magnetically active region on the stellar surface associated with
one of them. 
%It is not immediately clear how this might occur. One
%possibility is that the two X-ray--bright regions are in fact
%both associated with dark surface spots, but that viewing geometry
%presents us from seeing one of them. This might occur if,
%for example, the two active regions are not only opposite in longitude
%but are opposite in latitude as well, and are located near the stellar
%poles. Then, if our viewing angle is close to pole-on, we may fail to
%see the dark surface spot on the ``lower" hemisphere when it rotates
%to the front side of the star, but we may nonetheless be able to
%partially see the associated X-ray--bright region suspended above the
%photosphere in the corona.  
\object{COUP 1590} (Fig.~\ref{coup1590}),
which exhibits an X-ray light curve that appears to oscillate between
two different amplitudes, may be a case in point.

In addition to these stars with anti-correlated light curves,
there are 14 stars that exhibit positive correlations in
their X-ray/optical light curves. They are: \object{COUP 112},
\object[COUP 122]{122}, \object[COUP 147]{147}, 
\object[COUP 226]{226}, 
\object[COUP 250]{250},
\object[COUP 501]{501}, \object[COUP 566]{566}, \object[COUP 1071]{1071},
and \object[COUP 1292]{1292}. 
\citet{flac-spots} have searched for periodicity
in the X-ray light curves of eight of these 14 stars, and
have identified two of them (\object{COUP 226} and \object[COUP 250]{250}) 
as having definitive
periodic X-ray light curves, with periods equal to (or half of)
the published $\prot$. They also identified another three of these
stars (\object{COUP 112}, \object[COUP 122]{122}, and \object[COUP
501]{501}) as having ``likely" periodic X-ray light curves. Thus, the
incidence of periodic X-ray modulation among objects with positively
correlated optical/X-ray light curves appears to be high.

The positive correlation between the X-ray and optical light
curves suggests that these cases represent stable X-ray--bright
structures physically associated with photospheric hot spots,
rotating in and out of view. These hot spots may be the footpoints
of mass accretion streams on the stellar surface.  Alternatively,
the observed positively correlated X-ray/optical light curves may
represent examples of X-ray--faint coronal structures---i.e.\ coronal
holes---that are co-spatial with dark, magnetically active regions
on the stellar surface.  However, an interpretation of accretion
hot spots is more strongly supported by the \ewca\ properties of
these 14 stars, which include 
some of the strongest accretion signatures found in the entire study 
sample (Table~\ref{correlated-table}). Among these, 
\object{COUP 250} (Fig.~\ref{coup250}) is particularly interesting,
having been identified by \citet{flac-spots} as having a periodic
X-ray light curve. This star may
thus be a good example of a
star having stable structures that are physically associated with
accretion hot spots on their surfaces. Follow-up modeling and analysis
of the color information contained in our multi-band optical light
curves should help to clarify whether we are in fact observing hot
spots in this object.

\subsubsection{Group C: Fast X-ray with fast optical variations: 
White-light flares?}
Many of the X-ray flares seen in the COUP dataset resemble solar
long-duration flares seen with the {\it GOES} and {\it Yohkoh}
satellites which cover the {\it Chandra} spectral band.  The peak
luminosities and durations, and hence total energies, of the COUP
flare plasma emission are orders of magnitude higher
than seen in solar flares. But it is reasonable to suggest that,
as in solar flares, the long-duration soft ({\it Chandra} band)
X-ray flares are preceeded by brief impulsive phases when particle
acceleration, radio gyrosynchrotron, hard ($\sim 20-100$ keV) X-ray
emission, and optical ``white-light" continuum increases occur
\citep[e.g.][]{Brown71, Dennis85, Dennis93, Tandberg88, Warren04}. 
The impulsive phase has, on rare occasions, been seen in
magnetically active stars, notably the RS~CVn systems $\sigma$~Gem
\citep{Gudel02} and HR~1099 \citep{Osten04} and dMe stars Prox Cen
\citep{Gudel04} and EV~Lac \citep{Osten05}.

Two stars in our study sample---\object{COUP 250} and \object[COUP
816]{816}---exhibit rather irregular optical light curves that show
punctuated increases in brightness that are positively correlated
with short-duration flares in the X-ray light curves. 
While these correlations may
be coincidences, we think it is worthwhile to highlight these objects
since they may represent the best candidates for the occurrence of
white-light magnetic flares.

We note that \object{COUP 250} has also been identified as a candidate
in Groups A and B above due to the particularly complex nature
of the observed variability. In particular, we caution that the
optical brightening ``event" near day 10.5 may simply be related to
the longer-term, quasi-periodic optical variability of this object.
On the other hand, while we have very little data for \object{COUP
816}, what we do have is tantalizing. A single optical measurement,
brighter than the mean brightness level by $\sim 3\sigma$, occurs
precisely in the time bin of a very brief X-ray flare that at its peak
is $\sim 15$ times brighter than the mean X-ray flux level for this
object. We have carefully scrutinized the optical data for this star
and believe the observed optical brightening event to be credible.

\section{Discussion\label{discussion}}
In contrast to the strong ``rotation-activity
relationship" that is clearly present on the main sequence
\citep{pallavicini81,jeffries99,randich00,schrijver00,pizzolato03}, a series
of studies based on {\it Chandra} observations of a large sample
of PMS stars in the Orion Nebula Cluster (ONC) failed to find any
direct correlation between the stars' rotation rates and their X-ray
luminosities \citep{flac-time,feig03,stass04,preibisch05}.
A possible explanation for the lack of a PMS rotation-activity
relationship is that the X-rays originate not (or not only)
from dynamo-generated magnetic processes, but also (or instead)
from energetic processes related to accretion. 

If X-rays are
produced near the sites of accretion shocks, then we may expect that
X-ray and optical variability would be time-correlated.  However,
at most $\sim 5$\% of the stars in our sample show credible,
statistically significant, time-correlated optical and X-ray light
curves (\S \ref{corr-incidence}), and this time-correlated variability
is just as likely to occur among non-accreting stars as it is
among accretors (\S \ref{corr-accretion}). 
Our findings are consistent with previous studies of 
simultaneous X-ray and optical variability in PMS stars
\citep{feig94,gullbring97,stelzer04}, with a sample that is
hundreds of times larger.

Our results do not negate the possibility that accretion is indirectly
related to X-ray production at some level. For example, the presence
of an accretion disk may alter the stellar magnetospheric environment
\citep[e.g.\ via large star-disk magnetic loops;][]{favata05} 
such that large reconnection flares or other energetic processes take
place in a way that is not possible in the absence of a disk 
\citep[e.g.][]{jardine06}. Such
events, displaced and disconnected from magnetically active
regions on the stellar surface, might not be expected to produce
time-correlated behavior in X-rays and the optical. In addition, we
have in fact found a small number of stars that exhibit possibly
accretion-related X-ray emission, in the form of X-ray flares and
X-ray--bright features that may be co-spatial with accretion
hot spots on the stellar surface (\S \ref{credible}). 
The strong optical/X-ray brightening event
observed in the FU~Ori--like outburst of V1647~Ori (McNeil's Nebula)
serves as a particularly dramatic example of time-correlated
optical/X-ray variability that is likely to be accretion driven
\citep{kastner04b}.
Nonetheless, though intriguing, such cases are evidently very rare. 

Similarly, any physical connection of X-ray emission to magnetically
active surface regions (i.e.\ cool spots) does not for the vast 
majority of stars in our sample
manifest itself in a time-correlated way (\S \ref{credible}). This
may indicate that the X-ray emitting coronae in PMS stars have
spatial structures that little reflect the magnetic structure of
the underlying photosphere. A possible explanation for this is that
X-ray emission takes place on very large spatial scales relative
to the photosphere, and may not be confined to the small surface
regions (spots) from which the optical variability presumably
originates. While this picture is at odds with the small coronae
implicit in the finding of rotationally modulated X-ray emission
in $\sim 10$\% of COUP sources by \citet{flac-spots}, such an
interpretation could be consistent with the very large flaring loop
sizes inferred on a small number of COUP sources by \citet{favata05}.

More generally, our findings are consistent with recent analyses of 
the X-ray spectra of CTTSs, and with recent measurements of the 
magnetic field structures of WTTSs. In an analysis of
high-resolution X-ray spectra of the CTTSs BP~Tau,
CR~Cha, SU~Aur, and TW~Hya, \citet{robrade06} conclude that both
accretion shocks and coronal activity contribute to the observed X-ray
emission, but that coronal activity is by far the dominant contributor
in all cases except TW~Hya. 
These results, together with our finding that time-correlated optical
and X-ray variability occurs only very rarely, support
the general conclusion that neither surface accretion shocks nor 
magnetically active surface (cool) spots are the
exclusive sites of X-ray production in the vast majority of PMS stars.

\section{Summary and Conclusions\label{conclusions}}
We have presented a combined analysis of simultaneous optical and
X-ray light curves for a sample of more than 800 pre--main-sequence
(PMS) stars in the Orion Nebula Cluster (ONC). This dataset represents
an increase of several hundred-fold, both in the number of objects
studied and in the total duration of the X-ray and optical time-series
data, over all previous investigations of simultaneous optical and
X-ray variability in PMS stars.

We identify 40 stars that exhibit the first examples to our
knowledge of time-correlated
optical and X-ray variability in PMS stars. Among these 40 stars,
we find examples of both positive and anti-correlations. 
Positive correlations may represent X-ray--dark coronal
structures associated with cool, magnetically active surface spots, 
or may represent X-ray--bright structures associated with hot accretion
spots. Anti-correlations may represent, e.g., magnetic flaring events 
associated with cool surface spots. In any event, the implication
is that the X-ray emitting regions in these stars
are in close proximity to---and thus
likely to be physically linked with---cool and/or hot spots on their
surfaces. Follow-up modeling and analysis of the color information
contained in our multi-band optical light curves should help to
clarify whether we are observing cool or hot spots in each object.
In addition, we identify two possible examples of ``white-light"
flares, perhaps corresponding to the impulsive heating phase in
solar-analog flares. 

However, these cases represent at most $\sim 5$\% of our study sample. 
The remaining $\sim 95$\% of our study
sample exhibits no such time-correlated variability. Evidently,
time-correlated optical and X-ray variability is rare in PMS stars.
We conclude that the sites of optical and X-ray variability in PMS
stars are not---for the vast majority of stars in our study 
sample---instantaneously one and the same.
Surface accretion shocks and cool magnetic spots
are evidently not the exclusive or dominant sites of X-ray production 
in most PMS stars.

\acknowledgments
It is a pleasure to acknowledge the {\it Chandra} Orion Ultradeep
Project (COUP) team, supported by {\it Chandra} Guest Observer grant
SAO GO3--4009A (E.\ Feigelson, PI). This work is also supported by
NSF grant AST--0349075 to K.G.S. We wish to thank C.\
Deliyannis for generously sharing his WIYN 0.9-m time to make the
optical observations presented here possible. We also thank S.\
Wolk for assisting in the planning for the optical monitoring runs.
Facility: CXO(ACIS)

\clearpage

\begin{deluxetable}{rccl}
\tablecolumns{4}
\tablewidth{0pt}
\tabletypesize{\scriptsize}
\tablecaption{Observing log of optical data\label{obs-table}}
\tablehead{
\colhead{UT Date} & \colhead{Filter} & 
\colhead{$N_{\rm obs}$\tablenotemark{a}} & \colhead{Julian Dates} 
}
\startdata
\cutinhead{KPNO}
2003 Jan 15 & $B$ long & 4 &  2452654.69829 \\
            &          &   &  2452654.74710 \\
            &          &   &  2452654.78729 \\
            &          &   &  2452654.82935 \\
            & $B$ short& 4 &  2452654.69343 \\ 
            &          &   &  2452654.74240 \\ 
            &          &   &  2452654.78264 \\ 
            &          &   &  2452654.82473 \\ 
            & $V$ long & 4 &  2452654.68066 \\
            &          &   &  2452654.72999 \\
            &          &   &  2452654.77007 \\
            &          &   &  2452654.81229 \\
            & $V$ short& 4 &  2452654.67623 \\
            &          &   &  2452654.72604 \\
            &          &   &  2452654.76613 \\
            &          &   &  2452654.80839 \\
            & $R$ long & 4 &  2452654.68900 \\ 
            &          &   &  2452654.73799 \\ 
            &          &   &  2452654.77822 \\ 
            &          &   &  2452654.82024 \\ 
            & $R$ short& 4 &  2452654.68497 \\ 
            &          &   &  2452654.73398 \\ 
            &          &   &  2452654.77403 \\ 
            &          &   &  2452654.81634 \\ 
            & $I$ long & 4 &  2452654.70725 \\ 
            &          &   &  2452654.75627 \\ 
            &          &   &  2452654.79855 \\ 
            &          &   &  2452654.83816 \\ 
            & $I$ short& 5 &  2452654.70319 \\ 
            &          &   &  2452654.75232 \\ 
            &          &   &  2452654.79217 \\ 
            &          &   &  2452654.79445 \\ 
            &          &   &  2452654.83439 \\ 
2003 Jan 16 & $B$ long & 5 &  2452655.70634 \\ 
            &          &   &  2452655.75586 \\ 
            &          &   &  2452655.79487 \\ 
            &          &   &  2452655.83286 \\ 
            &          &   &  2452655.87096 \\ 
            & $B$ short& 5 &  2452655.70174 \\ 
            &          &   &  2452655.75128 \\ 
            &          &   &  2452655.79037 \\ 
            &          &   &  2452655.82835 \\ 
            &          &   &  2452655.86639 \\ 
            & $V$ long & 5 &  2452655.71509 \\
            &          &   &  2452655.76489 \\
            &          &   &  2452655.80322 \\
            &          &   &  2452655.84131 \\
            &          &   &  2452655.87962 \\
            & $V$ short& 5 &  2452655.71109 \\
            &          &   &  2452655.76104 \\
            &          &   &  2452655.79944 \\
            &          &   &  2452655.83746 \\
            &          &   &  2452655.87558 \\
            & $R$ long & 5 &  2452655.72068 \\
            &          &   &  2452655.77256 \\
            &          &   &  2452655.81079 \\
            &          &   &  2452655.84938 \\
            &          &   &  2452655.88714 \\
            & $R$ short& 5 &  2452655.72455 \\ 
            &          &   &  2452655.76876 \\ 
            &          &   &  2452655.80707 \\ 
            &          &   &  2452655.84568 \\ 
            &          &   &  2452655.88341 \\ 
            & $I$ long & 5 &  2452655.73077 \\
            &          &   &  2452655.78041 \\
            &          &   &  2452655.81872 \\
            &          &   &  2452655.85701 \\
            &          &   &  2452655.89490 \\
            & $I$ short& 5 &  2452655.72684 \\ 
            &          &   &  2452655.77667 \\ 
            &          &   &  2452655.81491 \\ 
            &          &   &  2452655.85327 \\ 
            &          &   &  2452655.89116 \\ 
2003 Jan 17 & $B$ long & 7 &  2452656.64191 \\
            &          &   &  2452656.68023 \\
            &          &   &  2452656.72500 \\
            &          &   &  2452656.76363 \\
            &          &   &  2452656.80270 \\
            &          &   &  2452656.84125 \\
            &          &   &  2452656.88129 \\
            & $B$ short& 7 &  2452656.63677 \\
            &          &   &  2452656.67572 \\
            &          &   &  2452656.72057 \\
            &          &   &  2452656.75909 \\
            &          &   &  2452656.79805 \\
            &          &   &  2452656.83668 \\
            &          &   &  2452656.87652 \\
            & $V$ long & 7 &  2452656.65095 \\
            &          &   &  2452656.69519 \\
            &          &   &  2452656.73391 \\
            &          &   &  2452656.77238 \\
            &          &   &  2452656.81125 \\
            &          &   &  2452656.85048 \\
            &          &   &  2452656.89037 \\
            & $V$ short& 7 &  2452656.64706 \\ 
            &          &   &  2452656.68480 \\ 
            &          &   &  2452656.72993 \\ 
            &          &   &  2452656.76837 \\ 
            &          &   &  2452656.80753 \\ 
            &          &   &  2452656.84657 \\ 
            &          &   &  2452656.88620 \\ 
            & $R$ long & 7 &  2452656.65643 \\
            &          &   &  2452656.70081 \\
            &          &   &  2452656.74172 \\
            &          &   &  2452656.78003 \\
            &          &   &  2452656.81670 \\
            &          &   &  2452656.85609 \\
            &          &   &  2452656.89592 \\
            & $R$ short& 7 &  2452656.66025 \\
            &          &   &  2452656.70467 \\
            &          &   &  2452656.73783 \\
            &          &   &  2452656.77625 \\
            &          &   &  2452656.82052 \\
            &          &   &  2452656.86029 \\
            &          &   &  2452656.90020 \\
            & $I$ long & 7 &  2452656.66661 \\
            &          &   &  2452656.71092 \\
            &          &   &  2452656.74953 \\
            &          &   &  2452656.78802 \\
            &          &   &  2452656.82649 \\
            &          &   &  2452656.86648 \\
            &          &   &  2452656.90659 \\
            & $I$ short& 7 &  2452656.66254 \\
            &          &   &  2452656.70706 \\
            &          &   &  2452656.74577 \\
            &          &   &  2452656.78419 \\
            &          &   &  2452656.82269 \\
            &          &   &  2452656.86271 \\
            &          &   &  2452656.90252 \\
2003 Jan 19 & $B$ long & 1 &  2452658.80562 \\
            & $B$ short& 3 &  2452658.80091 \\
            &          &   &  2452658.84655 \\
            &          &   &  2452658.88929 \\
            & $V$ short& 2 &  2452658.81034 \\
            &          &   &  2452658.85586 \\
            & $R$ short& 1 &  2452658.81810 \\
            & $I$ short& 2 &  2452658.82642 \\
            &          &   &  2452658.87448 \\
2003 Jan 20 & $B$ long & 3 &  2452659.69923 \\ 
            &          &   &  2452659.73927 \\ 
            &          &   &  2452659.78394 \\ 
            & $B$ short& 3 &  2452659.69449 \\ 
            &          &   &  2452659.73461 \\ 
            &          &   &  2452659.77938 \\ 
            & $V$ long & 3 &  2452659.70758 \\ 
            &          &   &  2452659.74787 \\ 
            &          &   &  2452659.79236 \\ 
            & $V$ short& 3 &  2452659.70375 \\
            &          &   &  2452659.74389 \\
            &          &   &  2452659.78848 \\
            & $R$ long & 3 &  2452659.71525 \\
            &          &   &  2452659.75586 \\
            &          &   &  2452659.80028 \\
            & $R$ short& 3 &  2452659.71140 \\
            &          &   &  2452659.75189 \\
            &          &   &  2452659.79629 \\
            & $I$ long & 3 &  2452659.72508 \\
            &          &   &  2452659.76987 \\
            &          &   &  2452659.80822 \\
            & $I$ short& 3 &  2452659.72127 \\
            &          &   &  2452659.76597 \\
            &          &   &  2452659.80436 \\
2003 Jan 21 & $B$ short& 1 &  2452660.87434 \\
            & $V$ short& 1 &  2452660.88349 \\
            & $I$ short& 1 &  2452660.89865 \\
2003 Jan 22 & $B$ long & 1 &  2452661.70622 \\
            & $B$ short& 2 &  2452661.70159 \\
            &          &   &  2452661.86023 \\
            & $V$ long & 1 &  2452661.83375 \\
            & $V$ short& 1 &  2452661.71098 \\
            & $R$ long & 1 &  2452661.84161 \\
            & $R$ short& 1 &  2452661.83778 \\
            & $I$ short& 1 &  2452661.84603 \\
2003 Jan 23 & $B$ long & 4 &  2452662.63928 \\
            &          &   &  2452662.67797 \\
            &          &   &  2452662.71604 \\
            &          &   &  2452662.75484 \\
            & $B$ short& 5 &  2452662.63462 \\
            &          &   &  2452662.67339 \\
            &          &   &  2452662.71156 \\
            &          &   &  2452662.75033 \\
            &          &   &  2452662.79970 \\
            & $V$ long & 6 &  2452662.64782 \\
            &          &   &  2452662.68635 \\
            &          &   &  2452662.72483 \\
            &          &   &  2452662.76400 \\
            &          &   &  2452662.81349 \\
            &          &   &  2452662.85621 \\
            & $V$ short& 6 &  2452662.64387 \\
            &          &   &  2452662.68253 \\
            &          &   &  2452662.72065 \\
            &          &   &  2452662.75986 \\
            &          &   &  2452662.80971 \\
            &          &   &  2452662.85213 \\
            & $R$ long & 5 &  2452662.65582 \\
            &          &   &  2452662.69412 \\
            &          &   &  2452662.73251 \\
            &          &   &  2452662.77315 \\
            &          &   &  2452662.82160 \\
            & $R$ short& 5 &  2452662.65175 \\
            &          &   &  2452662.69017 \\
            &          &   &  2452662.72875 \\
            &          &   &  2452662.76936 \\
            &          &   &  2452662.86015 \\
            & $I$ long & 4 &  2452662.66371 \\
            &          &   &  2452662.74026 \\
            &          &   &  2452662.78107 \\
            &          &   &  2452662.82975 \\
            & $I$ short& 4 &  2452662.65992 \\
            &          &   &  2452662.69818 \\
            &          &   &  2452662.73643 \\
            &          &   &  2452662.77733 \\
2003 Jan 25 & $B$ long & 1 &  2452664.81536 \\
            & $B$ short& 2 &  2452664.81050 \\
            &          &   &  2452664.85198 \\
            & $V$ long & 1 &  2452664.82516 \\
            & $V$ short& 1 &  2452664.82035 \\
            & $R$ long & 1 &  2452664.83348 \\
            & $R$ short& 1 &  2452664.82927 \\
            & $I$ long & 1 &  2452664.84101 \\
            & $I$ short& 1 &  2452664.83732 \\
\cutinhead{Loiano}
2003 Jan 15 & $B$ long & 3 &  2452655.31530 \\
            &          &   &  2452655.37790 \\
            &          &   &  2452655.43330 \\
            & $B$ short& 4 &  2452655.31070 \\
            &          &   &  2452655.31140 \\
            &          &   &  2452655.33080 \\
            &          &   &  2452655.43200 \\
            & $V$ long & 3 &  2452655.33870 \\
            &          &   &  2452655.40080 \\
            &          &   &  2452655.44700 \\
            & $V$ short& 2 &  2452655.39840 \\
            &          &   &  2452655.44550 \\
            & $R$ long & 3 &  2452655.34910 \\
            &          &   &  2452655.41040 \\
            &          &   &  2452655.45690 \\
            & $R$ short& 2 &  2452655.40820 \\
            &          &   &  2452655.45560 \\
            & $I$ long & 2 &  2452655.36310 \\
            &          &   &  2452655.42120 \\
            & $I$ short& 2 &  2452655.37050 \\
            &          &   &  2452655.41910 \\
2003 Jan 16 & $I$ long & 1 &  2452656.46640 \\
            & $I$ short& 1 &  2452656.46530 \\
2003 Jan 18 & $B$ long & 3 &  2452658.31120 \\
            &          &   &  2452658.36380 \\
            &          &   &  2452658.41440 \\
            & $B$ short& 3 &  2452658.31000 \\
            &          &   &  2452658.36230 \\
            &          &   &  2452658.41320 \\
            & $V$ long & 3 &  2452658.32980 \\
            &          &   &  2452658.37810 \\
            &          &   &  2452658.43280 \\
            & $V$ short& 3 &  2452658.32860 \\
            &          &   &  2452658.37690 \\
            &          &   &  2452658.42740 \\
            & $R$ long & 3 &  2452658.34020 \\
            &          &   &  2452658.38810 \\
            &          &   &  2452658.44180 \\
            & $R$ short& 3 &  2452658.33900 \\
            &          &   &  2452658.38670 \\
            &          &   &  2452658.44060 \\
            & $I$ long & 3 &  2452658.34880 \\
            &          &   &  2452658.39840 \\
            &          &   &  2452658.45470 \\
            & $I$ short& 2 &  2452658.39700 \\
            &          &   &  2452658.45020 \\
2003 Jan 19 & $B$ long & 4 &  2452659.28790 \\
            &          &   &  2452659.33810 \\
            &          &   &  2452659.38800 \\
            &          &   &  2452659.46500 \\
            & $B$ short& 1 &  2452659.46330 \\
            & $V$ long & 4 &  2452659.30670 \\
            &          &   &  2452659.35630 \\
            &          &   &  2452659.40160 \\
            &          &   &  2452659.48030 \\
            & $V$ short& 4 &  2452659.30560 \\
            &          &   &  2452659.35520 \\
            &          &   &  2452659.39990 \\
            &          &   &  2452659.47910 \\
            & $R$ long & 4 &  2452659.31510 \\
            &          &   &  2452659.36440 \\
            &          &   &  2452659.41030 \\
            &          &   &  2452659.48820 \\
            & $R$ short& 4 &  2452659.31400 \\
            &          &   &  2452659.36320 \\
            &          &   &  2452659.40920 \\
            &          &   &  2452659.48710 \\
            & $I$ long & 4 &  2452659.32480 \\
            &          &   &  2452659.37350 \\
            &          &   &  2452659.41900 \\
            &          &   &  2452659.49700 \\
            & $I$ short& 4 &  2452659.32300 \\
            &          &   &  2452659.37210 \\
            &          &   &  2452659.41790 \\
            &          &   &  2452659.49570 \\
2003 Jan 22 & $B$ long & 2 &  2452662.29290 \\
            &          &   &  2452662.36200 \\
            & $B$ short& 2 &  2452662.29150 \\
            &          &   &  2452662.35710 \\
            & $V$ long & 1 &  2452662.30880 \\
            & $V$ short& 1 &  2452662.30770 \\
            & $R$ long & 1 &  2452662.31990 \\
            & $R$ short& 1 &  2452662.31870 \\
            & $I$ long & 1 &  2452662.33070 \\
            & $I$ short& 1 &  2452662.32950 \\
\enddata
\tablenotetext{a}{Number of observations}
\end{deluxetable}

\clearpage

\begin{deluxetable}{rllrcrcclc}
\tablecolumns{10}
\tablewidth{0pt}
\tabletypesize{\scriptsize}
\tablecaption{Candidate cases of time-correlated 
X-ray and optical variability\label{correlated-table}}
\tablehead{
\colhead{COUP} & \colhead{LC$_{\rm opt}$\tablenotemark{a}} & 
\colhead{LC$_{\rm corr}$\tablenotemark{b}} &
\colhead{\ion{Ca}{2}\tablenotemark{c}} & \colhead{Sign\tablenotemark{d}} &
\colhead{$\prot$\tablenotemark{e}} & \colhead{$P_X$\tablenotemark{f}} &
\colhead{Group\tablenotemark{g}}
}
\startdata
  28 & $BVRI$ & $B$ &   1.6 & $+$ & 4.41 & \nodata & A \\
  54 & $BVRI$ & $B$ &  $-$1.0 & $-$ & \nodata & \nodata & A \\
  97 & \phm{$BVR$}$I$ & \phm{$BVR$}$I$ &  $-$3.4 & $-$ & \nodata & \nodata & \nodata \\
 112 & $BVRI$ & $VRI$ &  $-$0.7 & $+$ & 6.52 & Poss. & AB \\
 122 & $BVRI$ & $BVRI$ &   0.0 & $+$ & 9.19 & Poss. & AB \\
 139 & $BVRI$ & \phm{$BVR$}$I$ &   0.9 & $-$ & 9.04 & $\frac{1}{2}$ & B \\
 147 & \phm{$BV$}$RI$ & \phm{$BV$}$RI$ &  $-$1.0 & $+$ & \nodata & \nodata & B \\
 152 & $BVRI$ & \phm{$BV$}$R$ &  \nodata & $-$ & \nodata & \nodata & A \\
 161 & $BVRI$ & \phm{$BV$}$R$ &   1.4 & $-$ & 5.46 & 1 & B \\
 197 & $BVRI$ & \phm{$B$}$VR$ &  \nodata & $+$ & \nodata & \nodata & \nodata \\
 226 & $BVRI$ & $BVI$ &   1.2 & $+$ & 10.98 & 1 & B \\
 241 & $BVRI$ & \phm{$B$}$V$ &   2.8 & $+$ & 9.81 & \nodata & \nodata \\
 250 & $BVRI$ & \phm{$BVR$}$I$ &  $-$9.6 & $+$ & 6.76 & 1 & ABC \\
 298 & \phm{$B$}$VRI$ & \phm{$BVR$}$I$ &   0.0 & $-$ & 6.14 & \nodata & A \\
 309 & \phm{$BV$}$RI$ & \phm{$BV$}$RI$ &  \nodata & $+$ & \nodata & \nodata & \nodata \\
 325 & $BVRI$ & \phm{$BVR$}$I$ &   0.0 & $+$ & \nodata & \nodata & A \\
 346 & \phm{$B$}$VRI$ & \phm{$BVR$}$I$ &   0.0 & $+$ & \nodata & \nodata & \nodata \\
 501 & $BVRI$ & \phm{$B$}$V$ & $-$25.2 & $+$ & 16.33 & Poss. & AB \\
 566 & \phm{$B$}$VRI$ & \phm{$BVR$}$I$ &  \nodata & $+$ & \nodata & \nodata & AB \\
 597 & $BVRI$ & $BVRI$ &   4.5 & $+$ & \nodata & \nodata & A \\
 718 & $BVRI$ & \phm{$BV$}$R$ &  \nodata & $-$ & 5.74 & \nodata & AB \\
 816 & \phm{$BVR$}$I$ & \phm{$BVR$}$I$ &  \nodata & $+$ & \nodata & \nodata & C \\
1071 & $BVRI$ & \phm{$BVR$}$I$ &   1.6 & $+$ & 0.96 & \nodata & B \\
1076 & $BVRI$ & $B$ &  \nodata & $+$ & 11.68 & \nodata & \nodata \\
1143 & $BVRI$ & $B$ &  \nodata & $-$ & 13.00 & \nodata & B \\
1212 & \phm{$BV$}$RI$ & \phm{$BV$}$R$ &  \nodata & $+$ & \nodata & \nodata & \nodata \\
1252 & $BVRI$ & \phm{$BVR$}$I$ &   1.6 & $-$ & 11.68 & \nodata & \nodata \\
1264 & $BVRI$ & \phm{$BVR$}$I$ &  $-$1.8 & $-$ & 6.63 & \nodata & \nodata \\
1292 & $BVRI$ & $BI$ &  $-$3.0 & $+$ & \nodata & \nodata & AB \\
1309 & $BVRI$ & \phm{$B$}$V$ &  \nodata & $-$ & \nodata & \nodata & \nodata \\
1316 & $BVRI$ & \phm{$B$}$VR$ &  $-$1.8 & $-$ & \nodata & \nodata & A \\
1328 & \phm{$B$}$VRI$ & \phm{$B$}$V$ &  \nodata & $-$ & \nodata & \nodata & \nodata \\
1355 & \phm{$B$}$VRI$ & \phm{$BVR$}$I$ &   0.0 & $-$ & 10.36 & 1 & B \\
1384 & $BVRI$ & $BI$ &   1.9 & $-$ & 2.85 & 1 & A \\
1387 & $BVRI$ & $BVRI$ &   0.0 & $-$ & 6.42 & \nodata & A \\
1463 & $BVRI$ & \phm{$B$}$VR$ &  \nodata & $-$ & 10.66 & 1 & B \\
1521 & $BVRI$ & $BVRI$ &  \nodata & $-$ & \nodata & \nodata & B \\
1572 & \phm{$B$}$VRI$ & \phm{$BVR$}$I$ &   0.0 & $-$ & \nodata & \nodata & \nodata \\
1590 & $BVRI$ & \phm{$BVR$}$I$ &  $-$1.2 & $-$ & 7.10 & $\frac{1}{2}$ & B \\
1608 & $BVRI$ & \phm{$BV$}$R$ &  $-$1.3 & $+$ & 4.21 & \nodata & A \\
\enddata
\tablenotetext{a}{Optical light curves included in database.}
\tablenotetext{b}{Optical light curves time-correlated with
the X-ray light curve as indicated by a Kendall's $\tau$ test 
(\S \ref{corr}).}
\tablenotetext{c}{Equivalent width in \AA\ of \ion{Ca}{2} IR triplet lines.
Stars with \ewca\ $\le 1$\AA\ are identified 
as active accretors while those with \ewca\ $\ge 1$\AA\ are identified as
non-accretors (see \S \ref{corr-accretion}).}
\tablenotetext{d}{Sign of the Kendall's $\tau$ correlation statistic from
for those light curves with statistically significant correlations.}
\tablenotetext{e}{Variability period in days from previous optical monitoring 
reported in the literature. Typically, $\prot$ is assumed to reflect the 
rotation period of the star.}
\tablenotetext{f}{Flag from \citet{flac-spots} indicating period of X-ray
light curve relative to optical period: A value of 1 indicates that the 
X-ray period
is equal to the optical period, $\frac{1}{2}$ indicates that the X-ray period
is equal to one-half of the optical period. `Poss.' indicates
a possibly periodic X-ray light curve.}
\tablenotetext{g}{Classification of objects based on the nature of the
observed time-correlated X-ray/optical variability; see \S\ref{credible}.}
\end{deluxetable}

\clearpage

\clearpage

\clearpage

\clearpage

\clearpage

\clearpage

\clearpage

\clearpage

\clearpage

\clearpage

\clearpage

\clearpage

\clearpage

\clearpage

\clearpage

\clearpage

\clearpage

\clearpage

\clearpage

\clearpage

\clearpage

\clearpage

\clearpage

\clearpage

\clearpage

\clearpage

\clearpage

\clearpage

\clearpage

\clearpage

\clearpage

\clearpage

\clearpage

\clearpage

\clearpage

\clearpage

\clearpage

\clearpage

\clearpage

\clearpage

\clearpage

\clearpage

\clearpage

\clearpage

\clearpage

\clearpage

\clearpage

\end{document}